\documentclass[useAMS, usenatbib]{mn2e}

\usepackage{aasmacros}
\usepackage{amssymb} 
\usepackage{subcaption} 
\usepackage{booktabs} 
\usepackage{multirow} 
\usepackage{amsfonts}
\usepackage{lmodern}
\usepackage{amsmath}
\usepackage{float}
\usepackage{graphicx}

\title[Driving gas shells with radiation pressure on dust]{Driving gas shells with radiation pressure on dust in radiation-hydrodynamic simulations} \author[Costa et al.]{Tiago Costa$^{1}$\footnotemark[1], Joakim Rosdahl$^{1, 2}$, Debora Sijacki$^{3}$ and Martin G. Haehnelt$^{3}$\\\
  $^{1}$Leiden Observatory, Leiden University, PO Box $9513$, NL-$2300$ RA Leiden, the Netherlands \\
  $^{2}$CRAL, Universit\'e de Lyon I, CNRS UMR 5574, ENS-Lyon, 9 Avenue Charles Andr\'e, 69561, Saint-Genis-Laval, France\\
  $^{3}$Institute of Astronomy and Kavli Institute for Cosmology,
  University of Cambridge, Madingley Road, Cambridge CB$3$ $0$HA, UK}

\begin{document}
 
\maketitle

\begin{abstract}
We present radiation-hydrodynamic simulations of radiatively-driven gas shells launched by bright active galactic nuclei (AGN) in isolated dark matter haloes.
Our goals are (1) to investigate the ability of AGN radiation pressure on dust to launch galactic outflows and (2) to constrain the efficiency of infrared (IR) multi-scattering in boosting outflow acceleration. 
Our simulations are performed with the radiation-hydrodynamic code {\sc Ramses-RT} and include both single- and multi-scattered radiation pressure from an AGN, radiative cooling and self-gravity. 
Since outflowing shells always eventually become transparent to the incident radiation field, outflows that sweep up all intervening gas are likely to remain gravitationally bound to their halo even at high AGN luminosities.
The expansion of outflowing shells is well described by simple analytic models as long as the shells are mildly optically thick to IR radiation.
In this case, an enhancement in the acceleration of shells through IR multi-scattering occurs as predicted, i.e. a force $\dot{P} \approx \tau_{\rm IR} L/c$ is exerted on the gas.
For high optical depths $\tau_{\rm IR} \gtrsim 50$, however, momentum transfer between outflowing optically thick gas and IR radiation is rapidly suppressed, even if the radiation is efficiently confined. 
At high $\tau_{\rm IR}$, the characteristic flow time becomes shorter than the required trapping time of IR radiation such that the momentum flux $\dot{P} \ll \tau_{\rm IR} L/c$.
We argue that while unlikely to unbind massive galactic gaseous haloes, AGN radiation pressure on dust could play an important role in regulating star formation and black hole accretion in the nuclei of massive compact galaxies at high redshift.
\end{abstract}


\renewcommand{\thefootnote}{\fnsymbol{footnote}}
\footnotetext[1]{E-mail: costa@strw.leidenuniv.nl}

\section{Introduction}

In order to grow a supermassive black hole (SMBH) with mass $M_{\rm BH}$, an amount $\eta/(1 - \eta) M_{\rm BH} c^2$ of energy is released via gravitational accretion. 
If the SMBH grows with a mean radiative efficiency $\eta \, = \, 0.1$ at the centre of a galactic bulge with mass $M_{\rm *} \,=\, 500 \times M_{\rm BH}$ \citep{Kormendy:13} and velocity dispersion $\sigma_{\rm 250} \,=\, \sigma/ (250 \, \mathrm{km \, s^{-1}})$, the energy released exceeds the binding energy of the bulge by a factor $\approx 300 \sigma_{\rm 250}^{-2}$ \citep[e.g.][]{Fabian:12}.
Hence, the coupling between energy released through active galactic nucleus (AGN) activity and the interstellar/intergalactic medium (`AGN feedback') clearly has the potential to profoundly affect galaxy evolution.

Consequently, energy and momentum injection into the interstellar medium (ISM) through AGN feedback is often invoked to tackle a wide range of outstanding problems in galaxy formation. 
AGN feedback is thought to result in the suppression of star formation in massive galaxies and to explain the origin of the observed population of quiescent galaxies \citep{Scannapieco:04, Churazov:05, Bower:06}. In massive groups and galaxy clusters, AGN feedback is thought to regulate cooling flows \citep{Sijacki:06, Vernaleo:06, McCarthy:10, Gaspari:11, Teyssier:11, Dubois:11, Martizzi:12, Li:14, Barai:16}. Another potential role is to limit black hole growth \citep{Haehnelt:98, Wyithe:03, DiMatteo:05, Springel:05c}, possibly establishing the scaling relations that link $M_{\rm BH}$ to galactic properties \citep{Silk:98, King:03, Murray:05}.
AGN feedback is, hence, invoked in virtually every semi-analytical model \citep[e.g.][]{Kauffmann:00, Croton:06, Henriques:15} and large-scale cosmological simulation \citep[e.g.][]{Sijacki:07, DiMatteo:08, Booth:09, Dubois:12, Vogelsberger:13, Sijacki:15, Schaye:15, Khandai:15, Volonteri:16, Weinberger:17} to account for the properties of massive galaxies and SMBHs in the local and high redshift Universe.

In broad terms, AGN energy and momentum deposition into the vicinity of a rapidly accreting SMBH generates steep pressure gradients that can accelerate large masses of gas outwards.
A large number of such outflows has now been detected at a wide range of spatial scales.
At the smallest scales ($\lesssim 1 \-- 100 \, \rm pc$), detections of blue-shifted X-ray absorption lines indicate the existence of wide-angle winds with speeds $100 \-- 1000 \, \rm km \, s^{-1}$ \citep{Kaastra:00, McKernan:07, Gofford:11, Detmers:11, Reeves:13} and, in more extreme cases, up to a few tenths of the speed of light \citep{Pounds:03, Reeves:09, Cappi:09, Tombesi:12}.
At larger scales, the detection of broad emission and absorption features in the spectra of systems with luminosities dominated by AGN in ULIRGs/quasars \citep{Sturm:11, Aalto:12, Rupke:13, Liu:13, Genzel:14, Carniani:15, Tombesi:15, Zakamska:16, WIlliams:17, Bischetti:17} as well as in radio-loud AGN \citep{Tadhunter:14, Morganti:15} reveals the existence of mass-loaded ($M_{\rm out} \gtrsim 10^8 \, \rm M_\odot$), fast ($v_{\rm out}  \,=\, 1000 \-- 3000 \, \rm km \, s^{-1}$) AGN-driven outflows extended over up to several kpc or even up to tens of $\rm kpc$ \citep{Cicone:15}.
According to these observations, AGN-driven outflows typically have a multi-phase structure, comprising hot ionised, warm and cold molecular components. 
Importantly, the cold molecular phase often appears to contain momentum fluxes $> L/c$ and kinetic luminosities $\lesssim 0.05 L$ \citep[see][for a recent compilation of AGN-driven outflows]{Fiore:17}, where $L$ is the AGN luminosity \citep[though see][]{Husemann:16}.

Whether observed outflows couple to a sufficiently large volume of the ISM to directly suppress star formation remains an open question.
Indirect support is provided by the high inferred outflow rates in AGN hosts, which often exceed the star formation rates of host galaxies \citep[e.g.][]{Cicone:14, Fiore:17}. 
This scenario is also supported by the observed spatial overlap of gas in H$\alpha$ emission with broad emission lines tracing outflows in circumnuclear regions of high redshift quasars \citep{Cano-Diaz:12}. 
Though star formation might be suppressed, there are also indications that it may not be reduced to negligible levels and may even be enhanced in the outskirts of galactic nuclei \citep{Cresci:15, Carniani:16}.

The efficiency at which AGN feedback operates is largely unconstrained even from a theoretical point of view.
In the case of `radiative mode' feedback, this is mainly due to a poor understanding of the physical processes that govern the coupling between the AGN radiation field and the ISM.
Various theoretical models have been developed over the last two decades.
In one class of models, feedback relies on the interaction between a fast nuclear wind launched from accretion disc/dusty torus scales ($\lesssim 10 \, \rm pc$) and the ISM.
Such winds decelerate violently through a strong reverse shock, while driving a blast wave through the interstellar medium.
In cases in which the resulting outflow is energy-conserving (if the shocked wind preserves its thermal energy), such models are able to account for observations of momentum boosts $\sim 20 L/c$ and high kinetic luminosities $\lesssim 0.05 L$ \citep{King:03, King:05, Silk:10, Zubovas:12, Faucher-Giguere:12, Wagner:13, Costa:14, Hopkins:16}. 

However, AGN feedback may also proceed via the coupling of radiation to the ambient medium directly (i.e. without an intermediary wind).
Compton heating to super-virial temperatures of $\approx 10^{7} \, \rm K$ \citep{Ciotti:01, Sazonov:05} or even up to $\approx 10^{9} \, \rm K$ for a hard AGN spectrum \citep{Gan:14} has been shown to be effective at regulating accretion flows onto AGN and star formation in galactic nuclei \citep{Hambrick:11, Kim:11}. 

Another possibility is that radiation couples directly to the ISM through radiation pressure on dust at even larger ($\sim \, \rm kpc$) scales \citep{Fabian:99, 
Murray:05, Chattopadhyay:12, Novak:12, Thompson:15}. 
Indeed, most of the radiation generated during the growth of SMBHs is thought to be absorbed by a surrounding envelope of dusty gas \citep{Fabian:99b}.
The optical and ultraviolet (UV) radiation is absorbed and re-emitted at infrared (IR) wavelengths before escaping the galactic nucleus.  
In cases in which the gas configuration in the dusty nucleus has a very high column density ($N \gtrsim 10^{23 - 24} \, \rm cm^{-2}$), the gas becomes optically thick also in the IR. Instead of streaming out, the reprocessed IR photons undergo multiple scatterings.
In this scenario, the net momentum imparted by the AGN radiation field may exceed $L/c$.

Radiation pressure on dust as an AGN feedback mechanism has been investigated analytically \citep{Murray:05, Thompson:15, Ishibashi:15, Ishibashi:16} as well as in radiative transfer calculations in the context of dusty tori \citep[e.g.][]{Proga:04, Roth:12, Namekata:16} and isolated galaxies \citep{Chattopadhyay:12, Bieri:17}. This mechanism has also been explored in cosmological simulations \citep{Debuhr:11, Debuhr:12}, though without radiative transfer and using crude sub-grid approximations for the radiation force. 
In all cases, IR multi-scattering has been identified as essential to ensure enough momentum can be transferred to the ambient gas. 

However, some of the basic predictions of available analytic models have not yet been tested in more realistic simulations.
These predictions are often used to justify sub-grid prescriptions of radiation pressure in hydrodynamic simulations, particularly in the context of stellar feedback \citep[e.g.][]{Hopkins:12, Agertz:13, Aumer:13, Ceverino:14}.
This paper fills that gap.
Our strategy is to take analytic models in which IR trapping is assumed to lead to a radiation force $\tau_{\rm IR} L/c$ and compare their predictions against radiation-hydrodynamic simulations using identical initial conditions and gas configurations.
We thus start by reviewing analytic models for the expansion of radiatively-driven shells in galactic haloes (Section~\ref{sec_physradpress}). In Section~\ref{sec_idealsimulations}, we compare the predictions of analytic models with those of idealised simulations of radiation pressure-driven shells in Navarro-Frenk-White (NFW) haloes \citep{Navarro:97}. We explore the origin of differences identified between our simulations and the analytic models, particularly when the optical depth to the IR is high ($\tau_{\rm IR} \gtrsim 50$). The implications of our findings to the regulation of star formation and black hole accretion in massive galaxies, additional possible consequences of IR-driven winds as well as a discussion of the limitations in our modelling are discussed in Section~\ref{sec_discussion}. Finally, our conclusions are summarised in Section~\ref{sec_conclusions}.

\section{Radiation pressure-driven winds}
\label{sec_physradpress}

In order for radiation pressure to generate a powerful outflow, the central regions of the galaxy hosting the accreting black hole must, at the onset, be optically thick to incoming UV/optical photons.
High optical depths are easily achievable in the presence of dust, thanks to its high absorption cross section to UV/optical photons $\approx 500 \-- 1000 \kappa_{\rm T}$ \citep{Fabian:08} where $\kappa_{\rm T}$ is the Thomson opacity (cross-section per unit mass).  

The combination of high central gas concentrations and high dust production rates expected in starbursting galactic nuclei during rapid black hole growth means that the gas layers surrounding the AGN at $\lesssim 1  \, \rm kpc$ scales can be optically thick even to reprocessed IR radiation.
Interferometric observations of the nuclei of local ULIRGs have unveiled high IR optical depths \citep[e.g. $\tau_{\rm 434 \, \mu m} \,\gtrsim\, 5$ and $\tau_{\rm 2.6 \, mm} \, \approx\,1$ in Arp220,][]{Wilson:14, Scoville:17} 

Assuming spherical symmetry and an isotropic source of radiation with flux $F \, = \, L/(4 \pi R^2)$, where $L$ and $R$ denote the source luminosity and radial distance, respectively, the radiation force $F_{\rm rad}$ from trapped IR photons is
\begin{equation}
F_{\rm rad} \, =\, 4 \pi \int{\frac{\kappa_{\rm IR} F \rho}{c} R^2 dR} \, = \, \frac{L}{c} \int{\kappa_{\rm IR} \rho dR} \,=\, \tau_{\rm IR} \frac{L}{c} \, ,
\label{eq_frad}
\end{equation}
where $\rho$ is the density of the gas medium through which the radiation propagates, $\kappa_{\rm IR}$ is the dust opacity to IR radiation, $c$ is the speed of light in vacuum and we have assumed gas to be optically thick at all wavelengths.

In other words, the force exerted on the optically thick gas is boosted over the direct radiation force by a factor equal to the IR optical depth.
In Appendix \ref{sec_appendix1}, we present 2D Monte Carlo calculations of IR photons travelling through optically thick gas shells and spheres and develop an intuition of how IR multi-scattering is indeed expected to lead to an enhancement of $\tau_{\rm IR} \frac{L}{c}$, in both cases.

The result shown in Eq.~\ref{eq_frad} has been used to justify radiation forces in excess $L/c$ in numerical simulations of galaxy formation, resulting in strong stellar and AGN feedback. For instance, \citet{Hopkins:12a} measure the local gas column density in star forming clouds and impart a radial momentum $(1 + \tau_{\rm IR}) \frac{L}{c}$ to the surrounding gas. In their simulations, IR radiation pressure, which can exert a force as high as $\sim (50 \-- 100) \frac{L}{c}$ \citep[see Fig. 5 in][]{Hopkins:11}, constitutes a crucial stellar feedback mechanism in giant molecular cloud complexes, particularly in high-density regions of high-redshift galaxies and starbursts \citep[see also][]{Agertz:13, Aumer:13, Ceverino:14}. 
Following similar reasoning, \citet{Ishibashi:15, Ishibashi:16a, Ishibashi:16} explored, analytically, models of radiation pressure-driven shells in which AGN radiation exerts a force, which in their most optimistic estimates, can be as high as $(40 \-- 50) \frac{L}{c}$. If taken at face value, the corresponding outflows, which can propagate at speeds well in excess of $1000 \, \rm km \, s^{-1}$, have bulk properties such as outflow rate $\dot{M}$ and `momentum flux' $\dot{M} v_{\rm out}$ in good agreement with observations of cold AGN-driven winds \citep[e.g.][]{Cicone:14}.

\subsection{Energy- or momentum-driven?}

One of the main constraints on the nature of the physical mechanisms that lead to AGN-driven outflows is whether the flows are energy- or momentum-driven.
The observation of large-scale outflows with momentum fluxes $\dot{P}_{\rm out} \gg L/c$ \citep[e.g.][]{Cicone:14} has led to the conclusion that many AGN outflows are energy-driven.
In the wind picture proposed by \citet{King:03}, energy-driving occurs when the shocked wind fluid preserves its thermal energy.
It does `PdV' work on its surrounding medium, expanding adiabatically as the gas accumulated around its rim is pushed out.
With energy conservation between the fast nuclear wind and the large-scale outflow, the expression $\frac{\dot{P}_{\rm out}}{\dot{P}_{\rm wind}} \approx \sqrt{\frac{\dot{M}_{\rm out}}{\dot{M}_{\rm wind}}} \gtrsim 1$ is satisfied \citep[see also][]{Zubovas:12, Faucher-Giguere:12}, where $\dot{P}_{\rm wind}$ and $\dot{M}_{\rm wind}$ are the momentum flux and mass outflow rates of the inner wind, respectively.
If most of the thermal energy of the shocked wind fluid is radiated away, the only way in which the outflow can continue to accelerate is through the ram pressure of the incident nuclear wind itself. For such a momentum-driven flow, $\frac{\dot{P}_{\rm out}}{\dot{P}_{\rm wind}} \approx 1$ holds instead. 
Clearly, energy-driving results in more powerful large-scale outflows than momentum-driving \citep[see][for a detailed discussion]{Costa:14}.

The nomenclature `energy-' and `momentum-driven' is often incorrectly equated with `energy-' and `momentum injection' in the context of modelling AGN wind feedback in numerical simulations. We here briefly clarify why this can be a false equivalence. A wind may be initiated by imparting momentum on gas elements, as performed by e.g. \citet[][]{Choi:12} or \citet{Angles-Alcazar:17}, only to be thermalised at a high temperature. As long as the resulting over-pressurised bubble does not cool efficiently, its expansion leads to an energy-driven flow, despite it being achieved through `momentum injection'. Thermal energy injection, on the other hand, does not necessarily lead to a purely energy-driven flow if the heated bubble cools rapidly.  
Whether the flow is energy- or momentum-driven depends on the existence of an over-pressurised shocked wind component (as opposed to a shocked ISM component, which may exist in both cases), which retains at least part of its thermal energy.
Note also that the same outflow may undergo energy- and momentum-driven phases at different stages throughout its evolution.

How would radiation pressure-driven outflows compare to momentum- and energy-driven wind-based feedback?
In the picture addressed in this study, work is done by radiation; the hot bubble that leads to a momentum boost in energy-driven models is replaced by a relativistic fluid. 
A momentum flux larger than $L/c$ is achieved through multiple scatterings (see Fig.~\ref{fig_diffusion_schematic}) and can only take place if the gas is optically thick in the IR.
Unlike energy-driven flows of the type discussed in \citet{King:05}, where high momentum boosts can, in principle, occur at arbitrarily high radial distances from an AGN, high momentum boosts due to multi-scattered IR radiation are confined to the innermost regions of galaxies. The likely spatial scales are of the order $\lesssim 1\, \rm kpc$ \citep{Thompson:15}.
We should note that one might expect both nuclear winds and AGN radiation pressure, to some extent, to operate simultaneously, a scenario we investigate in a future study.

\subsection{Analytical background}
\label{sec_analyticalsolutions}

\subsubsection{Radiation pressure-driven wind equation}

We follow the analytical treatment presented in detail in \citet{Thompson:15} and \citet{Ishibashi:15}.
We consider a spherically symmetric galactic halo with total mass profile $M(R)$, a fraction $f_{\rm gas}$ of which is composed of gas.
The gas component is assumed to follow the dark matter distribution exactly.
We further assume that the halo is static and that an AGN is located at its origin, emitting radiation at a \emph{constant} luminosity $L$.
AGN radiation is assumed to couple to the ambient gas via radiation pressure on dust grains and we implicitly assume that gas and dust are coupled hydrodynamically \citep{Murray:05}.
For simplicity, dust is assumed to be evenly mixed with the gas at a fixed dust-to-gas ratio such that every shell of thickness $dR$ has an optical depth $d\tau \,=\, \kappa \rho(R) dR$, where $\kappa$ is the opacity to the incident radiation field.
Supposing that radiation pressure sweeps up all intervening gas into a thin shell of radius $R$, mass $M_{\rm sh}(R)$ and optical depth $\tau (R)$, the momentum equation for the expanding shell is
\begin{equation}
  \frac{d\left[ M_{\rm sh}(R) \dot{R} \right]}{dt} \, = \,  f \left[ \tau(R) \right] \frac{L}{c} - 4 \pi R^2 P_{\rm ext} - \frac{G M_{\rm sh}(R) M(R)}{R^2} \, ,
\label{equation_of_motion}
\end{equation}
where $P_{\rm ext}$ is the external pressure exerted by the ambient medium. 
The factor $f \left[ \tau(R) \right]$ encodes the fraction of momentum in the incident radiation field ($L/c$) that can be transferred to the expanding gas shell.
As in \citet{Thompson:15}, we take 
\begin{equation}
f \left[ \tau(R) \right] \, = \, (1 - e^{-\tau_{\rm UV}}) + \tau_{\rm IR} \, ,
\label{eq_radforce}
\end{equation}
where the first term accounts for single-scattering (UV) momentum transfer (i.e. at most $L/c$ is transferred if the shell is optically thick) and the second term allows for momentum exchange between trapped IR radiation and outflowing gas (see Eq.~\ref{eq_frad}).

The external pressure term can be dropped if $f[\tau] \frac{L}{c} \gg 4 \pi R^2 P_{\rm ext}$.
If $f[\tau] \frac{L}{c} \approx 4 \pi R^2 P_{\rm ext}$, as eventually occurs when the shell decelerates, the `thin shell' assumption breaks down. The shell's outer boundary instead propagates as a weak shock while its inner surface stalls\footnote{As a consequence, analytic solutions based on Eq.~\ref{equation_of_motion} cannot, in general, be exactly reproduced in hydrodynamic simulations, particularly when the halo is set up in hydrostatic equilibrium and has a high sound speed.}. 

If the external pressure is assumed to be negligible, the final equation of motion reads
\begin{equation}
  \frac{d\left[f_{\rm gas} M(R) \dot{R} \right]}{dt} \, = \,  f[\tau (R)] \frac{L}{c} - f_{\rm gas} \frac{G M^2(R)}{R^2} \, ,
\label{equation_of_motion_sim}
\end{equation}
where we have made use of the relation $M_{\rm sh}(R) \,=\, f_{\rm gas} M(R)$ which holds for a halo in which gas follows the dark matter distribution exactly. 

In the following, we examine solutions to Eq.~\ref{equation_of_motion_sim} for isothermal and NFW profiles.

\subsubsection{Wind solutions for an isothermal profile}

We start by examining wind solutions assuming an isothermal profile for both gas and dark matter.
Isothermal profiles are particularly instructive since they are simple to treat mathematically, while providing a good match with observational constraints for the inner density profiles of early-type galaxies \citep[e.g.][]{Treu:10} and with the halo profiles that arise in the rapid gas cooling regime in cosmological simulations \citep[e.g.][]{Duffy:10}.
In the case of a singular isothermal sphere, the total radial density profile is 
\begin{equation}
\rho (R) \, = \, \frac{\sigma^2}{2 \pi G R^2} \, ,
\end{equation}
where $\sigma$ is the velocity dispersion of the halo.
The mass enclosed at radius $R$ is simply
\begin{equation}
M(<R) \, = \, \int_{0}^{R}{4 \pi R^2 \rho(R) dR} \, = \, \frac{2 \sigma^2 R}{G} \, .
\end{equation}

For the gas and dark matter components, the above expressions gain extra factors $f_{\rm gas}$ and $(1 - f_{\rm gas})$, respectively.

\begin{figure}
\centering
\includegraphics[scale = 0.43]{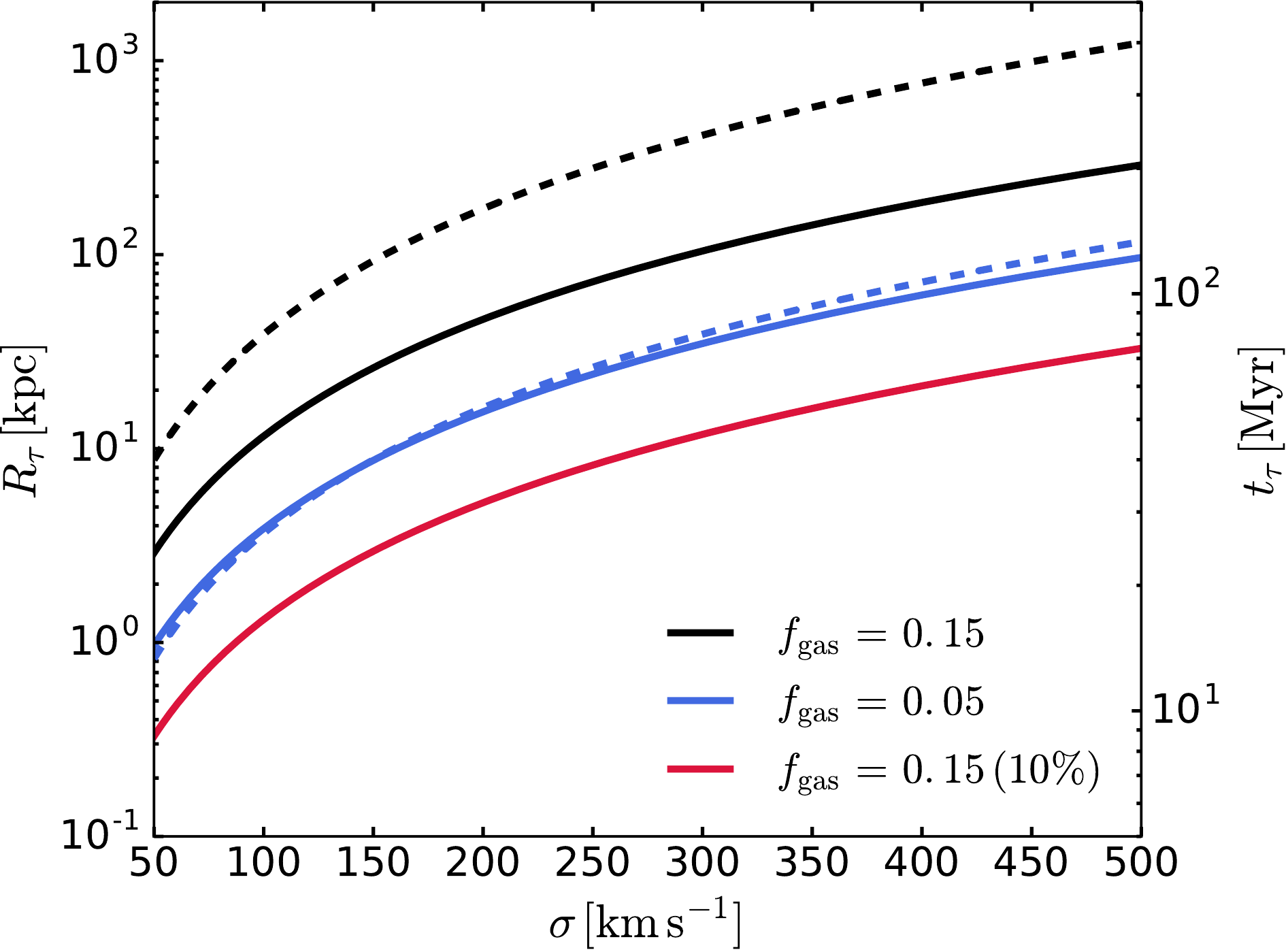}
\caption{Physical values for the transparency radius (solid lines, left-hand y-axis, Eq.~\ref{eq_r_tau}) and the dynamical time measured at the transparency radius (dashed lines, right-hand y-axis, Eq.~\ref{eq_t_tau}) as a function of velocity dispersion for an isothermal halo. Both quantities are used to cast the equation of motion of radiation pressure-driven shells into dimensionless form (Eq.~\ref{equation_of_motion_iso}). We show the transparency radius for a high gas fraction $f_{\rm gas} \,=\, 0.15$ (black) and low gas fraction $f_{\rm gas} \,=\, 0.05$ (blue). The red solid line shows the value corresponding to $10\%$ of the transparency radius, which gives a more realistic measure of the spatial scale at which radiation pressure-driven shells stall (see text). The extremely large transparency radii ($> 100 \, \rm kpc$) seen here result from the unrealistic assumption of constant (high) dust to gas ratio out to arbitrarily large radii (see text).}
\label{fig_physical}
\end{figure}

We now non-dimensionalise the equation of motion.
Isothermal haloes lack a characteristic spatial scale.
Thus, we express the radius of the shell in units of the transparency radius $R_{\rm \tau}$, i.e., the radius for which the UV optical depth $\tau_{\rm UV} \,=\, 1$, such that we write the non-dimensional radius as $R^\prime \, = \, R / R_{\rm \tau}$.
Accordingly, we scale time to units of the dynamical time at $R_{\rm \tau}$ as  $t_{\rm \tau} \, = \, R_{\rm \tau}/(\sqrt{2} \sigma)$, such that the dimensionless time is $t^\prime \, = \, t / t_{\rm \tau}$.
For an isothermal profile, we have
\begin{eqnarray}
R_{\rm \tau} & \,=\, & \frac{f_{\rm gas} \kappa_{\rm UV} \sigma^2}{2 \pi G} \label{eq_r_tau}\\ 
          & \,\approx\, & 26.1 \left( \frac{f_{\rm gas}}{0.15} \right) \left( \frac{\kappa_{\rm UV}}{10^3 \mathrm{cm^2 \, g^{-1}}} \right) \left( \frac{\sigma}{150 \mathrm{km \, s^{-1}}} \right)^2 \, \rm kpc \, , \nonumber
\end{eqnarray}

\begin{figure*}
\includegraphics[scale = 0.47]{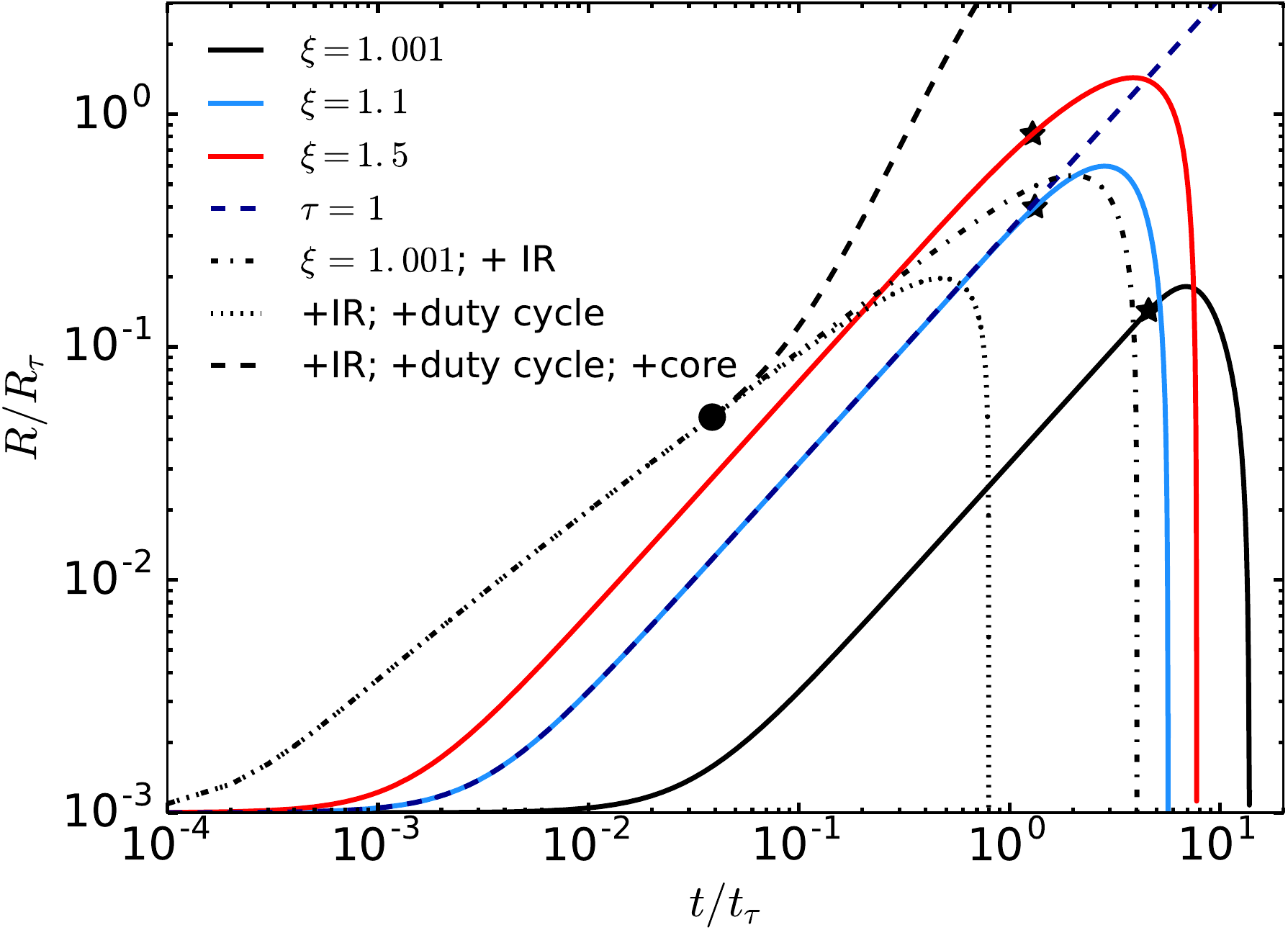}
\includegraphics[scale = 0.47]{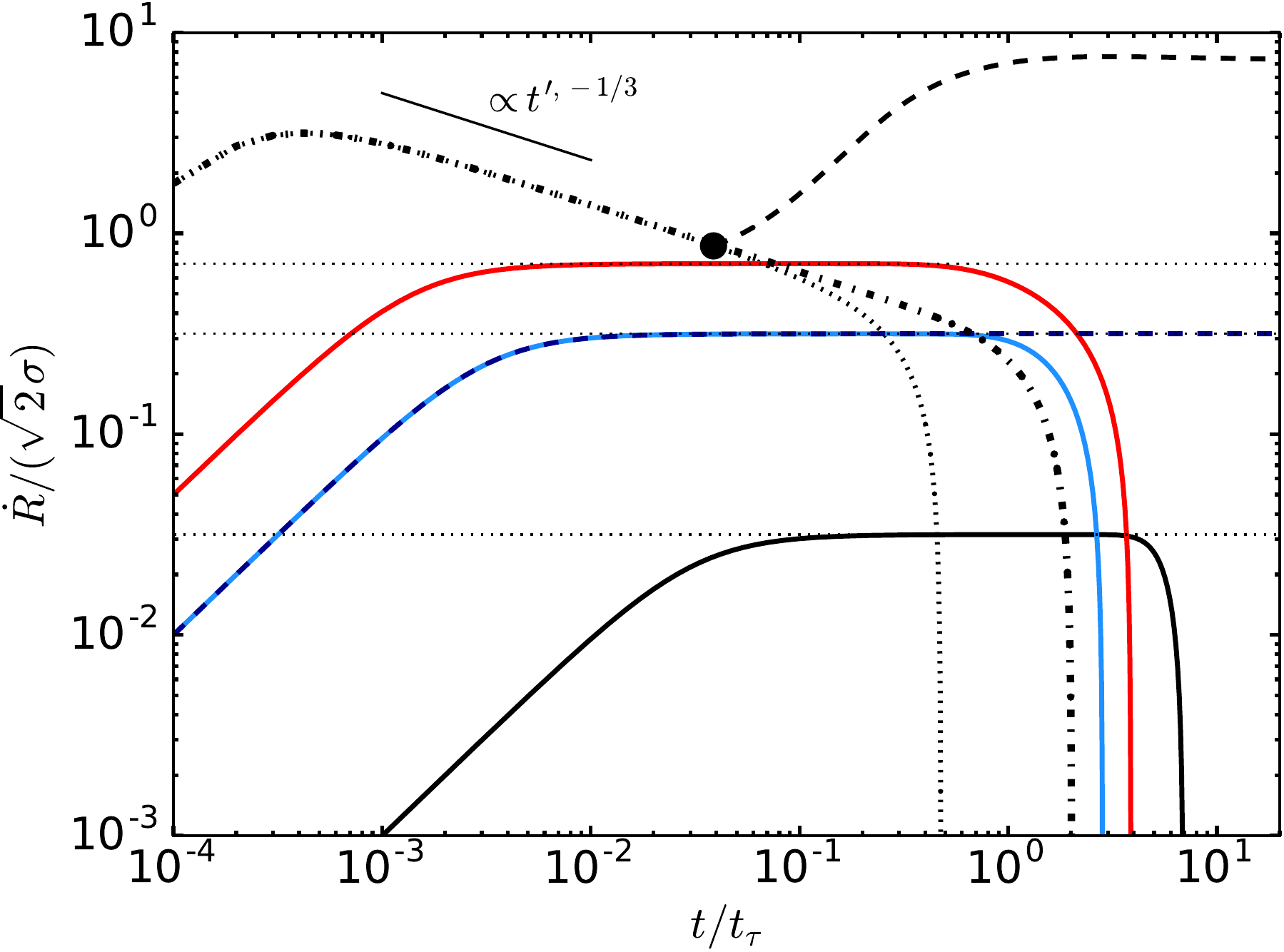}
\caption{Solutions to the equation of motion of radiation pressure-driven shells in an isothermal potential for AGN luminosities $L \,=\, \xi L_{\rm crit}$ (see text). On the left, we show the dimensionless radius of shells as a function of dimensionless time. On the right, we show shell velocities (in units of $\sqrt{2} \sigma$) as a function of time. If a shell was always optically thick to UV radiation and IR radiation force could be neglected, it would be accelerated with a fixed radiation force $L/c$. In this case, the equation admits strictly unbound solutions if $\xi > 1$ ( dark blue dashed line). However, the optical depth of any expanding shell eventually drops below unity and the radiation force declines. All radiation pressure-driven solutions (solid lines) accordingly remain bound to the galactic halo. The black dot-dashed line shows a solution with $\xi \, = \, 1.001$ for which the radiation force is boosted due to multi-scattered IR radiation proportionally to the IR optical depth. The shell expands far more quickly in the central regions, reaching its turn-over radius well before the corresponding single-scattering solution. If the AGN lifetime is taken into account, shells turn around at an even smaller radius, as shown by the dotted black curve, where we assume $L \propto e^{-t/t_{\tau}}$. The dashed black line shows a solution for which the shell sweeps up no mass beyond a radius $R_{\rm c} \,=\, 0.05 R_{\rm \tau}$ (shown by the black circle). Thin dotted lines on the right-hand panel show constant velocity solutions to the equation of motion.}
\label{fig_shell_analytic}
\end{figure*}

\begin{equation}
t_{\rm \tau} \,\approx\, 120 \left( \frac{f_{\rm gas}}{0.15} \right) \left( \frac{\kappa_{\rm UV}}{10^3 \mathrm{cm^2 \, g^{-1}}} \right) \left( \frac{\sigma}{150 \mathrm{km \, s^{-1}}} \right) \, \rm Myr \, .
\label{eq_t_tau}
\end{equation}
Numerical values for the transparency radius and dynamical time as a function of halo velocity dispersion are provided in Fig.~\ref{fig_physical}.
Taken at face value, the solid lines in Fig.~\ref{fig_physical} would imply extremely high transparency radii $\gtrsim 100 \, \rm kpc$, though only because the models presented in \citet{Ishibashi:15} and \citet{Thompson:15} assume, optimistically, the gas to remain dusty to arbitrarily high radial distances from the central galaxy.
This assumption, which should break down due to a drop of gas metallicity and dust in the outer halo and likely due to in-shock dust destruction and thermal sputtering, is favourable to the model because it boosts the optical depths and, hence, the radiation force.
Note, in addition, that for parameters adequate for massive galaxies ($\sigma \,\gtrsim\, 200 \, \rm km \, s^{-1}$), the timescale $t_{\rm \tau} \sim 60 \-- 120 \, \rm Myr$ is much longer than the typical AGN lifetime $\lesssim 1 \, \rm Myr$ \citep[see e.g.][]{Schawinski:15}.

We now define a critical luminosity $L_{\rm crit}$ given by
\begin{equation}
L_{\rm crit} \,=\, \frac{4f_{\rm gas}c\sigma^4}{G} \,\approx\, 4.1 \times 10^{46} \left( \frac{f_{\rm gas}}{0.15} \right) \left( \frac{\sigma}{200 \, \mathrm{km \, s^{-1}}} \right)^4 \, \rm erg \, s^{-1} \, ,
\label{eq_critlum}
\end{equation}
which compares well with the Eddington luminosity for a supermassive black hole lying on the $M_{\rm BH} \-- \sigma$ relation presented in Eq. 7 in \citet{Kormendy:13}, which is
\begin{equation}
L_{\rm Edd} \,=\, \frac{4 \pi G M_{\rm BH} c}{\kappa_{\rm T}} \,\approx\, 4 \times 10^{46} \left( \frac{\sigma}{200 \, \mathrm{km \, s^{-1}}} \right)^{4.38} \, \rm erg \, s^{-1} \, ,
\end{equation}
though note the dependence of Eq.~\ref{eq_critlum} on the gas fraction $f_{\rm gas}$.
Thus, the dimensionless form of Eq. \ref{equation_of_motion_sim} can be expressed as 
\begin{equation}
\frac{d}{dt^\prime} \left[R^\prime \frac{dR^\prime}{dt^\prime} \right] \,=\, f[ \tau] \xi - 1 \, ,
\label{equation_of_motion_iso}
\end{equation}
where we have defined a luminosity ratio $\xi \,=\, L/L_{\rm crit}$.

In isothermal profiles, solutions depend only on the two dimensionless numbers $\xi$ and $\tau(R^{\prime})$. 
The significance of $L_{\rm crit}$ (and hence $\xi$) is clear; supposing  $f[\tau] \, = \, 1$ and $\xi > 1$, Eq.~\ref{equation_of_motion_iso} admits expanding shell solutions with a radius given by $R^\prime \,=\, \sqrt{\left( \xi - 1 \right) t^{\prime,\,2} + 2 \dot{R}^\prime_0 R^\prime_{0} t^{\prime} + R^{\prime, 2}_0}$ \citep[see also][]{King:05}.
Thus, in the absence of variations in the radiation force $f \left[ \tau(R) \right] L/c$, the parameter $\xi$ separates gravitationally bound and unbound shell solutions.
The critical luminosity $L_{\rm crit}$ is thus the luminosity required to launch an unbound shell from an isothermal halo with velocity dispersion $\sigma$.

In Fig.~\ref{fig_shell_analytic}, a family of solutions to Eq.~\ref{equation_of_motion_sim} is shown for representative values\footnote{Note that solutions with $\xi \leq 1$ are omitted. The corresponding shells do not propagate outwards in the absence of a positive initial velocity, which would be difficult to justify if we exclude additional feedback processes.} of the luminosity ratio $\xi$.
In all cases, an initial launch radius of $R^\prime_{\rm 0} \,=\, 10^{-3}$ and a zero initial velocity are assumed.
The dashed dark blue curve shows the time evolution of the shell radius for a solution with $f[\tau] \,=\,1$ and $\xi \, = \, 1.1$; the shell expands indefinitely, as expected.

The drop in column density of expanding shells and the corresponding decrease in optical depth, however, means that even super-critical shells (with $\xi \, > \, 1$) will turn around.
Neglecting IR multi-scattering, $f[\tau]$ falls off at large radii as
\begin{equation}
\tau(R^\prime) \,=\, \frac{\kappa_{\rm UV} M_{\rm sh}(R)}{4 \pi R^2}  \, = \, \frac{1}{R^\prime} \, ,
\label{eq_tauir}
\end{equation}
where $\kappa_{\rm UV}$ is the dust opacity in the UV.
Thus, the right-hand side of Eq.~\ref{equation_of_motion_iso} always drops below zero.
This behaviour is illustrated with solid lines in Fig.~\ref{fig_shell_analytic}, where a family of solutions to Eq.~\ref{equation_of_motion_sim} is shown for a range of AGN luminosities in the case of a spatially varying optical depth $\tau \,=\, (1 - e^{-\tau_{\rm UV}})$.
In all cases, the shells reach a maximum radial distance $(0.1 \-- 1) R_{\rm \tau}$ from the AGN and stall.

It might appear counter-intuitive that shells driven by higher luminosity AGN stall in a shorter time, e.g. the shell driven in the case of $\xi \,=\, 1.001$ stalls only after a time $t \gtrsim 8 t_{\rm \tau}$, while in the case of $\xi \,=\, 1.1$ it does so already at $t \,\approx\, (3 \-- 4) t_{\rm \tau}$.
This behaviour can be understood  by inspecting the right-hand side of Eq.~\ref{equation_of_motion_iso}.
The expression $\left( 1 - e^{-1/R^\prime } \right) \xi - 1$ that governs the effective force exerted on the shell always drops below zero, at which point the shells decelerate.
The stars shown in Fig.~\ref{fig_shell_analytic} mark the time\footnote{The time at which the outward radiation force and the inward gravitational pull balance is, to good approximation, given by $- \frac{1}{\log{(1 - 1/\xi)} \sqrt{\xi - 1}}$. This expression has a minimum at $\xi \,\approx\, 1.24$, which explains why the solutions with $\xi \,=\, 1 \-- 1.5$ collapse roughly at the same time.} at which the net force becomes attractive. At low AGN luminosities, the shell propagates more slowly, but remains in the optically thick regime for a longer time period. On the other hand, at higher luminosities, the shell reaches the transparency radius more rapidly such that it turns over after a shorter time. 

As shown on the right-hand panel of Fig.~\ref{fig_shell_analytic}, the speeds of the outflows for $\xi \approx 1$ are generally lower or comparable to the velocity dispersion $\sigma$ of the isothermal potential if only single-scattering radiation pressure is assumed. 
After a phase of initial acceleration, these shells reach a constant expansion speed, which is maintained for as long as the optical depth $\tau_{\rm UV} \gg 1$, i.e. $f \left[ \tau \right] \, = \, 1$.
A closer look at Eq.~\ref{equation_of_motion_iso} shows that the associated speeds correspond to constant velocity solutions, i.e. $\dot{R} \,=\, \sqrt{2 (\xi - 1)} \sigma$.
The respective solutions are shown on the right-hand panel of Fig.~\ref{fig_shell_analytic} with thin dotted lines.

We now consider solutions for the case in which the momentum thrust onto the expanding shell is boosted by multi-scattered IR radiation. 
In our calculations, we liberally take $\kappa_{\rm IR} \,=\, \kappa_{\rm UV}/50$, i.e. the IR transparency radius $R_{\rm IR} \, = \, R_{\rm \tau} / 50$, in order to obtain an upper limit on the effect of multi-scattered IR emission. 
For a typical UV opacity $\kappa_{\rm UV} \,=\, 1000 \, \rm cm^2 \, g^{-1}$, this would imply $\kappa_{\rm IR} \, = \, 20 \, \rm cm^2 \, g^{-1}$, i.e., a factor $2 \-- 4$ greater than typically assumed in numerical simulations that attempt to model IR radiation pressure \citep[e.g.][]{Hopkins:12, Agertz:13, Rosdahl:15b}.
Introducing an IR radiation term means introducing an additional dimensionless quantity that is given by the ratio $\kappa_{\rm IR} / \kappa_{\rm UV}$ such that, from Eqs.~\ref{eq_radforce} and~\ref{eq_tauir}, we have
 \begin{equation}
f \left[ \tau(R) \right] \, = \, (1 - e^{-1/R^\prime}) + \frac{\kappa_{\rm IR}}{\kappa_{\rm UV}} \frac{1}{R^\prime} \, ,
\label{eq_motion_ir}
\end{equation}
Given our choice of $\kappa_{\rm IR} \,=\, \kappa_{\rm UV}/50$, the IR optical depth of the shell at the launch radius is $\kappa_{\rm IR} / (\kappa_{\rm UV} R_{\rm 0}^\prime) \, = \, 20$.

The resulting solution, shown in Fig.~\ref{fig_shell_analytic} with a dot-dashed line for $\xi \,=\,1.001$, illustrates that the shell reaches larger radial distances from the AGN at an earlier time than in the single-scattering case.
As in other solutions, the shell turns around after a few dynamical times as the column density drops and the shell becomes transparent to the radiation.
The peak shell velocity is higher by about two orders of magnitude than for the single-scattering case at matching $\xi$.
For a gas rich halo with $\sigma \, = \, 250 \, \rm km \, s^{-1}$, the dash-dotted line on the right-hand panel of Fig.~\ref{fig_shell_analytic} indicates outflow speeds $(750 \-- 1000) \, \rm km \, s^{-1}$ within $\approx 1 \, \rm kpc$.
Clearly, if efficiently trapped, IR radiation has the ability to generate very fast outflows in the velocity range that is typically observed \citep{Thompson:15, Ishibashi:15}.
IR radiation pressure boosts the speed of the shell during its entire evolution, as seen by comparing the dash-dotted and the lowest solid black lines on the right-hand panel of Fig.~\ref{fig_shell_analytic}. In contrast to single-scattering solutions, however, the shell starts decelerating at much smaller radii (with $\frac{dR^\prime}{dt^\prime} \propto t^{\prime , -1/3}$), since the outward radiation force drops rapidly, while the shell still sweeps up mass at a high rate.

It is important to note that the typical time lapse until the shells reach the turn-around radius is $\sim t_{\rm \tau} \,\approx\, 10^8 \, \rm yr$ (see Eq.~\ref{eq_t_tau}), likely much longer than the typical AGN lifetime. In Fig.~\ref{fig_shell_analytic}, the dotted black line shows a solution for a $\xi \,=\, 1.001$ model including IR multi-scattering and using an AGN light curve $L \,=\, L_{\rm 0} \exp{-t/t_{\tau}}$. 
Eq.~\ref{eq_t_tau} shows that, for typical values, $t_{\rm \tau}$ is higher than the typical Salpeter time $t_{\rm s} \,=\, 4.5 \times 10^7 \rm yr$ by a factor of a few.
In order to avoid breaking the dimensionless character of Eq.~\ref{eq_motion_ir}, we use $t_{\rm \tau}$ as an optimistic choice for the AGN lifetime and in the next section investigate a number of solutions with more realistic AGN light-curves. As shown by the dotted black line in Fig.~\ref{fig_shell_analytic}, the shell stalls at a significantly smaller radius $\approx (0.1 \-- 0.2) R_{\rm \tau}$. 

If the density profile in the inner regions of galaxies hosting rapidly growing supermassive black holes is well described by an isothermal profile, this analysis shows that radiation pressure alone is expected to efficiently expel large quantities of gas from the \emph{galactic nucleus}.
Radiation pressure on dust can therefore play an important role in regulating the growth of the central black hole as well as the star formation history of the galaxy (see Section~\ref{sec_discussion}) as long as the AGN luminosity exceeds the critical value given in Eq.~\ref{eq_critlum}.
However, this analysis also shows that mass-loaded outflows initially driven out by radiation pressure are unlikely to clear the galactic halo or to displace large masses beyond $R \approx R_{\rm \tau}$ \citep[see also][]{King:14}, or arguably $R \approx 0.1 R_{\rm \tau} \, \approx \, 2.6 ( \sigma/150 \mathrm{km \, s^{-1}})^2 \, \rm kpc$.
Instead, large quantities of gas might be left within the galactic halo or, possibly, re-accreted onto the central galaxy at a later stage.

So far, we have focussed strictly on isothermal haloes for which all the gas content is swept out in an outflow.
It becomes easier to radiatively drive matter out to large radii if the density distribution profile becomes steeper than $\propto R^{-2}$ or if it sweeps no mass after reaching a radial scale $R_{\rm c}$.  
We close this section by considering the case in which no mass is swept up beyond $R_{\rm c}$.
The dashed black line in Fig.~\ref{fig_shell_analytic} shows the corresponding solution, including IR momentum transfer and a duty cycle, and assuming that $R_{\rm c} / R_{\tau} \,=\, 0.05$ and $\xi \,=\,1.001$.
The shell now accelerates beyond $R_{\rm c}$ because the radiation force is constant at $L/c$ (while the shell remains optically thick to the UV), while the gravitational force decreases as $R^{-2}$. The speed peaks at $\approx 10 \sigma$ and falls slowly thereafter, showing AGN radiation pressure is, at least in principle, able to generate extremely fast outflows even at large ($\sim \rm kpc$) scales, though the velocity peak is sensitive to the scale at which the outflow stops sweeping up mass.

\subsubsection{Wind solutions for an NFW profile}

We now examine solutions for a Navarro-Frenk-White halo \citep{Navarro:97}. The gas component has a mass profile 
\begin{equation}
\begin{split}
M(<x^\prime) \, & = \, f_{\rm gas} M_{\rm vir} \left[ \frac{\ln{(1 + \mathcal{C} x^\prime)} - \mathcal{C} x^\prime/(1 + \mathcal{C} x^\prime)}{\ln{(1 + \mathcal{C})} - \mathcal{C}/(1 + \mathcal{C})} \right] \\ \, & \equiv f_{\rm gas} M_{\rm vir} \left[ \frac{g(\mathcal{C} x^\prime)}{g({\mathcal{C}})} \right] \, , \\
\end{split}
\end{equation}
where $M_{\rm vir}$ denotes the virial mass, $\mathcal{C} \, = \, R_{\rm vir}/R_{\rm s}$ the concentration parameter relating the scale radius of the halo $R_{\rm s}$ to the virial radius $R_{\rm vir}$, $x^\prime \equiv R/R_{\rm vir}$ the dimensionless radius and $g( x ) \, \equiv \, \log{(1 + x)} - x/(1 + x)$. The virial radius $R_{\rm vir}$ is here defined as the radius for which the mean enclosed density is $200$ times the critical density.

A characteristic timescale can be constructed from the ratio of $R_{\rm vir}$ to the virial velocity $V_{\rm vir} \, = \, \sqrt{G M_{\rm vir} / R_{\rm vir}}$. This is
\begin{equation}
\begin{split}
t_{\rm vir} \, & = \, \frac{R_{\rm vir}}{V_{\rm vir}} \\ \, & \approx \, 1.3 \left( \frac{R_{\rm vir}}{200 \mathrm{kpc}} \right) \left( \frac{V_{\rm vir}}{150 \mathrm{km \, s^{-1}}} \right)^{-1} \rm \, Gyr \, .
\end{split}
\end{equation}

\begin{figure*}
\includegraphics[scale = 0.48]{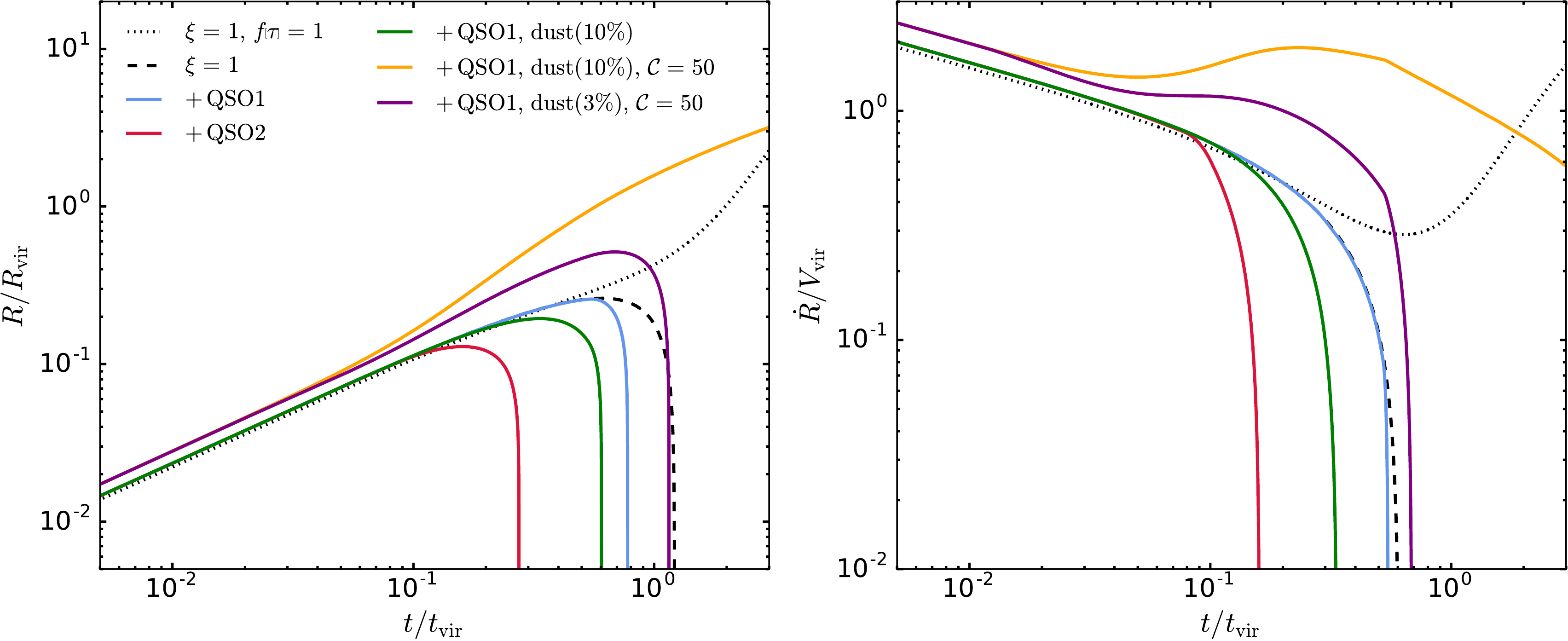}
\caption{Solutions to the equation of motion of radiation pressure-driven shells in an NFW potential. The shell radius in units of the virial radius is shown on the left, while the shell velocity in units of the virial velocity is shown on the right. In the single-scattering regime, i.e. assuming a constant radiation force $L/c$, the shell accelerates after it overcomes the peak of the circular velocity profile and escapes the halo (dotted black line). The drop in optical depth in the outer regions of the halo leads to shell deceleration and prevents it from escaping the halo (dashed black line). AGN lifetime effects always lead to earlier shell deceleration (blue and red lines) and the evolution is particularly sensitive to whether $\xi \, = \, 1$ while the shell is within the halo's circular velocity peak. The green line shows an example for which we impose the condition that no dust is swept up beyond a scale of $0.1 R_{\rm vir}$. The orange line shows a case in which we use light-curve QSO1, a zero dust fraction beyond $0.1 R_{\rm vir}$, but a concentration parameter of $\mathcal{C} \, =\, 50$ (instead of $\mathcal{C} \, =\, 10$), while the purple line shows how this solution changes if the dust-to-gas ratio drops to zero beyond $0.03 R_{\rm vir}$. Since the peak of the circular velocity profile occurs at much smaller scales for higher gas concentrations, the evolution of the radiatively-driven shell is less sensitive to the declining dust-to-gas ratio and AGN duty cycle; by the time these two effects play a role in reducing the radiation force, the shell is sufficiently far outside the region in which gravity has a chance of keeping them bound to the halo (orange line). The sharp kinks in the shell velocity curves are due to sudden bursts of high AGN luminosity.}
\label{fig_shell_analytic_nfw}
\end{figure*}

The optical depths now, however, depend on the halo parameters. For a thin shell that has swept up all gas mass up to a radius $x^\prime$, we have
\begin{equation}
\begin{split}
\tau \, & = \, \kappa f_{\rm gas} \frac{M_{\rm vir}}{4 \pi g(\mathcal{C}) R_{\rm vir}^2} \frac{g (\mathcal{C} x^\prime)}{x^{\prime, 2}} \\ \, & = \, \kappa f_{\rm gas} \frac{M_{\rm vir}^{1/3}}{{4 \pi } g(\mathcal{C})} \frac{g (\mathcal{C} x^\prime)}{x^{\prime, 2}}  \left( \frac{100 H(z)^2}{G} \right)^{2/3} \,
\end{split}
\end{equation}
where we use the virial relation $M_{\rm vir} \, = \, 100 H^2(z) R_{\rm vir}^3 / G$.
In order to mimic the case of a high-redshift, gas-rich, massive galaxy, where gas column densities should be highest, we assume $M_{\rm vir} \, = \, 10^{12} \, \rm M_\odot$, $z \, = \, 2$, $f_{\rm gas} \, = \, 0.1$. In line with the previous section, we also take $\kappa_{\rm UV} \, = \, 10^3 \, \rm cm^2 \, g^{-1}$ and $\kappa_{\rm IR} \, = \, 20 \, \rm cm^2 \, g^{-1}$

Using $t^\prime \, = \, t / t_{\rm vir}$, the equation of motion (Eq.~\ref{equation_of_motion_sim}) can be written as
\begin{equation}
\frac{d}{dt^\prime} \left[ g(\mathcal{C} x^\prime) \frac{d x^\prime}{dt^\prime} \right] \,= \, f[\tau] g(\mathcal{C}) \frac{G L}{f_{\rm gas} c} V_{\rm vir}^{-4} - \frac{g^2(\mathcal{C} x^\prime)}{g({\mathcal{C}}) x^{\prime, 2}} \, .
\label{eq_shellmotion_app}
\end{equation}

As previously, we look for a critical luminosity; we set $f \left[ \tau \right] \, = \, 1$ and the left-hand side of Eq.~\ref{eq_shellmotion_app} to zero. This gives
\begin{equation}
L_{\rm crit} \, = \, \frac{f_{\rm gas} V_{\rm vir}^4 c}{G} \left[ \frac{g(\mathcal{C} x^\prime)}{g(\mathcal{C}) x^{\prime}} \right]^2 \, .
\end{equation}
The bracketed expression on the right-hand side of the last equation is simply the functional form of the circular velocity for an NFW halo (to the fourth power). This has a maximum at  a radius of  $R \, \approx \, 2.16 R_{\rm s}$ with a value of $\approx 0.047 [\mathcal{C}/g(\mathcal{C})]^2$.
A sufficient condition for escape is that the shell overcomes the peak $V_{\rm p}$ of the circular velocity curve of its halo, i.e. that the AGN luminosity exceeds
\begin{equation}
\begin{split}
L_{\rm crit}^{\rm max} \, & = \, 0.047 \left[ \frac{\mathcal{C}}{g(\mathcal{C})} \right]^2 \frac{f_{\rm gas} V_{\rm vir}^4 c}{G} \\ \, & = \, \frac{f_{\rm gas} V_{\rm p}^4 c}{G} \\ \, & \approx \, 4.8 \times 10^{45} \left( \frac{f_{\rm gas}}{0.1} \right) \left( \frac{V_{\rm vir}}{150 \mathrm{km \, s^{-1}}} \right)^4 \, \rm erg \, s^{-1} ,
\end{split}
\label{eq_appcritlum}
\end{equation}
where $\mathcal{C} \, = \, 10$ was assumed in the last step. The dependence on the concentration parameter $\mathcal{C} \, = \, 10$ in Eq.~\ref{eq_appcritlum} is expected since more centrally concentrated haloes are more tightly bound, such that a higher AGN luminosity is required. 
Note also that the critical luminosity derived for isothermal haloes (Eq.~\ref{eq_critlum}) can be recovered by taking $\sigma \, = \, V_{\rm p} / \sqrt{2}$ in Eq.~\ref{eq_appcritlum}.

The final dimensionless equation then reads 
\begin{equation}
\frac{d}{dt^\prime} \left[ g(\mathcal{C} x^\prime) \frac{d x^\prime}{dt^\prime} \right] \,= \, 0.047 f[\tau] \frac{\mathcal{C}^2}{g(\mathcal{C})} \xi - \frac{g^2(\mathcal{C} x^\prime)}{x^{\prime, 2} g(\mathcal{C})} \, ,
\label{eq_shellmotion_dim_nfw}
\end{equation}
where $\xi \, = \, L/L_{\rm crit}^{\rm max}$, as in the case for an isothermal halo.

In Fig.~\ref{fig_shell_analytic_nfw}, we plot solutions to Eq.~\ref{eq_shellmotion_dim_nfw} assuming a dimensionless AGN luminosity $\xi \, = \, 1$.
We show the evolution of the shell radius in units of the virial radius on the left-hand panel and the shell velocity in units of the virial velocity on the right-hand side.
In the absence of optical depth effects, a shell with $\xi \, = \, 1$ is critical and, hence, completely ejected from the halo, a scenario illustrated by a dotted black line in Fig.~\ref{fig_shell_analytic_nfw}. 
The right-hand panel shows that the corresponding solution in fact accelerates after the shell overcomes the peak of the circular velocity profile at $R \, \approx \, 0.1 R_{\rm vir}$.

Even under the unrealistic assumption that the gaseous halo remains dusty out to arbitrarily high radii, a critical shell cannot escape the halo if the optical depth declines, as is indeed shown by the dashed black line for which $f \left[ \tau \right] \, = \, 1 - e^{-\tau_{\rm UV}}$. The solution diverges from that for which $f \left[ \tau \right] \, = \, 1$ already within a virial time, decelerating for most of its evolution. 

\begin{figure}
\includegraphics[scale = 0.43]{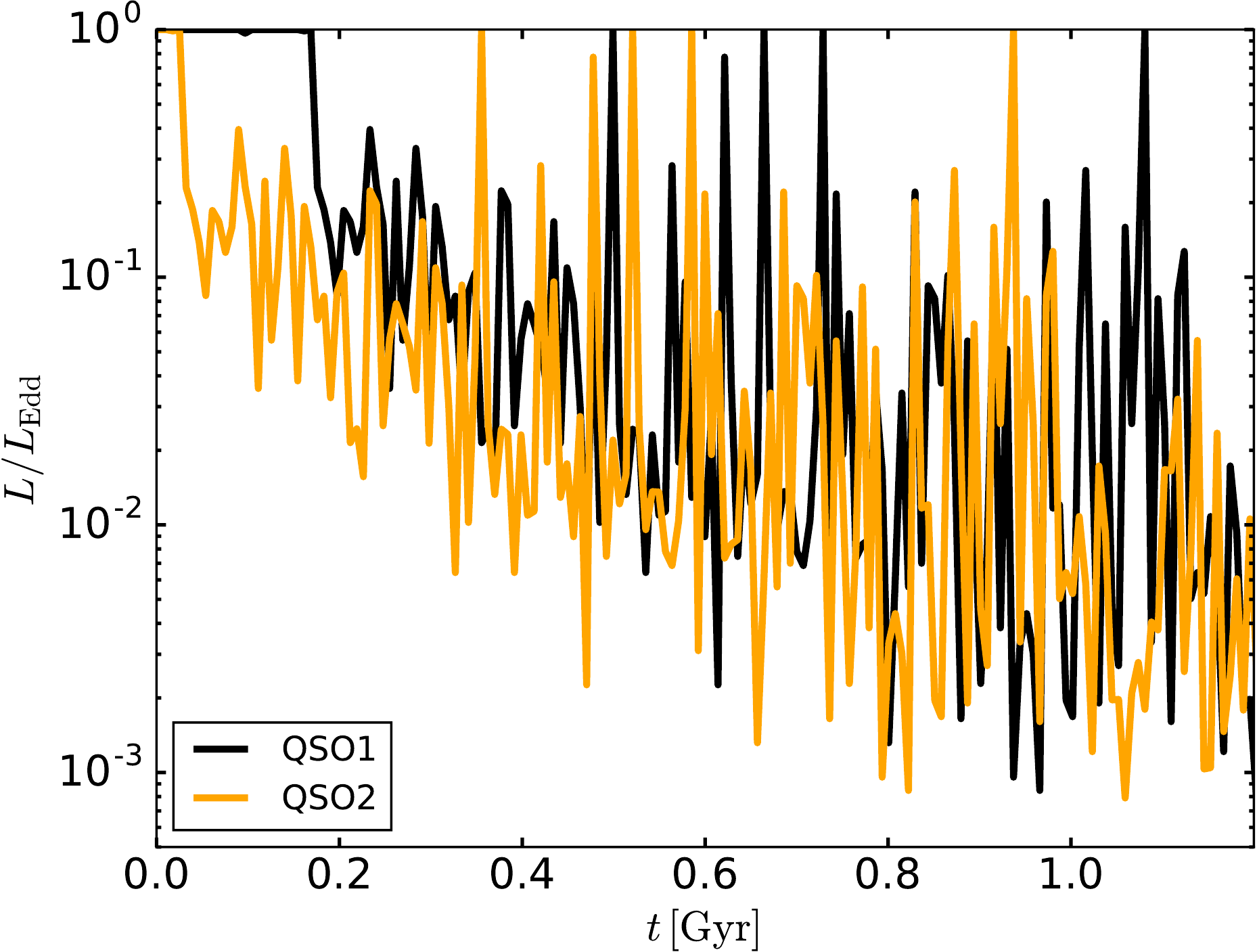}
\caption{AGN light-curves extracted from the cosmological simulations of \citet{Costa:14a}. The black and orange lines show the same light-curve, but shifted in time such that the duration of the initial Eddington-limited burst is $\approx 170 \, \rm Myr$ in the case of the black line (QSO1), but only $\approx 25 \, \rm Myr$ in the case of the orange line (QSO2).}
\label{fig_lightcurve}
\end{figure}

As before, an additional limitation to the expansion of shells through radiation pressure is the AGN lifetime.
For illustrative purposes, we employ the light-curve of a rapidly growing supermassive black hole as extracted from a cosmological simulation of a $z \, = \, 6$ quasar \citep{Costa:14a}.
This is shown in terms of the Eddington ratio as a function of time with a black line in Fig.~\ref{fig_lightcurve}. 
For light-curve `QSO1', the Eddington ratio is close to unity during the first $\approx 170 \, \rm Myr$.
In the simulations of \citet{Costa:14a}, such a period of Eddington-limited accretion is required in order to grow the black hole to $\approx 10^9 \, \rm M_\odot$ by $z \, = \, 6$. The Eddington ratio then gradually drops to values as low as $< 10^{-3}$ when the black hole growth becomes regulated by AGN feedback.
Growth rates, however, are not required to be as extreme for supermassive black holes assembling at lower redshift.
The orange line in Fig.~\ref{fig_lightcurve}, which is used to mimic the latter scenario, shows the identical light-curve, but shifted in time such that the initial Eddington-limited burst lasts only for $\approx 25 \, \rm Myr$ (`QSO2'). 

Solutions using the cosmological light-curves are shown by the blue and red lines in Fig.~\ref{fig_shell_analytic_nfw}. 
The outcome is clearly sensitive to the adopted light curve and is particularly dependent on whether the AGN goes through an initial phase of prolonged Eddington-limited emission.
This result is linked to the shape of the NFW potential and should hold in general for haloes with peaked circular velocity profiles; the inward gravitational force on the shell increases with radius up until it reaches a radius $\approx R_{\rm s}$, i.e. $x^\prime \, \approx \, \mathcal{C}^{-1}$.
It is crucial for the AGN luminosity to be close to $\xi \, \approx \, 1$ for as long as the shell remains within this region if it is to have a chance of escaping the halo.
While for light-curve QSO1 (blue line) the shell follows an evolution which is similar to that without an AGN duty cycle, it collapses at scales of $0.1 R_{\rm vir}$ in the case of light-curve QSO2 (red line).

An additional limiting factor is the expected drop in dust-to-gas ratio away from the central galaxy.
The models presented in \citet{Thompson:15, Ishibashi:15} do not account for such a spatially varying dust-to-gas ratio. Given the choice of $\kappa_{\rm UV} \, = \, 10^3 \rm \, cm^2 \, g^{-1}$, which is appropriate for a dust-to-gas ratio of about $f_{\rm d} \, = \, 0.01$, the models would, at face value, predict total dust masses within the virial radius on the order of $M_{\rm d} \, = \, f_{\rm d} f_{\rm gas} M_{\rm vir} \, \approx \, 10^9 \-- 10^{10} \, \rm M_\odot$ for haloes with mass $M_{\rm vir} \, = \, 10^{12} \-- 10^{13} \, \rm M_\odot$, respectively. 
Dust masses inferred by observations of high redshift starburst galaxies and quasars are, however, likely to be much lower. Recent ALMA observations targeting the dust continuum of a sample of $z \, = \, 2.5$ massive ($M_{\rm *} \approx 10^{11} \, \rm M_\odot$) star forming galaxies, for instance, yield dust masses in the range $10^8 \-- 10^{9.3} \, \rm M_\odot$ \citep{Barro:16}. The inferred dust masses for the observed population of reddened quasars at $z \, = \, 2.5$ are also in the range of $10^8 \-- 10^9 \, \rm M_\odot$ \citep{Banerji:17} so that the high dust masses implicitly assumed in \citet{Thompson:15, Ishibashi:15} are difficult to justify.

We therefore assess wind solutions for which we limit the total dust masses to values in the range $10^{8 \-- 9} \, \rm M_\odot$. 
Specifically, we assume the additional swept up mass to be dust-free after the shell reaches a radius $x_{\rm 0}$, i.e. the optical depth scales with $M(< x^\prime)$ or $M(< x_{\rm 0})$ depending on the shell location as
\begin{equation}
\tau \, = \, \begin{cases} 
      f_{\rm gas} M_{\rm vir} \kappa \frac{g(\mathcal{C} x^\prime)}{4 \pi g(\mathcal{C}) x^{\prime\, 2}}, & x^\prime < x_{\rm 0} \, , \\
      f_{\rm gas} M_{\rm vir} \kappa \frac{g(\mathcal{C} x_{\rm 0})}{4 \pi g(\mathcal{C}) x^{\prime\, 2}}, & x^\prime \geq x_{\rm 0} \, ,
   \end{cases}
\end{equation}

where the dependence on $x^{\prime}$ in the denominator in the second case is required to account for the geometric expansion of the shell and the consequent drop in column density.
In our models for an NFW halo with $\mathcal{C} \, = \, 10$ and $M_{\rm vir} \, = \, 10^{12} \, \rm M_\odot$, we assume no dust is swept up beyond $x_{\rm 0} \, =\, 0.1 R_{\rm vir}$. This gives a total dust mass of $\approx 1.3 \times 10^8 \, \rm M_\odot$ in line with observational constraints.  For models with $\mathcal{C} \, = \, 50$ and $M_{\rm vir} \, = \, 10^{12} \, \rm M_\odot$, we take $x_{\rm 0} \, =\, 0.03 R_{\rm vir}$ and $x_{\rm 0} \, =\, 0.1 R_{\rm vir}$, giving dust masses of $\approx 10^8 \, \rm M_\odot$ and $\approx 3 \times 10^8 \, \rm M_\odot$, respectively.

Despite the long initial period of Eddington-limited accretion (light-curve QSO1), the shell driven out in the case of $\mathcal{C} \, = \, 10$ (shown in green) now reaches a maximum radius of only $\approx 0.02 R_{\rm vir}$.
Increasing the concentration, which could mimic the scenario in which the gas component accumulates in the central regions due to very efficient cooling, means that the peak of circular velocity profile occurs at smaller radii.
The dynamical time up to the circular velocity peak radius is shorter and the outflow is less sensitive to both duty cycle and declining dust-to-gas ratio effects.
Consequently, the shell has a higher chance of reaching larger radii.
The orange line shown in Fig.~\ref{fig_shell_analytic_nfw} shows a solution for a concentration parameter $\mathcal{C} \, = \, 50$, representing a shell that escapes the halo completely despite duty cycle and declining dust-to-gas ratio effects. 
Reducing the dust mass (by a factor of 3) in the case of $\mathcal{C} \, = \, 50$ even mildly, however, results in a shell which remains trapped within the halo, as shown by the purple line. The latter is possibly the most likely scenario as a significant portion of dust should be expected to be destroyed in the outflow both collisionally and through thermal sputtering \citep[e.g.][]{Draine:95}.

The evolution of radiatively-driven shells in NFW profiles appears to be very sensitive to the choice of AGN luminosity, light-curve, gas concentration as well as the dust-to-gas ratio profile.
Clearly, the number of parameters is sufficiently high that they can be selected to result in very different shell evolutions and we have only examined a few possible combinations.

AGN radiation pressure on dust, however, seems unlikely to eject a majority of baryons from the galactic halo as a combination of very high AGN luminosities, long duty cycles and high dust masses are required.
We also expect a fraction of gas in the halo to be flowing in towards the central galaxy and not to be static,  such that the radiation force is required to revert the inflowing gas in addition to ejecting it.
Anisotropy in the gas distribution and the AGN radiation field is likely to further limit the volume of gas that is exposed to the AGN radiation field.
In \citet{Costa:17}, we explore in detail these regimes with fully cosmological radiation-hydrodynamic simulations.
Our current analysis, however, allows us to identify the regime in which radiation pressure on dust should be most efficient: (1) at high redshift, when galaxies are more compact and gas column densities are higher, (2) in phases in which the AGN luminosity is at $\xi \gtrsim 1$ for several tens of Myr, (3) in compact galaxies with high circular velocity peaks.

\section{Idealised simulations of radiation pressure-driven outflows}
\label{sec_idealsimulations}

In this section, we make the same assumptions as made in the analytic models described in the previous sections and perform radiation-hydrodynamic simulations in order to test their validity as well as to identify their main limitations.

\subsection{The simulations}

\subsubsection{The code}

Our simulations are performed with the radiation-hydrodynamic code {\sc Ramses-RT}\footnote{See https://bitbucket.org/rteyssie/ramses for the publicly available code (including the RHD used here).} \citep{Rosdahl:13}, an extension of the hydrodynamic adaptive mesh refinement (AMR) code {\sc Ramses} \citep{Teyssier:02}. 

In {\sc Ramses}, the flow is discretised onto a Cartesian grid and the Euler equations are solved using a second-order Godunov scheme.
The grid can be dynamically refined in order to obtain higher numerical resolution within sub-regions of the simulation domain satisfying a specified refinement criterion.
Shock capturing relies on a Riemann solver and does not require the addition of artificial viscosity, unlike `smoothed-particle hydrodynamics' (SPH) methods.

In {\sc Ramses-RT}, radiation transport is computed on-the-fly using a first-order moment method. The set of radiation transport equations are closed using the M1 relation for the Eddington tensor.
Radiation is allowed to couple to gas through photoionisation, photoheating and radiation pressure from ionising photons.
Additionally, the code follows energy and momentum transfer between radiation and dust grains \citep{Rosdahl:15a} in both single- and multi-scattering regimes.
We note that the dynamics of dust is not followed self-consistently in our simulations. Like in the analytic models presented in Section~\ref{sec_analyticalsolutions}, we here make the assumption that dust and gas are coupled hydrodynamically.

Since the radiative transfer is solved explicitly, the time-step size is limited by the speed of light.
In order to avoid prohibitively small time-steps, the `reduced speed of light approximation' \citep{Gnedin:01} is used, though only in a subset of our simulations. When this approximation is employed, we verify that our results are converged with respect to the speed of light.

\subsubsection{The setup}

The setup of the simulations is kept deliberately simple in order to compare the numerical solutions with the analytic models reviewed in Section~\ref{sec_analyticalsolutions}. 
Our initial conditions consist of an NFW profile with virial velocity $V_{\rm vir} \, = \, 150 \, \rm km \, s^{-1}$, concentration $\mathcal{C} \,=\, 10$ and gas fraction $f_{\rm gas} \,=\, 0.1$ (see Table~\ref{table1} for all relevant parameters). 
Virial quantities are evaluated at the radius at which the enclosed mean density is a factor $200$ times the critical density at $z \,=\, 0$.
The simulation box width is chosen to be $30$ times the halo scale length, i.e. about $640 \, \rm kpc$, and the base grid is set at level 8 (i.e. $256^3$ cells).
The grid is dynamically refined using a Lagrangian strategy up to a maximum level 13 (i.e. $8,192^3$ cells), with a cell refined, i.e. split into 8 equal size child cells, when its gas mass exceeds $10^6 \, \rm M_\odot$. The minimum cell width is thus  $\approx 80 \, \rm pc$.
We, however, also perform a higher resolution simulation in order to verify the convergence properties of our main results. For this simulation (see halo-IR46-high-res in Table~\ref{table_sims}), we use a maximum refinement level of 14, such that the minimum cell size is $\approx 40 \, \rm pc$ and a cell is refined if its gas mass exceeds $1.25 \times 10^5 \, \rm M_\odot$, such that the resolution increases not only in the central regions but also throughout the galactic halo as a whole. This simulation is far more computationally expensive and is therefore run for $30 \, \rm Myr$ only, i.e. sufficient to probe the early optically thick stages, which are the most critical for the shell evolution.

Gas self-gravity is followed self-consistently, whereas the gravitational potential of dark matter is treated as a static source term in the momentum and energy equations.
This simplification enables a closer comparison with solutions to Eq.~\ref{eq_shellmotion_dim_nfw}, where a possible response of dark matter to the outflow is neglected.
As in the analytic model, no stellar component is included.

The gas halo is initially set up in hydrostatic equilibrium and truncated at a radial distance of about $240 \, \rm kpc$ (i.e. $75 \, \%$ of half a box side length).
At this radial distance, the gas density profile is artificially reduced to a value of $4.13 \times 10^{-31} \, \rm g \, cm^{-3}$.
In order to minimise grid artefacts in our simulations, we introduce isobaric perturbations in the gas density distribution as
\begin{equation}
\frac{\delta \rho}{\rho} \, = \, 1 + \chi \,
\end{equation}
where $\chi$ is a random number\footnote{Note that the gas density is never allowed to fall below a floor value of $4.13 \times 10^{-31} \, \rm g \, cm^{-3}$, though, in practice, such low densities only occur for gas beyond a radius greater than $75 \, \%$ of the box width.} in the range $[-1, 1]$.
We have experimented performing simulations in which we employed a random number range of $[-0.5, 0.5]$, finding no difference in their behaviour with respect to the simulations we present here.

We verify that the halo remains relaxed for the entire duration of our simulation runs. In particular, we compare the density profile in the initial conditions and at $t \,=\, 100 \, \rm Myr$ in a purely hydrodynamical simulation, i.e. without a radiation source, finding a mean ratio between initial and final densities of $1.02$, indicating that the gas halo remains stable for the entire duration.
We adopt outflow boundary conditions for both gas and radiation.

In the analytic model presented in Section~\ref{sec_analyticalsolutions}, two types of radiation pressure are considered: single-scattering UV radiation and multi-scattering IR radiation.
In order to establish a close comparison, we consider only two frequency bins.
Thus, we have a UV bin with mean energy $18.85 \, \rm eV$ and an IR bin with mean energy $1 \, \rm eV$, though, since we disable gas photoionisation, the values of the radiation band frequency are immaterial in this study. We choose not to adopt an AGN spectrum in order to maintain a close link with the analytic models reviewed in Section~\ref{sec_analyticalsolutions}.

Radiation is injected at every fine AMR step into the cell closest to the centre of the halo. We have employed photon rates corresponding to AGN bolometric luminosities of $10^{45}$,  $10^{46}$ and $10^{47} \, \rm erg \, s^{-1}$ in separate simulations.
Given the selected halo parameters (Table~\ref{table1}), the critical luminosity required to launch unbound shells in the single-scattering regime is about $5 \times 10^{45} \, \rm erg \, s^{-1}$ (see Eq. \ref{eq_appcritlum}).
Thus, the selected AGN luminosities probe both the `sub-critical' scenario, for which the full force of radiation pressure is insufficient to overcome gravitational pull, as well as the `super-critical' case\footnote{In practice, since we assume hydrostatic equilibrium, the main reason why low luminosity shells fail to escape the halo is due to the confining pressure of the ambient medium.}. We remind that even for high AGN luminosities, the declining optical depth means that shells propagate less quickly and even turn-over at large radii.
In order to ensure a close comparison with Section~\ref{sec_analyticalsolutions}, we fix the AGN luminosity to a constant value.

The `critical luminosity' can always be expressed in terms of a critical supermassive black hole mass. Assuming the AGN cannot radiate at luminosities higher than its Eddington limit, Eq.~\ref{eq_appcritlum} yields a critical mass
\begin{equation}
M_{\rm BH}^{\rm crit} \, \geq \, 4 \times 10^7 \left( \frac{f_{\rm gas}}{0.1} \right) \left( \frac{V_{\rm vir}}{150 \mathrm{km \, s^{-1}}} \right)^4 \, \rm M_\odot \, .
\end{equation}
This black hole mass is that which is expected to lead to self-regulation of black hole accretion in our chosen NFW halo and also that which would place the black hole and its halo on the observed $M \-- \sigma$ relation \citep{Larkin:16}.

In most of our simulations, radiation is injected entirely in the UV band. 
We shall refer to those simulations as `halo-UV' if only UV radiation is considered and `halo-UVIR', if radiation is initially injected in the UV, but allowed to be reprocessed in the IR.
However, in cases in which we are exclusively interested in the efficiency of IR trapping, we actually inject all radiation already in the IR band, ignoring the UV band completely, such that, at matching AGN luminosity, the total power injected is the same as in the `halo-UV' and `halo-UVIR'.
We shall refer to those simulations as `halo-IR'.
For each simulation name, we include also the selected AGN luminosity. Thus, for example, we shall have `halo-IR46' for an IR only simulation with an AGN luminosity of $10^{46} \rm \, erg \, s^{-1}$.
We list the main simulations investigated in this study in Table~\ref{table_sims} together with the assumed AGN luminosity, the initial IR optical depth (when IR radiative transfer is followed), the reduced speed of light factor and the minimum cell size.
 
In this study, we focus exclusively on the role of radiation pressure on dust and, accordingly, set the ionisation cross-sections to zero. 
There is therefore neither photo-heating nor ionisation radiation pressure from photons in any of our simulations.
The effects of these processes are also ignored in the analytic model reviewed in Section~\ref{sec_analyticalsolutions}.

We also consider the effects of radiative cooling (including primordial and metal-line radiative processes) in some of our simulations.
The non-equilibrium cooling of hydrogen and helium, coupled to the local radiation field, is described in \citet{Rosdahl:13}.
For the contribution of metals, above temperatures of $10^4 \, \rm K$, which has been drawn from {\sc Cloudy} \citep{Ferland:98}, we assume photoionisation equilibrium with a redshift zero UV background.
For the metal contribution at $T < 10^4 \, \rm K$, we use the fine structure cooling from \citet{Rosen:95}. 
The inclusion of cooling only affects the hydrodynamical response of the gas in terms of shocks and compression, but not the dust opacity, which, for simplicity and ease of comparison with analytic models, is kept constant in each simulation.

We adopt a reduced speed of light $\tilde{c} \, = \, 0.1c$ for our halo-UV and halo-UVIR simulations, but use the full speed of light in our halo-IR simulations.
We also perform one of our halo-UVIR simulations at full speed of light (halo-UVIR46-c) in order to verify the convergence properties of this suite of simulations with respect to the speed of light.
While results for single-scattering radiation pressure converge rapidly with the reduced speed of light value, we find that a high speed of light is required in order to avoid overestimating the diffusion time of IR radiation in the highly optically thick regime (see Section~\ref{sec_momenergy}).

\begin{table}
\centering
\begin{tabular}{c*{5}{c}r}
\hline
$V_{\rm vir} \, \rm [km \, s^{-1}]$ & $M_{\rm vir} \, \rm [M_\odot]$ & $R_{\rm vir} \, \rm [kpc]$ & $\mathcal{C}$ & $f_{\rm gas}$\\
\hline
$150$   & $1.1 \times 10^{12}$ & $214.26$ & $10$ & $0.1$\\
\hline
\end{tabular}
\caption{Parameters describing the NFW halo modelled in our simulations. From left to right, we show the virial velocity, the virial mass, the virial radius, the concentration parameter and the gas fraction. The virial radius is defined as the radial scale enclosing a mean density a factor $200$ the critical density at $z \, = \, 0$.}
\label{table1}
\end{table}

\begin{table}
\centering
\begin{tabular}{c*{5}{c}r}
\hline
Name & $L_{\rm AGN}$        & $\tau_{\rm IR}$ & $\tilde{c}/c$  & $\Delta x$ \\
           & $\rm [erg \, s^{-1}]$ &                           &                   & $\rm [pc]$ \\
\hline
halo-UV45         & $10^{45}$ & $-$     & $0.1$   & $80$\\
halo-UV46         & $10^{46}$ & $-$     & $0.1$   & $80$\\
halo-UV47         & $10^{47}$ & $-$     & $0.1$   & $80$\\
halo-UVIR45     & $10^{45}$ & $11$     & $0.1$ & $80$\\
halo-UVIR46     & $10^{46}$ & $11$     & $0.1$ & $80$\\
halo-UVIR46-c  & $10^{46}$ & $11$     & $1$    & $80$\\
halo-UVIR47     & $10^{47}$ & $11$     & $0.1$ & $80$\\
halo-IR46          & $10^{46}$ & $11$     & $1$    & $80$\\
halo-IR46-med  & $10^{46}$ & $110$   & $1$    & $80$\\
halo-IR46-high  & $10^{46}$ & $1100$ & $1$    & $80$\\
halo-IR47-med  & $10^{47}$ & $110$  & $1$     & $80$\\
halo-IR46-high-res  & $10^{46}$ & $1100$  & $1$   & $40$\\
\hline
\end{tabular}
\caption{The main simulations presented in the paper shown with the corresponding AGN luminosities, initial IR optical depths, reduced speed of light factor $\tilde{c}/c$ and the minimum cell size.}
\label{table_sims}
\end{table}

 \subsubsection{Selecting dust opacities}
 \label{sec_seldust}
 
For NFW haloes, the dynamics of radiation-pressure driven shells is governed by the luminosity ratio $\xi \, = \, L / L_{\rm crit}^{\rm max}$, the optical depth $\tau$ and the concentration parameter $\mathcal{C}$.
Due to the shallower density profile ($\rho \propto R^{-1}$) of their inner regions, it is far more difficult to achieve high IR optical depths in NFW haloes than in isothermal spheres. 
The difference between the dotted black line (for which $f \left[ \tau \right] \, = \, 1$ is assumed), and the coloured lines in Fig.~\ref{fig_shell_analytic_nfw} shows that the IR boost is modest even in the case for which $\mathcal{C} \, = \, 50$.

In particular, the column density for a thin shell that has swept up all gas up to a radius $R$ in an NFW halo is given by
\begin{equation}
\Sigma_{\rm shell} \, = \, \frac{M_{\rm gas}(<R)}{4 \pi R^2} \, = \, f_{\rm gas} \frac{M_{\rm vir}}{4 \pi R_{\rm vir}^2} \frac{g( \mathcal{C} x^\prime)}{g(\mathcal{C}) x^{\prime, 2}} \, ,
\label{eq_sigmagas1}
\end{equation}
where $x^\prime \, \equiv \, R/R_{\rm vir}$ and $g(x^\prime)$ is given by
\begin{equation}
g(x^\prime) \,=\, \log(1 + x^\prime) - \frac{x^\prime}{1 + x^\prime} \, .
\end{equation}
Unlike in isothermal haloes, where shells can have arbitrarily high column densities ($\Sigma_{\rm shell} \propto R^{-1}$), the column density of shells in NFW haloes flattens out below the scale radius, converging to a maximum value given by
\begin{equation}
\Sigma_{\rm shell}^{\rm max} \, = \, \frac{f_{\rm gas}}{8 \pi} \frac{\mathcal{C}^2}{g(\mathcal{C})} \frac{M_{\rm vir}}{R_{\rm vir}^2} \, = \, \frac{f_{\rm gas}}{8 \pi} \frac{\mathcal{C}^2}{g(\mathcal{C})} M_{\rm vir}^{1/3}\left( \frac{100 H^2(z)}{G} \right)^{2/3} \, ,
\label{eq_sigmagas}
\end{equation}
where the virial relation $M_{\rm vir} \, = \, 100 H^2(z) R_{\rm vir}^3 / G$ is used in the last step.
For convenience, we evaluate the expression {$\mathcal{C}^2/g(\mathcal{C})$} appearing in Eq.~\ref{eq_sigmagas} for a range of concentration parameters and present them in Table~\ref{table_gc}.
Note that the concentration parameters shown in the table, which reach $\mathcal{C} \,=\, 100$, are not to be interpreted as realistic dark matter halo concentrations.
Instead, they are chosen to indicate how concentrated the baryonic component would have to be in order to result in high IR optical depths.

Using $H(z)\,=\, 70 \, \rm km \, s^{-1} \, Mpc^{-1}$ and $\tau_{\rm IR}^{\rm max} \, = \, \kappa_{\rm IR} \Sigma_{\rm shell}^{\rm max}$ gives
\begin{equation}
\tau_{\rm IR}^{\rm max} \, \approx \, 1 \left( \frac{f_{\rm gas}}{0.17} \right) \left( \frac{\kappa_{\rm IR}}{10 \, \mathrm{cm^2 \, g^{-1}}} \right)  \left( \frac{\mathcal{C}^2/g(\mathcal{C})}{2758.60} \right) \left( \frac{M_{\rm vir}}{10^{12} \, \mathrm{M_\odot}} \right)^{1/3} \, ,
\end{equation}
where the value $2758.60$ is used for $\mathcal{C} \, = \, 100$.
We see that in order to obtain IR optical depths greater than unity for thin shells in NFW haloes, a combination of high gas concentration, high redshift and high halo mass would have to be assumed.
Even for a gas concentration of $100$ and a halo with virial mass of $10^{12} \, \rm M_\odot$, the IR optical depth is still of order unity for an IR opacity of $\kappa_{\rm IR} \, = \, 10 \, \rm cm^2 \, g^{-1}$ (which is already on the high end of IR opacities adopted in the literature).

One could instead opt for simulating an isothermal profile.
However, resolving regions of high IR optical depths $\tau_{\rm IR} \gtrsim 10$ even in massive isothermal haloes requires very high numerical resolution. For instance, in order to capture scales for which $\tau_{\rm IR} \, = \, 50$ in an isothermal halo, we would have to resolve the spatial scale
\begin{equation}
\begin{split}
R\left( \tau_{\rm IR} \, = \, 50 \right)\, & = \, \frac{f_{\rm gas} \sigma^2}{2 \pi G} \kappa_{\rm IR}\\ \, & \approx \, 16 \left( \frac{f_{\rm gas}}{0.17} \right) \left( \frac{\sigma}{250 \, \mathrm{km \, s^{-1}}} \right)^2 \left( \frac{\kappa_{\rm IR}}{10 \, \mathrm{cm^2 \, g^{-1}}}  \right)\, \rm \mathrm{pc}.
\end{split}
\end{equation}
Since we here focus on performing a large number of simulations, each of which is fully radiation-hydrodynamic and, in many cases, performed using the full speed of light, such high resolution simulations are infeasible. 

\begin{table}
\centering
\begin{tabular}{c*{6}{c}r}
\hline
$\mathcal{C}$ & $1$ & $5$ & $10$ & $50$ &  $100$ \\
\hline
$\mathcal{C}^2 / g(\mathcal{C})$ & $5.18$  & $26.08$ & $67.17$ & $ 847.04$ & $2758.60$\\
\hline
\end{tabular}
\caption{Function $\mathcal{C}^2 / g(\mathcal{C})$ from Eq.~\ref{eq_sigmagas} evaluated for various choices of NFW halo concentration parameters $\mathcal{C}$. Concentrations $> 50$ are unrealistic for the dark matter component, but could mimic a scenario in which the gas component cools efficiently to form a compact core. Such high concentrations are required to render gas following an NFW density profile optically thick in the IR.}
\label{table_gc}
\end{table}

Various features of NFW haloes make them more desirable to simulate than isothermal spheres. 
One of the disadvantages of exploring the problem of driving shells with radiation pressure in isothermal haloes is that solutions to the equation of motion (Eq.~\ref{equation_of_motion_iso}) are very sensitive to the shell launch radius. In NFW haloes, thanks to flatness of the shell column density (Eq.~\ref{eq_sigmagas}) below the scale radius, solutions are much less dependent on uncertainties in the launch radius and, hence, the resolution of the simulation.
Based on hydro-only test simulations of isothermal haloes, we also find simulations of haloes following NFW profiles to be more stable and to suffer less from relaxation effects, mainly due to their shallower inner density profiles.

\begin{figure*}
\begin{minipage}{0.5\textwidth}
	\centering
	\includegraphics[scale = 0.5]{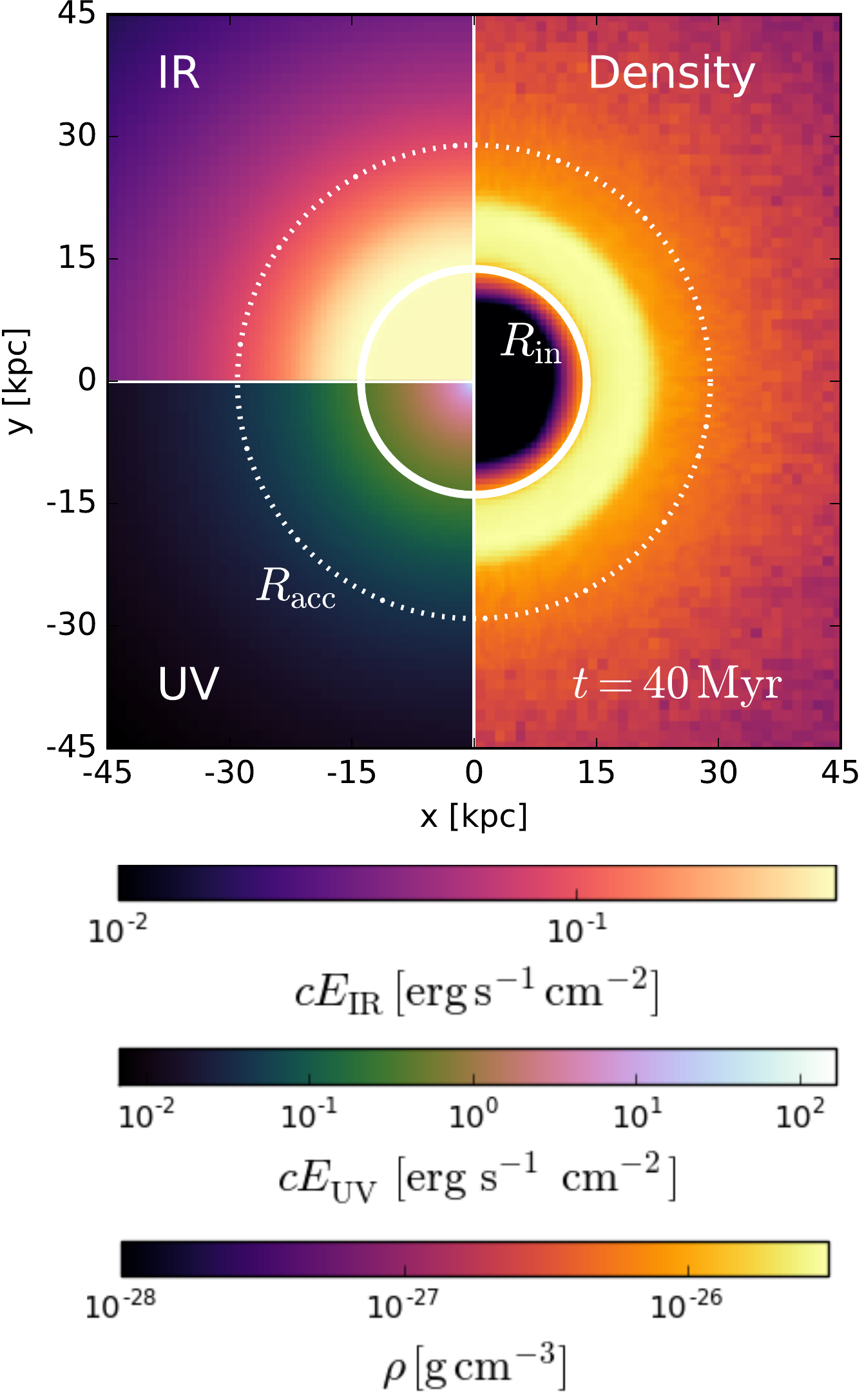}
\end{minipage}%
\begin{minipage}{0.5\textwidth}
	\centering
	\includegraphics[scale = 0.7]{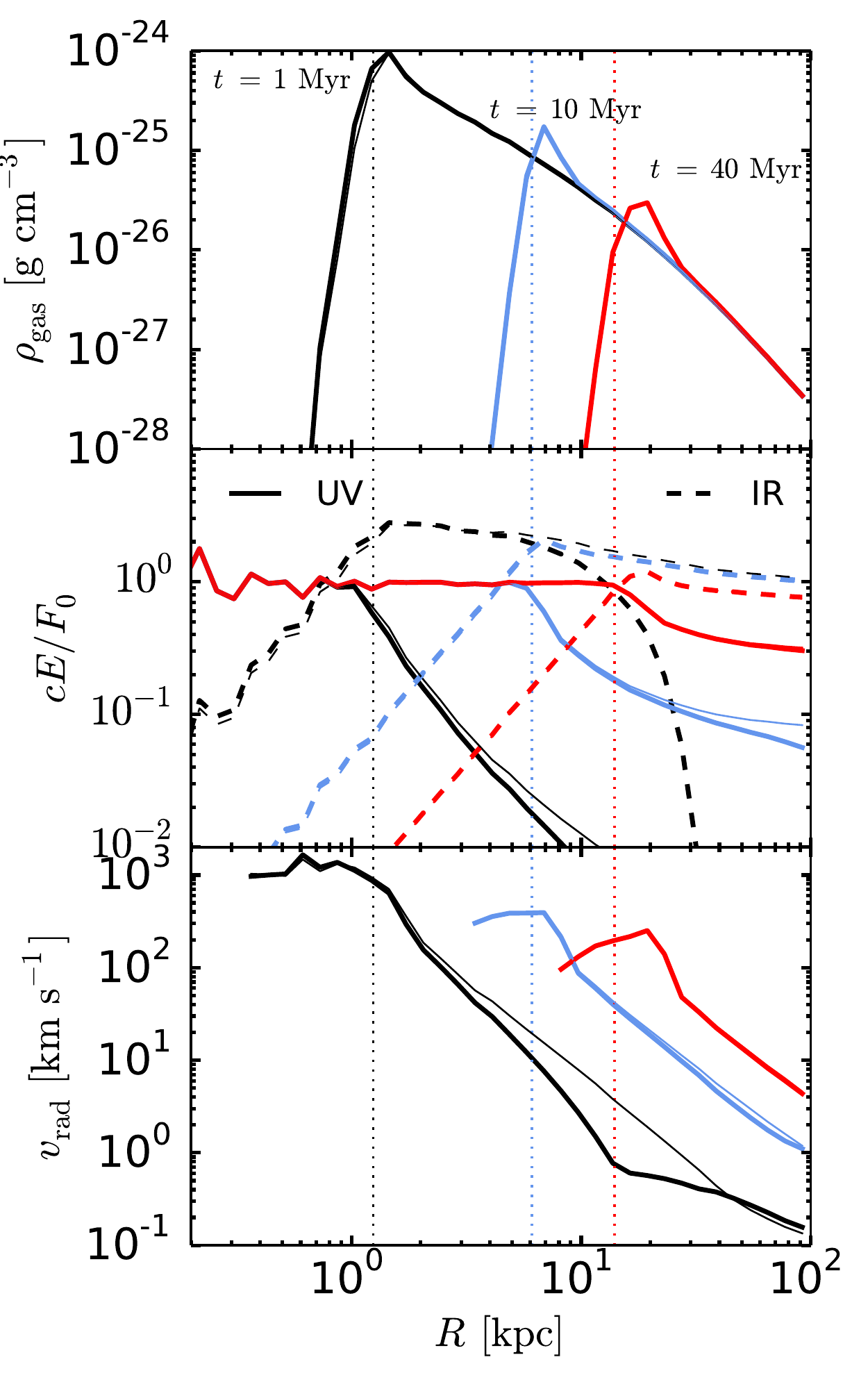}
\end{minipage}
\caption{Left: The density and radiation energy density distribution in simulation halo-UVIR46 at $t \,=\, 40 \, \rm Myr$. Volume-weighted projected UV and IR energy densities are shown on the bottom and top left, respectively, and gas density on the right. All quantities are projected along a slab of thickness $\approx 13 \, \rm kpc$. The solid line shows the location of the tangential discontinuity ($R_{\rm in}$), which is slightly broadened due to resolution and finite mean free path effects. The forward shock can be seen as a sharp density jump at $R \, \approx \, 20 \, \rm kpc$. The dotted circle marks the radial scale $R_{\rm acc}$ at which the gas radial velocity is $100 \, \rm km \, s^{-1}$. This region, which lies beyond the shock radius, contains gas that has been pre-accelerated by IR radiation emitted from the outer surface of the shell. Right: Profiles of various important gas and radiation properties as a function of time in our halo-UVIR46 simulation (thick lines) and also for simulation halo-UVIR46-c (thin lines).The top panel shows density profiles in simulation halo-UVIR46 at different times (as labeled). A shell of swept-up material is driven outwards as expected. The middle panel shows radial profiles of UV and IR radiation energy densities divided by $F_{\rm 0} \, = \, L/(4 \pi R^2)$, at the corresponding times. UV radiation is efficiently absorbed at its inner boundary and its flux is, correspondingly, low at large radii. As the shell expands, its optical depth drops and the UV radiation leaks out, raising the radiation energy density at larger radii. Reprocessed IR photons are trapped in the shell, leading to a boost with respect to $L/(4 \pi R^2)$. They are re-emitted inwards towards the source as well as outwards away from the source. The bottom panel shows mass-weighted radial velocity profiles for gas with density $> 10^{-30} \, \rm g \, cm^{-3}$. The shells slow down as they sweep up mass. The thin dotted lines show the location of $R_{\rm in}$.}
\label{fig_shell}
\end{figure*}

We thus adopt the following strategy: we retain the advantages of simulating NFW haloes by instead assuming (unrealistically) high values for the IR dust opacity. This procedure artificially increases the IR transparency radius of the gaseous halo out to scales at which it can be resolved with a large number of cells.
Our choice of dust opacities is thus motivated entirely by the aim of exploring the hydrodynamical response of the gas to IR radiation in media of different optical depths and comparing the simulated solutions with analytical expectations.
Simulations performed with realistic IR dust opacities and dark matter haloes will be presented in a joint paper using fully radiation-hydrodynamic cosmological simulations (Costa et al., in prep.).

Thus, while for UV radiation we choose a realistic $\kappa_{\rm UV} \, = \, 10^3 \, \rm cm^2 \, g^{-1}$, for the IR we take $\kappa_{\rm IR} \, = \, 10^3 \-- 10^5 \, \rm cm^2 \, g^{-1}$, i.e. several orders of magnitude above realistic IR opacities.
Given this choice, the initial optical depths measured out to the virial radius are, approximately, $11.3 \-- 1130$, for the lowest and highest choice of IR opacities, respectively (see Table~\ref{table_sims} for a list of the initial optical depths in all simulations).
Another relevant quantity, however, is the time-varying optical depth measured out to the outer edge of the radiatively-driven shells.
In practice, these are much lower, starting from typical values $\approx 5 \-- 500$ and reaching about $0.5 \-- 50$ at later times.

We now present the results of our suite of simulations.
We start by providing a qualitative discussion of the outflow structure and then carry out a detailed comparison with solutions to the equation of motion of radiatively-driven shells in NFW potentials.

\subsection{Expanding, radiation pressure-driven shells}
\label{sec_exprad}

\begin{figure*}
\includegraphics[scale = 0.6]{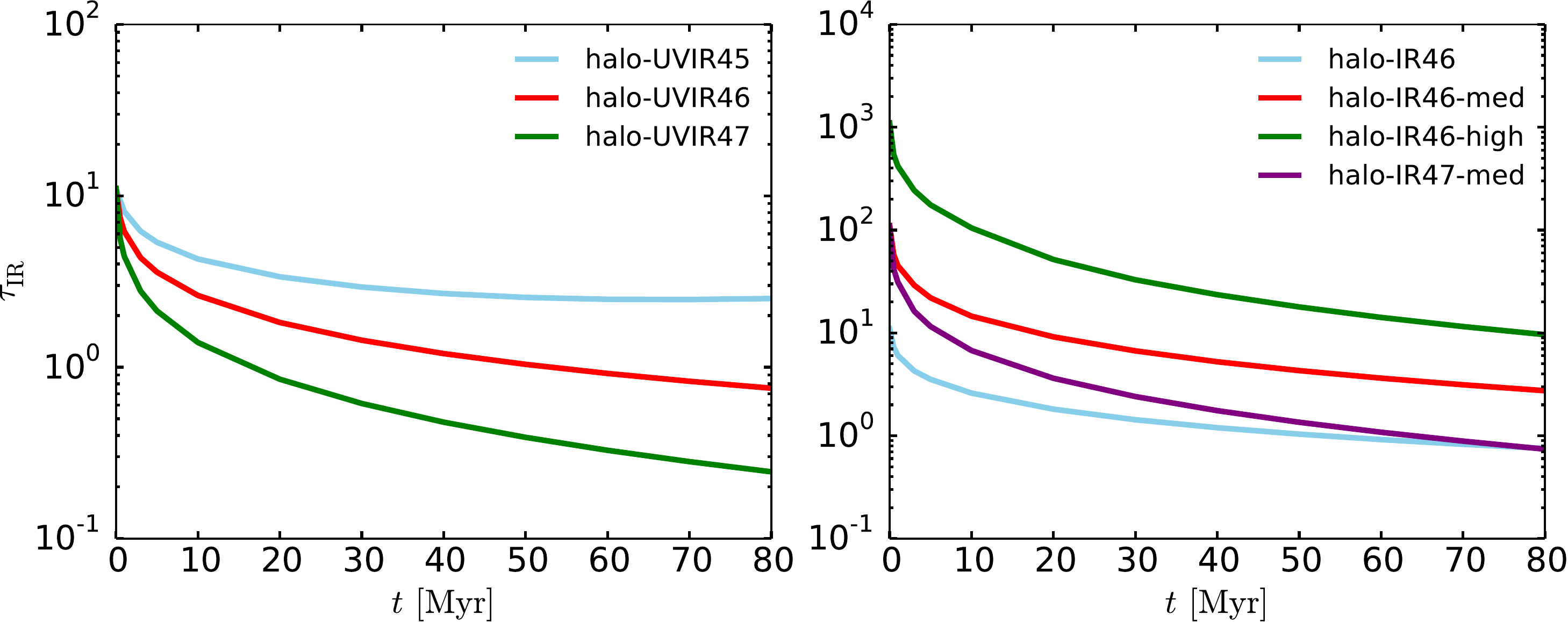}
\caption{The time evolution of the IR optical depth $\tau_{\rm IR}$, measured out to the virial radius of the halo, in the main simulations of our suite. On the left, we show the optical depth evolution in our halo-UVIR set of simulations. In this case, $\tau_{\rm IR}$ lies mostly within the range $(1 \-- 10)$, though it can drop to $\sim 0.1$ in our halo-UVIR47 simulation, where the shell reaches very high radii. On the right, we show the evolution of $\tau_{\rm IR}$ in our halo-IR suite of simulations, where we investigate the effect of very high IR optical depths. In the most extreme case, halo-IR47-high, $\tau_{\rm IR}$ is in the range $(10 \-- 1000)$.}
\label{fig_tauevolv}
\end{figure*}

As envisaged in analytic models, radiation pressure sweeps up intervening material into outwardly expanding shells. 
The outflow structure is illustrated on the left-hand panel of Fig.~\ref{fig_shell}, where projected (volume-weighted) UV- and IR energy densities as well as the gas density are shown for simulation halo-UVIR46 at $t \,=\, 40 \, \rm Myr$.
Despite the Cartesian geometry of the {\sc Ramses} grid cells, the shell retains a remarkably spherical geometry. 
Its inner boundary\footnote{The boundary is computed by taking the gradient of the density profile and locating the radius corresponding to the maximum value.} (shown as a solid white line) separates the shock-heated ambient medium from a radiation-filled cavity, while its outer boundary (seen as a sharp contrast at the outer edge of the shell) constitutes a shock front propagating into the ambient medium.

At the top of the right-hand panel of Fig.~\ref{fig_shell}, gas density profiles are shown for simulations halo-UVIR46 (thick lines) at various times.
For reference, we also show profiles for simulation halo-UVIR46-c, which is identical to halo-UVIR46 with the exception that it is performed using the full speed of light (thin lines).
The flow structure can be clearly divided into different sections.
Starting at large radii, these include the mildly perturbed ambient medium (a portion of which is, as will be shown, `pre-accelerated'), the shell of shock-heated swept-up gas at intermediate radii and the low density cavity at the centre.
Dotted lines show the position of the maximum density gradient, which was found by taking the gradient of the density profile, and traces the approximate location at which the inner low density cavity and the shell meet.
This is also the location at which most UV radiation is absorbed and, hence, where most of the initial acceleration takes place.
The discontinuity is broadened as a consequence of finite resolution, but also due to the finite mean free path of the impinging UV photons.
This is simply given by $l \,=\, (\rho \kappa)^{-1}$ and amounts to $\approx 0.3 \, \rm kpc$ at $R \,\approx\, 1.3 \, \rm kpc$ and $\approx 3.2 \, \rm kpc$ at $R\,\approx\, 6 \, \rm kpc$ in good agreement with the trend of an increasing discontinuity width seen in Fig.~\ref{fig_shell} (note that the axes are logarithmic). 
The waves seen at low radii ($\lesssim 500 \, \rm pc$) in the profiles, particularly in the radiation energy density curves shown in the central panel, are caused by finite resolution; the bin size becomes comparable to the cell sizes as the grid is de-refined in these low density regions. The results for simulation halo-UVIR46-c (thin lines) show that the gas density profile is fully converged with respect to the speed of light.
 
The middle panel shows UV (solid lines) and IR (dashed lines) energy density profiles (calculated as $cE$, where $E$ is the radiation energy density) for the same simulation times as in the top panel, normalised by the flux $\mathbf F_{\rm 0}$ that would be expected for unobscured radiation emitted by a point source, i.e. by $F_{\rm 0} \,=\, L/(4 \pi R^2)$.
The UV flux drops as $R^{-2}$ at radii $R < R_{\rm in}$ as expected, but falls more steeply beyond the inner boundary of the shell, where the UV photons are absorbed most efficiently. 
The decline in UV flux across the shell is greater at early times, when the optical depth is highest. 
The optical depth, however, drops as the shell expands to larger radii (see Fig.~\ref{fig_tauevolv}).
Thus, at later times, a higher proportion of UV radiation leaks out of the shell into the ambient medium, bringing the flux into better agreement with $F_{\rm 0}$.

As the UV radiation is re-emitted in the IR, part of the IR flux is directed towards the source, crossing the optically thin region before it is re-absorbed at some other location in the shell.
Thus, while the UV band energy density falls off with increasing radius (as $R^{-2}$) within the central cavity, IR radiation fills the same region homogeneously (as illustrated Fig.~\ref{fig_diffusion_schematic} in the Appendix).
IR radiation emitted outwardly from the shell's outer boundary can then couple to gas ahead of the forward shock, pre-accelerating it to high speeds $\lesssim 100 \, \rm km \, s^{-1}$ \citep[see also][]{Ishiki:16}, i.e. about $3 \-- 5$ times lower than the speed of the shell itself, in the case shown.
For halo-UVIR46, the radius $R_{\rm acc}$ at which the outward radial velocity drops below $100 \,\rm\, km \, s^{-1}$ is indicated in Fig.~\ref{fig_shell} with a white dotted line. 
The size of the volume enclosed gives a visual impression of the region within which pre-acceleration takes place.
Radiation pressure thus couples to gas both contained in the main outflow (the shell) and in the ambient medium beyond.
The main effect of the reduced speed of light approximation, shown by comparing thin and thick lines on the right-hand side of Fig.~\ref{fig_shell}, is the moderately less spatially extended pre-accelerated region at the beginning of the simulation, which is caused by the somewhat shorter light-crossing times.

The IR energy density (shown in Fig.~\ref{fig_shell} with dashed lines) is to good approximation constant up to a radius $\approx R_{\rm in}$ and the ratio $cE/F_{\rm 0}$ therefore rises as $R^2$.
This peaks at a value a factor $\approx 4$ greater than that for UV radiation, because IR radiation is trapped within the expanding shell.
IR radiation trapping is more efficient at earlier times when the shell is closer to the source and occurs also beyond the shock radius.
The shell's column density drops as the shell moves to larger radii and more of the IR flux escapes\footnote{Note that the steep drop in IR flux shown in Fig.~\ref{fig_shell} at $t \, = \, 1 \, \rm Myr$ is caused by the finiteness of the speed of light rather than photon absorption. This feature is, accordingly, absent in halo-UVIR46-c.}.
As long as the shell remains optically thick in the UV, the radiation energy density is dominated by the IR at radii larger than the shock radius.

The bottom panel of Fig.~\ref{fig_shell} shows mass-weighted radial velocity profiles. 
For this panel, we take into account only gas with density greater than $10^{-30} \, \rm g \, cm^{-3}$ in order to exclude the low density cavity at the centre, where most gas has reached the density floor allowed by {\sc Ramses-RT}. 
The also high, though gradually declining, outflow speeds seen at radii greater than the shell's outer boundary are caused by local interactions between the ambient medium and the portion of IR radiation that leaks out of the shell.
This `pre-acceleration' effect \citep{Ishiki:16} is most pronounced at early times when the optical depth in the region just outside the shell is still very high.
Comparison with simulation halo-UVIR46-c indicates that the velocity of the pre-accelerated gas is moderately underestimated in simulations performed with a reduced speed of light of $0.1 c$, though only at early times.
Note also that the peak velocity of the outflows, which is $\lesssim 1000 \, \rm km \, s^{-1}$, decreases with time. This is caused both by the declining optical depth, but also due to momentum conservation; the shell becomes mass-loaded as it expands and must decelerate.
Already at $t \, = \, 40 \, \rm Myr$, the shell has slowed down to propagate at $\approx 200 \, \rm km \, s^{-1}$, which is about twice as high as the speed of the sound of the gas in the halo. In this regime, the confining pressure of the ambient gas medium also contributes to the deceleration of the outflow.

In the next section, we compare the propagation of the radiation-driven shells with the analytic models presented in Section~\ref{sec_analyticalsolutions} in detail.

\subsection{Comparison with analytic models}

\begin{figure*}
\centering 
\includegraphics[scale = 0.6]{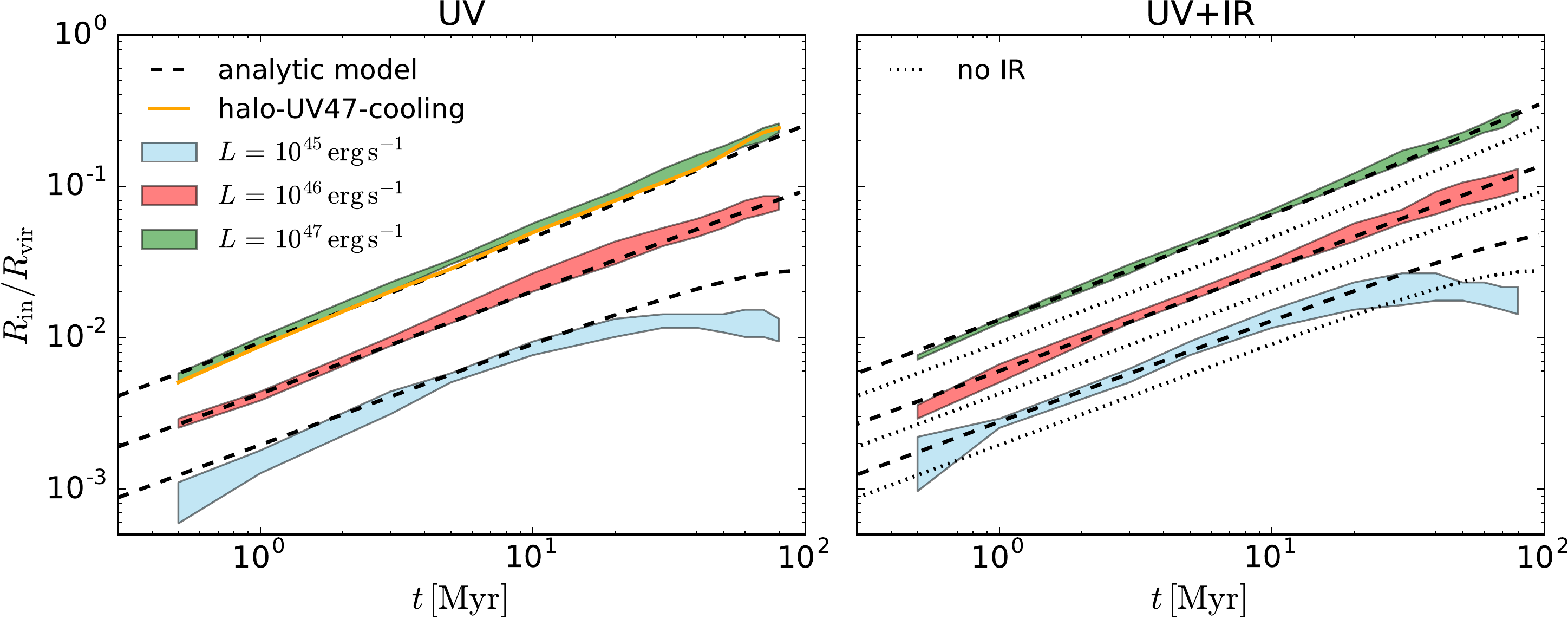}
\caption{Comparison between numerical (coloured bands) and analytic (dashed lines) radiation pressure-driven shell solutions for setups including only a UV band (left) and including reprocessing of UV photons to IR (right). We show the single scattering solutions on the right-hand panel as dotted lines in order to highlight the effects of the IR boost. We compare the location of the tangential discontinuity $R_{\rm in}$ with predictions based on the analytic model of \citet{Thompson:15} as a function of time. We investigate AGN luminosities in the range $10^{45} \-- 10^{47} \, \rm erg \, s^{-1}$, finding remarkably close agreement with the analytic solutions in all our simulations as long as the AGN luminosity is high $L \gtrsim 10^{46} \, \rm erg \, s^{-1}$. Radiative cooling affects the structure and compressibility of the outflowing shell but does not affect its propagation history (orange line). For lower luminosities $L \, = \, 10^{45} \, \rm erg \, s^{-1}$, the simulations generate transsonic shells that are efficiently confined by the thermal pressure of the ambient medium. This analysis indicates that as long as $\tau \lesssim 10$, radiation pressure indeed imparts a momentum flux of $(\tau_{\rm IR} + (1 - e^{-\tau_{\rm UV}}))  L/c$ as expected. }
\label{fig_shellsolutions}
\end{figure*}

In order to compare the time evolution of the shell to predictions based on Eq.~\ref{equation_of_motion_sim}, we must track the shell's position.
For simulations in which the outward force is dominated by single-scattering radiation pressure and/or mildly boosted by IR radiation, our results are not very sensitive to this definition and agree with each other regardless of whether we define the location shell as the radius of steepest density gradient or of maximum density.
However, as we shall see, shells driven out due to strong IR radiation pressure are not thin and the results become more sensitive to our definition.

While the location of steepest density gradient is a more suitable measure of the shell location, since it traces the region in which the radiation force is applied, tracking the shell according to the location of steepest gradient is unreliable since there is a chance that waves propagating within the low density cavity are selected instead.
We therefore opt for a straightforward definition that is more easily reproducible.
We first construct radial profiles of the density for all our snapshots using $200$ logarithmic bins in the radial range $[0.1 \, \mathrm{kpc}, 200 \, \mathrm{kpc}]$.
The uncertainty in the shell's location is then quantified by considering the region bounded by the location of the density peak and of the radius at which the density rises to $25 \%$ of its maximum value.

\subsubsection{The case of $\tau \lesssim 10$}
\label{sec_lowtau}

We start by investigating simulations for which the initial optical depth in the UV (and IR, if included), measured out to the virial radius, is $\tau_{\rm 0} \approx 11$. 
The relevant simulations are therefore the halo-UV and the halo-UVIR set (see Table~\ref{table_sims}). 
For reference, we plot the evolution of the IR optical depth for the halo-UVIR simulations on the left-hand panel of Fig~\ref{fig_tauevolv}.

The shell radius as a function of time is shown for all simulations with coloured bands in Fig.~\ref{fig_shellsolutions}.
The left-hand panel shows results for simulations for which reprocessing into IR radiation is disabled (halo-UV simulations), while the right-hand panel shows results for simulations in which re-emission in the IR and multi-scattering radiation pressure are allowed (halo-UVIR simulations).

Solutions to Eq.~\ref{eq_shellmotion_dim_nfw} for an identical NFW profile and matching AGN luminosities are shown with dashed lines in all panels.
When computing the solutions to the equation of motion, we assumed an initial radius of $80 \, \rm pc$ equal to our spatial resolution.
Overall, the agreement in shell radius evolution between the analytic models and the simulations is very close, although this is particularly the case when AGN luminosities are high ($L \gtrsim 10^{46} \, \rm erg \, s^{-1}$).
For our lowest luminosity simulation ($L \,=\, 10^{45} \, \rm erg \, s^{-1}$), the shell radius, however, falls systematically below the analytic solution and the shell reaches a turn-around radius just over a factor of two below that predicted in the analytic models, at a time about twice as early.
Despite this mismatch, the behaviour of shells is qualitatively similar, i.e. they either expand to large radii or they stall early on if the AGN luminosity is not sufficiently high.

At lower luminosities, the propagation of the shell is resisted by the confining pressure of the gas halo, as was also seen in the momentum-driven solutions of \citet{Costa:14}.
For completeness, we verify that the shell's density contrast increases with AGN luminosity in line with the expectation that a higher momentum input rate generates stronger shocks. 
Notably, the density contrast falls off with time for halo-UV46 and approximately satisfies $\Delta \rho/\rho \,=\, 1$ at $t \,\gtrsim\, 50 \, \rm Myr$, a time when the shell in the simulations starts to slow down with respect to the analytic solution.
For halo-UV45, the density contrast satisfies $\Delta \rho/\rho \,\lesssim\, 1$ for almost the entire duration of the simulation.

If radiative cooling is turned on, as is the case for halo-UV47-cooling, then the material compressed in the strong shock radiates away its thermal energy and collapses to form an even thinner shell.
The orange line on the left-hand panel of Fig.~\ref{fig_shellsolutions} shows the evolution of $R_{\rm in}$ with time in simulation halo-UV47-cooling.
The match with the simulation without cooling is very close.
Radiative cooling merely modifies the shell's structure; the expansion of the shell as a whole is driven by the pressure exerted by radiation when this is absorbed in its interior.
A comparable situation arises in the context of wind-based feedback models; even if an outflowing shell is momentum-driven, there could be a hot component of shocked ambient gas that does not affect its overall dynamics. 

\begin{figure*}
\includegraphics[scale = 0.475]{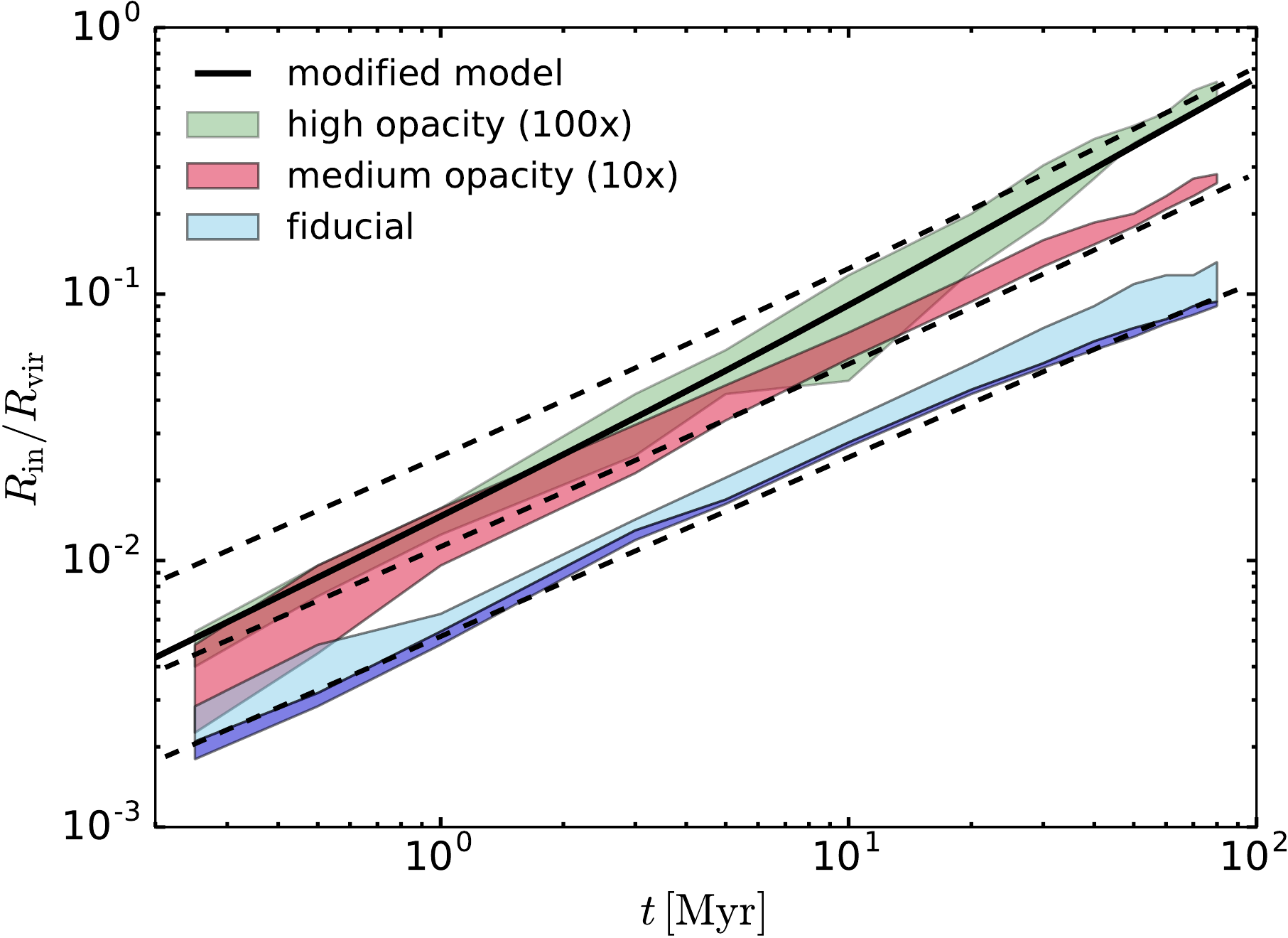}
\includegraphics[scale = 0.475]{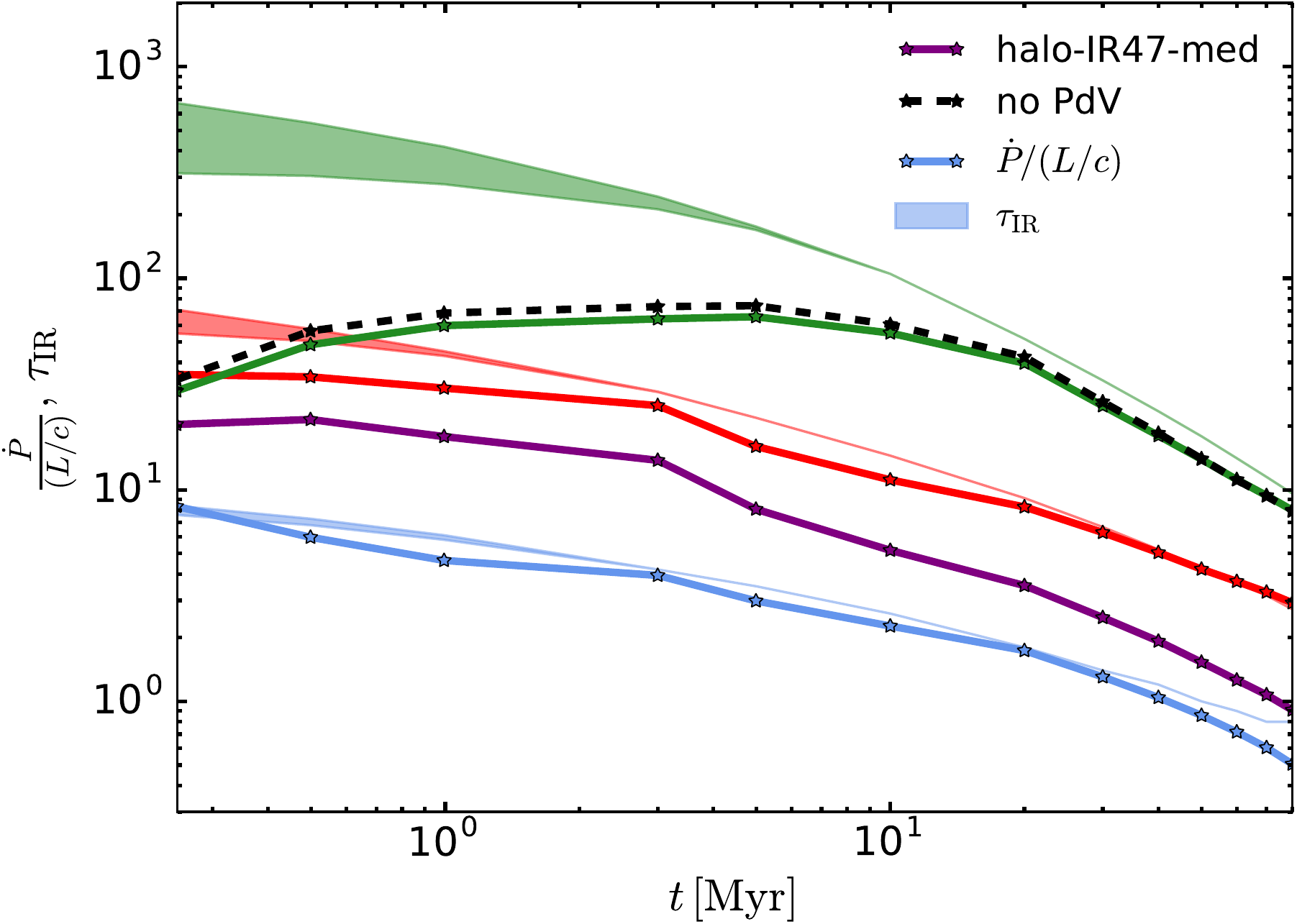}
\caption{Left: The time evolution of radiation pressure-driven shells in highly optically thick conditions. The dashed black lines provide the solution to the equation of motion of radiatively-driven shells, while the coloured bands show the simulated results. While the optical depths are low $\lesssim 10$ (as in the case of halo-IR46), the match between analytic and numerical solutions remains close. A good agreement exists also in simulations halo-IR46-med, though not for halo-IR46-high, for which the simulated shell is always at lower radii than predicted. The thick black line shows the result of a modified equation of motion in which we attempt to model the suppression of momentum transfer between optically thick gas and IR radiation in the case of long diffusion times. Its predictions compare much better with the numerical solution. The dark blue shaded region for simulation halo-IR46 shows the region bounded by the radii corresponding to $10 \% \rho_{\rm max}$ and $25 \% \rho_{\rm max}$. Right: The momentum flux in our simulations including multi-scattering radiation pressure (solid lines) and optical depth measured out to the edge of the accelerated region (lower bound of shaded region) and to the virial radius (upper bound of shaded regions). While the boost in the momentum flux closely follows the optical depth in our simulations with $\tau_{\rm IR} \lesssim 50$, it is significantly lower in our simulations with $\tau_{\rm IR} \gtrsim 50$. In particular, we never find momentum boosts greater than $\approx 60 L/c$ despite initial optical depths of $\gtrsim 500$. We conclude that $\dot{P} / (L/c) < \tau_{\rm IR}$ for high $\tau_{\rm IR}$. The black dashed line shows the result of a simulation identical to halo-IR46-high, but performed without the `PdV' term that governs transfer of energy between the trapped radiation fluid and the gas. The purple line shows the result for simulation halo-IR47-med, showing that the maximum momentum boost drops with increasing AGN luminosity.}
\label{fig_shellopacity}
\end{figure*}

The agreement between analytic models and simulations is remarkably close also for simulations including scattering of IR photons, as shown in the right-hand panel of Fig.~\ref{fig_shellsolutions}. 
Since the analytic models take into account an additional pressure term $\tau_{\rm IR} L/c$ for IR radiation pressure, the close agreement in itself shows that IR radiation efficiently boosts the outflow momentum in our simulations.
In order to ease comparison with the case in which multi-scattering is not accounted for, we also show solutions without IR radiation as dotted lines.
All simulated shells propagate more rapidly than would have been expected if IR multi-scattering was neglected.
The slight slow-down seen in simulation halo-UVIR46 at late times also occurs when the shell slows down to about the gas sound speed through which it propagates. 

We conclude that, as long as $\tau_{\rm IR} \lesssim 10$, trapped IR radiation can build up a significant pressure gradient and boost the momentum injection into the material surrounding the AGN.
In this case, the match with the predictions of the analytic models reviewed in Section~\ref{sec_analyticalsolutions} appears to be excellent, provided that we are in the strong shock regime.

This agreement, however, breaks down when we consider optical depths that are much higher ($\tau_{\rm IR} \gtrsim 50$), as we show next. 

\subsubsection{The case of $\tau \gg 10$}

We focus here on simulations in which the dust IR opacity is increased by factors of $10$ (halo-IR46-med) and $100$ (halo-IR46-high) compared to the fiducial value (halo-IR46). 
The right-hand panel of Fig.~\ref{fig_tauevolv} provides the IR optical depths probed in this section, which are all in the range of ${\lesssim 0.8 \-- 1000}$. 
Other properties of the simulations investigated in this section are summarised in Table~\ref{table_sims}.
Note that for all simulations discussed in this section, we take a fixed AGN luminosity of $L \, = \, 10^{46} \, \rm erg \, s^{-1}$ and use the full speed of light.

We start by comparing the simulated results to solutions to the equation of motion (Eq.~\ref{equation_of_motion_sim}).
On the left-hand-side of Fig.~\ref{fig_shellopacity}, we show the radius of the outflowing shell as a function of time in our various simulations.
Shells are much thicker than they were in our simulations performed with mild IR optical depths, as can be seen from the thicker shaded regions.
This is not surprising since a majority of the AGN radiation is now trapped inside the shell, pressurising it from within.
As a consequence, predictions based on the analytic models of Section~\ref{sec_analyticalsolutions}, which assume the shell width to be negligible, have to be compared only crudely with the results of our simulations.

For simulation halo-IR46 ($\tau_{\rm IR} < 10$), the agreement with the analytic solution is, as before, close.
The numerical solution is slightly above the analytic solution, though both could be reconciled by changing the definition of shell radius, which, as we have seen, is ill-defined in simulations with IR radiation pressure.
The dark blue band in Fig.~\ref{fig_shellopacity}, which encompasses the region for which the gas density is between $10 \% \rho_{\rm max}$ and $25 \% \rho_{\rm max}$ indeed shows a better match with the analytic solution.

For the two simulations with enhanced optical depths, the expanding shells initially go through a period during which their inner radii are systematically lower than in the analytic model.
For halo-IR46-med, this period lasts only less than 1 Myr, while for halo-IR46-high, it lasts about 30 Myr.
At late times, and in both cases, the shells `catch-up' with the analytic solution following it closely thereafter. 

The slower propagation of the shells at early times occurs because they fail to absorb a momentum $\tau_{\rm IR} L/c$ as assumed in Eq.~\ref{equation_of_motion_sim}.
On the right-hand panel of Fig.~\ref{fig_shellopacity}, we provide an estimate of the momentum flux between IR radiation and outflowing gas.
We first measure the total radial momentum in the domain in our various snapshots and then take the time derivative, normalising the result to $L/c$.
The maximum momentum flux is $\approx 7 L/c$ in our fiducial run\footnote{For the fiducial simulation, the momentum flux is lower than the optical depth at late times, because the shell decelerates as it becomes confined by the pressure of the ambient medium. We have checked that the actual radiation force is still $\tau_{\rm IR} L/c$.} with $\mathbf \tau_{\rm 0} \approx 11$, $\gtrsim 30 L/c$ in our simulation with $\mathbf \tau_{\rm 0} \approx 110$ up to just above $60 L/c$ for the run with $\mathbf \tau_{\rm 0} \approx 1100$.
The scaling of momentum flux with optical depth is therefore sub-linear, i.e. $\dot{P} / (L/c) < \tau_{\rm IR}$. 
Moreover, the momentum boost achieved depends on the AGN luminosity. The purple line on the right-hand panel of Fig.~\ref{fig_shellopacity} shows the resulting momentum flux for a simulation identical to halo-IR46-med but with an AGN luminosity 10 times higher; the maximum momentum flux is now $\approx 20 L/c$ instead of $\approx 30 L/c$ \citep[see also][]{Bieri:17}.

Since the expanding shells retain their geometry and efficiently confine the radiation, the origin of this effect cannot be attributed to radiation leakage through low density channels caused by instabilities \citep[see e.g.][]{Krumholz:09, Krumholz:12, Rosdahl:15a, Bieri:17}. Nor is the apparent inefficiency of momentum transfer in the highly optically thick regime caused by the declining optical depth; this effect has already been included in the analytic solutions.

We also rule out the additional limitation that arises if radiatively-driven shells enter the `photon tiring' regime \citep{Owocki:04}. This occurs when the work done in accelerating the shell out of its halo becomes comparable to the energy available in the radiation field. We verified that our results (particularly those with highest optical depth) are not sensitive to this effect by (1) performing one simulation identical to halo-IR46-high, however excluding the `PdV' work term that transfers energy between the radiation field and the fluid (result shown as dashed black line on the right-hand panel of Fig.~\ref{fig_shellopacity}) and (2) by verifying that the kinetic and gravitational potential energy of shells with the energy deposited by the AGN are always lower by a factor of a few than the net energy injected by the AGN. 

Instead, we argue that a momentum boost lower than $\tau_{\rm IR}$ occurs due to the too long diffusion times associated with the radiation transport through the outflowing shell. If the diffusion time becomes comparable or exceeds the shell's flow time, it becomes impossible for it to receive a momentum boost of $\tau_{\rm IR} L/c$.

\subsubsection{The momentum imparted by IR radiation and the upper limit of the IR boost factor}
\label{sec_momenergy}

Since it diffuses through the outflowing gas, the trapped IR radiation front travels at an effective speed which is lower than the speed of light $c$.
In highly optically thick conditions, the flux $\vec{F}$ of trapped radiation depends on the radiative energy density gradient $\vec{\nabla} E$ as
\begin{equation}
\vec{F} \, \approx \, - \frac{cl}{3} \vec{\nabla}{E} \, ,
\end{equation}
where $l$ is the mean free path of the IR photons \citep[see][]{Rosdahl:15a}. 
The expression $\frac{cl}{3} \equiv D$ gives the diffusion coefficient and can be used to obtain an approximate estimate for the characteristic `diffusion' or `trapping timescale' as $t_{\rm trap} \,=\, \delta^2 / D$, where $\delta$ is the width of the shell, as well as an effective propagation speed $v_{\rm eff}$.
The latter is approximately given by
\begin{equation}
v_{\rm eff} \, \sim \, D/\delta \, = \, \frac{c}{3\tau_{\rm IR}} \,\approx \, 10^3 \left( \frac{\tau_{\rm IR}}{100} \right)^{-1} \, \rm km \, s^{-1} \, ,
\label{eq_effspeed}
\end{equation}
where we have used $\tau_{\rm IR} \,=\, \delta /l$.
For very high optical depths $\gtrsim 100$\, the effective speed of the trapped radiation front drops to $\approx 1000 \, \rm km \, s^{-1}$ and becomes lower than the typical propagation speed of AGN-driven outflows.

Efficient momentum coupling between trapped radiation and the outflowing material is to be expected when the outflow time $t_{\rm flow} \approx \frac{\delta}{v_{\rm out}} \gg t_{\rm trap}$, such that a high enough number of scatterings can take place. In other words, the interaction is efficient as long as the trapped radiation fluid has enough time to build up a high enough radiation pressure.
The maximum optical depth $\tau_{\rm max}$ at which efficient coupling can take place at a given outflow velocity $v_{\rm out}$ is therefore
\begin{equation}
\tau_{\rm max} \lesssim \frac{c}{3v_{\rm out}} \,=\, 100 \left( \frac{v_{\rm out}}{10^3 \mathrm{km \, s^{-1}}} \right)^{-1} \, .
\label{eq_couplinglow}
\end{equation}
For typical outflow speeds $1000 \-- 3000 \, \rm km \, s^{-1}$, efficient momentum transfer (high momentum boosts) can only take place at optical depths lower than $ \approx 100 \-- 30$, respectively. 

\begin{figure}
\includegraphics[scale = 0.44]{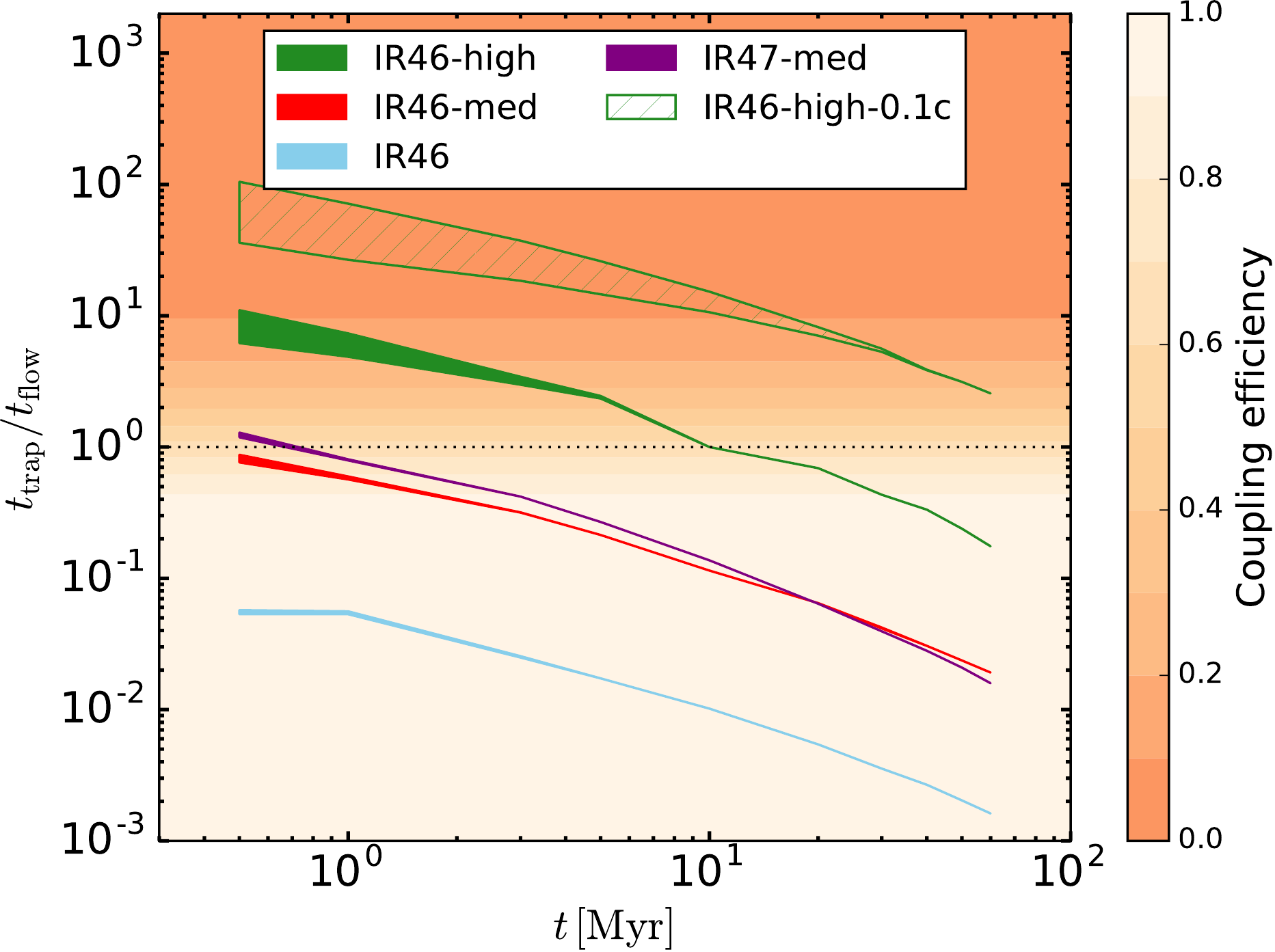}
\caption{The time evolution of the ratio of trapping (diffusion) time to flow time in our simulations of IR optically thick gaseous haloes. Bands are colour-coded as in Fig.~\ref{fig_shellopacity}. The coloured shade identifies the coupling efficiency, i.e. the fraction of $\tau_{\rm IR} L/c$ that can be used to impart outwardly radial momentum. Both high velocity shells and very high IR optical depths lead to low coupling efficiency. Even for simulation halo-IR46-med, which yields realistic IR optical depths, the initial coupling efficiency $\approx 50\%$.}
\label{fig_coupling}
\end{figure}

In Fig.~\ref{fig_coupling}, we plot the ratio between diffusion and outflow times $t_{\rm trap}/t_{\rm flow} \, = \, 3 (v/c) \tau$ in our various simulations.
For the shell's radial velocity, we have taken the mean mass-weighted radial velocity of gas between radii of $25 \% \rho_{\rm max}$ and $\rho_{\rm max}$.  
For simulation halo-IR46-high, momentum transfer between radiation and gas is suppressed until a time $t \approx 10 \rm Myr$, roughly when the simulated and analytic solutions converge and the momentum boost is close to its peak value (see Fig.~\ref{fig_shellopacity}).
The effect of AGN luminosity becomes clear by comparing the curves for simulations halo-IR46-med and halo-IR47-med (red and purple bands, respectively). The ratio $t_{\rm trap} / t_{\rm flow}$ is systematically higher for higher AGN luminosities since the shell is accelerated to higher speeds.
While the offset in ($t_{\rm trap} / t_{\rm flow}$) for halo-IR47-med and halo-IR46-med appears to be small, we note that in the former, the shell is initially in the regime in which $t_{\rm trap} \gtrsim t_{\rm flow}$, just over the boundary over which the coupling efficiency is rapidly suppressed. As shown by the colour-scale, the difference in coupling efficiency between these two cases is about $20 \%$.

We hence attempt to modify the equation of motion (Eq.~\ref{equation_of_motion_sim}) to include the suppression of radiation-gas momentum transfer in this regime.
If we assume that the coupling is exponentially suppressed as $t_{\rm trap} > t_{\rm flow}$, i.e. 
\begin{equation}
\tau_{\rm IR} \frac{L}{c} \rightarrow \tau_{\rm IR} \frac{L}{c} \left(1 - e^{\frac{-c/v_{\rm out}}{3 \tau_{\rm IR}}}\right) \, ,
\end{equation}
we now find much better agreement between our halo-IR46-high simulation and the analytic model (thick solid line on the left-hand panel of Fig.~\ref{fig_shellopacity}).
The colour scale in Fig.~\ref{fig_coupling} gives the value of the suppression factor $\left(1 - e^{\frac{-c/v_{\rm out}}{3 \tau_{\rm IR}}} \right)$ and can therefore be used to read off the coupling efficiency.
We estimate this to drop to values as low as $\lesssim 10\%$ at early times in simulation halo-IR46-high, which explains why the momentum flux in this simulation is, in fact, comparable to that of simulation halo-IR46-med.

One might argue that optical depths as high as $\tau_{\rm IR} \, = \, 30 \-- 100$ are unlikely to come about.
Measurements of IR optical depth in simulations of galaxy formation, however, yield various individual lines-of-sight along which $\tau_{\rm IR} \, = \, 30 \-- 100$ is possible \citep[e.g.][]{Hopkins:11, Bieri:17}, suggesting that high IR optical depths may indeed occur in a realistic setting.
Various simulations show that radiation leakage through low density channels in systems in which there is substantial density inhomogeneity is a stringent limitation to IR radiation pressure as a feedback mechanism \citep{Krumholz:12, Krumholz:13, Bieri:17}, though the extent to which trapping efficiencies are affected remains highly debated \citep[see e.g.][]{Davis:14, Tsang:15, Zhang:16}.
The main conclusion that can be drawn from our simulations is that in the high optical depth regime $\tau_{\rm IR} \sim 100$, in situations in which the distribution of dense gas has a short response time (the gas has a short flow time), the coupling between trapped radiation and the outflowing column may already be suppressed. The formation of low density channels, in this scenario, would intensify the degree by which the coupling is reduced, but could not be said to be the sole source of inefficiency. Both processes likely play a role.

We also emphasise that, while the flow time is the natural timescale which to compare against the diffusion time in the problem of an outflowing shell, different timescales should be taken into account if relevant for the problem at hand.
For the question of the interaction between IR radiation and optically thick interstellar clouds, for instance, another relevant timescale is the cloud destruction time.
If this is shorter than the diffusion time, we, again, cannot expect a high coupling efficiency between IR radiation and interstellar gas. 

\begin{figure*}
\includegraphics[scale=0.6]{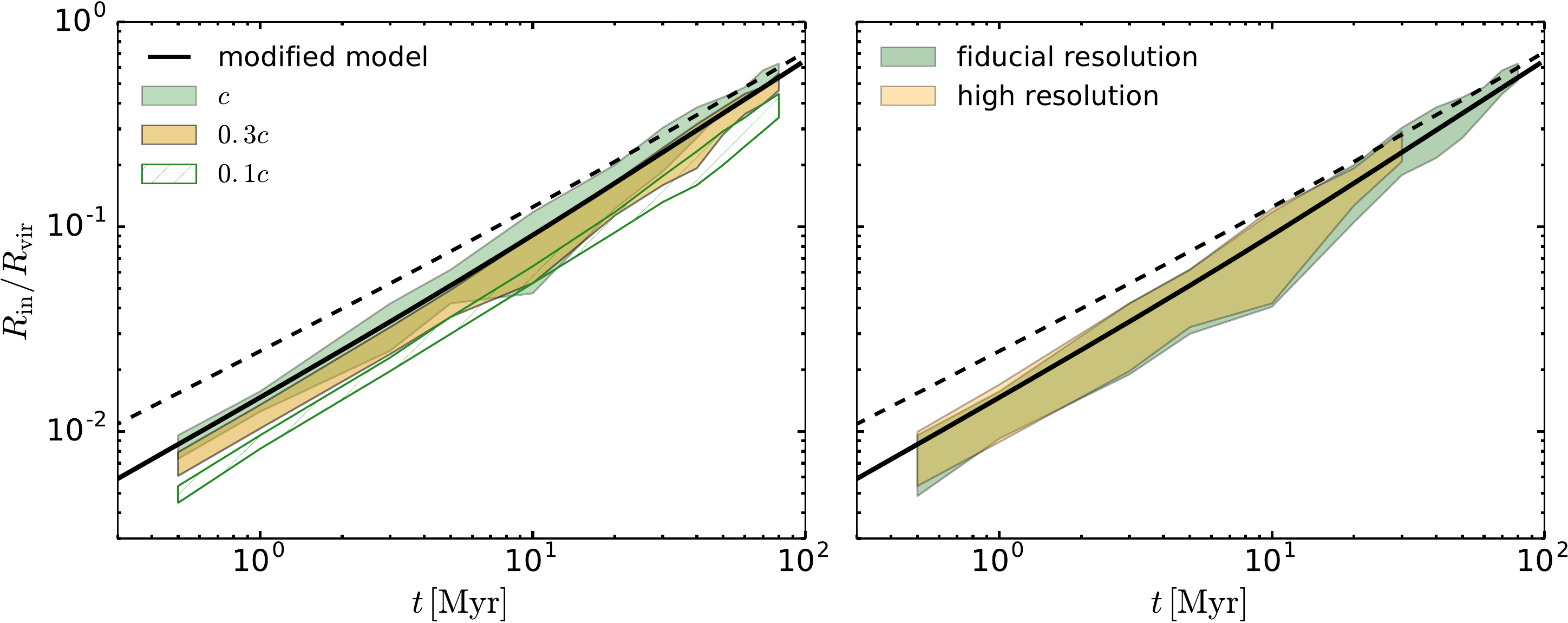}
\caption{Time evolution of highly optically thick radiatively-driven shells using different reduced speed of light values (left panel) and for two different numerical resolutions (right panel). The dashed black line gives the prediction of the analytic model, while the solid black line gives the prediction of the model modified to account for long radiation diffusion times. At full speed of light, the numerical solution agrees with the modified model better, while simulations using reduced speed of light approximations fall below it at all times. The solutions are robust to changes in numerical resolution.}
\label{fig_shell_c}
\end{figure*}

We conclude this section with a brief discussion on how sensitive our results are to changes in reduced speed of light factor as well as numerical resolution.
Perhaps the most worrying consequence of Eq.~\ref{eq_couplinglow} from the numerical standpoint is that it implies that momentum transfer between trapped IR radiation and optically thick gas can be \emph{artificially} suppressed by the use of a reduced speed of light approximation. 
We performed simulations identical to our halo-IR46-high, but this time with reduced speeds of light of $0.1 c$ and $0.3 c$. 
The resulting solutions are shown in Fig.~\ref{fig_shell_c}.
At all times, the shell radius is lower than the simulation with full speed of light by factors of $2 \-- 3$.
Momentum transfer between gas and radiation is suppressed to a great degree purely due to the overestimated diffusion times, as shown by the green hatched region of Fig.~\ref{fig_coupling} for a simulation with a reduced speed of light of $0.1 c$.
Accordingly, the match between simulations performed at low speed of light with the modified analytic model is poor.

On the other hand, we find that our results are robust to changes in numerical resolution. On the right-hand panel of Fig.~\ref{fig_shell_c}, we compare the shell radius evolution using our halo-IR46-high setup between our fiducial simulation and a higher resolution simulation in which the minimum cell size is now $\approx 40 \, \rm pc$, as opposed to $80 \, \rm pc$ (halo-IR46-high-res). Both the radius and the width of the expanding shells in our higher resolution simulation (shown with the yellow shade) are in excellent agreement with our fiducial resolution simulation. This agreement is expected since (i) the IR transparency radius is well resolved in our fiducial simulations and (ii) because the initial optical depth of shells driven in NFW haloes is roughly independent of launch radius for very small radii (see Section~\ref{sec_seldust}).

\section{Discussion}
\label{sec_discussion}

\subsection{Star formation and black hole growth self-regulation}

Assuming the main source of radiation to be a central AGN, we have verified that high IR optical depths imply high outflow momentum fluxes provided that $\tau_{\rm IR} \lesssim 50$ and that optically thick gas has a high covering fraction. 
The necessary conditions might be met in starbursting galaxies in the aftermath of a major merger or, alternatively, as the result of direct cold gas inflow from the cosmic web. The latter has been shown to generate extremely tightly bound gaseous and stellar bulges (with circular velocities $\gtrsim 500 \, \rm km \, s^{-1}$ at scales $\lesssim 1 \, \rm kpc$) in massive galaxies at high redshift \citep{Dubois:12b, Costa:15}.

During a high AGN luminosity event, we would expect a radiation pressure-driven outflow to develop, efficiently expelling the bulk of cold gas from within, at least, the IR transparency radius ($\approx 1 \, \rm kpc$).
The sudden removal of large portions of cold gas, potentially aided by feedback from stars, could efficiently reduce the star formation rate in these systems.
Based on analytical arguments, we also expect the most extreme examples of AGN feedback through radiation pressure on dust to take place at high redshift since AGN luminosities are more likely to be high and also because galaxies are more compact and have higher circular velocities and, hence, a higher critical AGN luminosity for self-regulation.
Once cleared off their gas reservoir, the galaxies would become redder as they age, potentially giving rise to the population of quenched compact ellipticals observed at high redshift \citep[][]{Toft:07, Szomoru:12, Straatman:14, vanDokkum:14, Glazebrook:17}.

Due to the ram pressure confinement caused by cosmological infall, which we have neglected in this study, and AGN duty cycle effects (see Section~\ref{sec_analyticalsolutions}), we might, however, expect a significant portion of outflowing material to stall in the outskirts of the host galaxy. The gas resulting from such a `failed outflow' could be re-accreted onto the central galaxy, where star formation would be revived. However, a possible fate of, at least, part of the ejected nuclear gas is its prolonged presence in the circumgalactic medium of the AGN host halo \citep[see also][]{Ishibashi:16}.
If the ejecta remain cold, or if part of the shocked component cools through a thermal instability \citep[e.g.][]{Zubovas:14}, it might partially account for observations of large deposits of cold gas and metals in the circumgalactic medium of quasar host galaxies \citep[e.g.][]{Hennawi:06, Prochaska:14} as probed by quasar absorption lines.  
A rigorous estimate of the fate of radiation pressure-driven outflows from compact massive galaxies at high redshift, however, requires detailed cosmological radiation-hydrodynamic simulations .

It seems plausible that the efficient removal of cold gas from the nuclei of dense high redshift galaxies would ultimately result in the suppression of black hole accretion.
The simultaneous regulation of star formation and black hole accretion could establish `scaling relations' between the mass of supermassive black holes and the properties of their host galaxies \citep[e.g.][]{Tremaine:02, Gultekin:09, McConnell:13, Kormendy:13, Shankar:16, vandenBosch:16} as proposed by \citet{Murray:05}.
However, the dependence of the IR optical depth on the halo's velocity dispersion means, as we now show, that a close fit to the observed $M_{\rm BH} \-- \sigma$ relation can no longer be straightforwardly achieved.

\subsection{Implications for the $M_{\rm BH} \-- \sigma$ relation}

Consider the hypothetical case in which a radiatively-driven dust shell in a isothermal halo remains optically thick to UV emission throughout its propagation. We, for now, neglect IR multi-scattering.
The equation of motion (Eq.~\ref{equation_of_motion_iso}) gives
\begin{equation}
\frac{d}{dt^\prime} \left[R^\prime \frac{dR^\prime}{dt^\prime} \right] \,=\, \xi - 1 \, ,
\end{equation}
thus when the AGN luminosity is `critical', i.e. $L \,=\, 4f_{\rm gas}c\sigma^4 / G$, the dusty shell is ejected out of its dark matter halo.
Rewriting the AGN luminosity in terms of the `critical mass' $M_{\rm BH, crit}$ for a supermassive black hole radiating at its Eddington limit, i.e. setting
\begin{equation}
L \,=\, L_{\rm Edd} \,\equiv\, \frac{4 \pi G M_{\rm BH, crit}c}{\kappa_{\rm T}} \, ,
\end{equation}
where $\kappa_{\rm T}$ is the Thomson opacity, gives 
\begin{equation}
M_{\rm BH, crit} \,=\, \frac{f_{\rm gas} \kappa_{\rm T}}{\pi G^2} \sigma^4 \, \approx \, 3.8 \times 10^8 \left( \frac{f_{\rm gas}}{0.17} \right) \left( \frac{\sigma}{200 \mathrm{km \, s^{-1}}} \right)^4 \, \rm M_\odot .
\end{equation}

The resulting $M_{\rm BH} \-- \sigma$ relation \citep[see also][]{King:03, Murray:05} has a scaling of $M_{\rm BH} \propto \sigma^4$ and a normalisation close to that reported in observations \citep[e.g.][]{Kormendy:13}.
Note that a scaling with halo velocity dispersion as $\propto \sigma^4$ is, in fact, quite general and remains even for NFW profiles (Eq.~\ref{eq_appcritlum}) and, indeed, for any halo with a peaked circular velocity profile \citep{McQuillin:12}.

However, the origin of an $M_{\rm BH} \-- \sigma$ relation in line with that observed becomes more complicated when including IR multi-scattering. 
In this case, Eq.~\ref{equation_of_motion_iso} becomes
\begin{equation}
\frac{d}{dt^\prime} \left[R^\prime \frac{dR^\prime}{dt^\prime} \right] \,=\, \left( 1 + \frac{1}{R^\prime} \frac{R_{\rm IR}}{R_{\rm \tau}} \right) \xi - 1 \, ,
\label{eq_msigma}
\end{equation}
where $R_{\rm IR} \, \approx 0.5 \, \left( \frac{f_{\rm gas}}{0.17} \right) \left( \frac{\kappa_{\rm IR}}{10 \, \mathrm{cm^2 \, g^{-1}}} \right) \left( \frac{\sigma}{200 \, \mathrm{km \, s^{-1}}} \right)^2 \, \rm kpc$ is the IR transparency radius.
The $1/R^\prime$ dependence of the driving term on the right-hand side means that the scaling of the resulting $M_{\rm BH} \-- \sigma$ relation now depends on the physical scale out to which gas has to be ejected in order for both black hole accretion and star formation to be quenched. In other words, in the presence of IR multi-scattering, the precise scaling between black hole mass and halo velocity dispersion depends on the spatial scale at which the feedback loop is closed.
Using Eq.~\ref{eq_msigma} and the definition of $\xi$, the critical black hole mass required for a dusty shell to reach a radius $R$ is
\begin{equation}
M_{\rm BH, crit} \propto L_{\rm crit} (R) \, = \, \frac{4 f_{\rm gas} \sigma^4 c}{G} \left( 1 + \frac{R_{\rm IR}}{R} \right)^{-1}
\end{equation}
Since $R_{\rm IR} \propto \sigma^2$ in an isothermal halo, we have two limiting cases:
\[ \begin{cases} 
      M_{\rm BH} \propto \sigma^2, & R \ll R_{\rm IR} \\
      M_{\rm BH} \propto \sigma^4, & R \geq R_{\rm IR}.
   \end{cases}
\]
Intriguingly, $R_{\rm IR}$ is of roughly the same spatial scale as that of very compact massive galaxies at $z \gtrsim 2$, suggesting that a $M_{\rm BH} \propto \sigma^4$ scaling could potentially be achieved through this mechanism.

However, we stress that effects such as the need to revert gas inflows at small scales, density inhomogeneity, spatially varying dust-to-gas ratio, the inclusion of additional feedback effects, the shape of the potential well and the cosmological evolution of galaxies should be expected to greatly complicate the simple picture outlined here.
Investigating the scaling relations between black hole mass and galaxy properties clearly deserves to be investigated using radiation-hydrodynamic simulations.

\subsection{IR-driven nuclear winds?}

Another potential consequence of efficient IR trapping is the launching of a nuclear wind from scales just greater than the dust sublimation radius ($\gtrsim 1 \, \rm pc$). Similarly to the scenario proposed by \citet{King:03}, such a hypothetical IR-driven small-scale wind might be expected to collide with the ISM at larger scales, resulting in the formation of an over-pressurised bubble and, possibly, an energy-driven (large-scale) outflow phase.
IR-driving of small-scale winds and their possible consequences on AGN host galaxies have, respectively, been explored with radiative transfer simulations of a dusty torus in \citet{Roth:12} and, implicitly, in the subset of models for slower ($\approx 1000 \, \rm km \, s^{-1}$) nuclear winds presented in \citet{Faucher-Giguere:12}.

The momentum flux carried by an IR-driven wind is, at most, $\dot{P}_{\rm w} \, = \, \dot{M}_{\rm w} v_{\rm w} \, = \, \tau_{\rm IR} (L/c)$. Thus, note that the wind speed decreases with increasing mass outflow rates.
For a homogenous cone with opening angle $\Omega$, we can use the above expression for the momentum flux to obtain the expected launch velocity at a given distance from the AGN as
\begin{eqnarray}
v_{\rm w}  & = &  \sqrt{\frac{\kappa_{\rm IR}}{\Omega R} \frac{L}{c}}\\  & \approx & 3400 \left( \frac{\Omega}{\pi} \right)^{-1/2} \left( \frac{R}{3 \, \mathrm{pc}} \right)^{-1/2} \left( \frac{L}{10^{46} \, \mathrm{erg \, s^{-1}}} \right)^{1/2} \, \mathrm{km \, s^{-1}} , \nonumber
\end{eqnarray}

where we have assumed $\kappa_{\rm IR} \, = \, 10 \, \rm cm^2 \, g^{-1}$.
Note here that a lower covering fraction results in a higher IR optical depth (at fixed mass) and hence a higher launch velocity.
We have additionally assumed that the optical depth is sufficiently low that the IR radiation fluid and the optically thick gas are in equilibrium, an assumption that might break down if $ \tau_{\rm IR} \, \gtrsim \, (30 \-- 100)$, in which case the hydrodynamic response time of the gas (the `flow time') may become comparable to the diffusion time of the IR photons (Eq.~\ref{eq_couplinglow}).

We would expect such a wind to decelerate in a reverse shock as it collides against the ambient medium. The post-shock wind temperature would be $T_{\rm r} \approx10^7 \left( \mu / 0.59 \right) \left( v_{\rm wind} / 1000 \mathrm{\, km \, s^{-1}} \right)^2 \, \rm K$, i.e. much lower than that obtained for an ultra-fast outflow ($T_{\rm r} \gtrsim 10^{10} \, \rm K$). Consequently, radiative cooling through thermal Bremsstrahlung could progress rapidly over a flow time. In this case, an energy-driven phase should be short-lived.
A potentially observational consequence of such an event would be a spatially extended X-ray corona tracing the shock-heated wind fluid.

\subsection{Limitations of our modelling}

In order to establish a close link between our simulations and available analytic models, we had to exclude various physical processes from our modelling.
Notably, we excluded photoionisation as well as thermal energy exchange between the coupled dust and gas fluids, also neglected in most existing analytical treatments of radiative AGN feedback. 
Photo-ionisation, in particular, may result in additional momentum transfer as photo-heated gas expands, boosting the acceleration of an outflow, though we expect this effect to be important mainly for a low dust-to-gas ratio \citep[e.g.][]{Ishiki:16, Bieri:17}.
The inclusion of X-ray radiation, which accounts for as much as $\sim 20 \%$ of the AGN bolometric luminosity, could result in vigorous heating in the central regions of the galaxy, even if the AGN is accreting at a low rate \citep[e.g.][]{Xie:17} and should be addressed in future studies.
Other potentially important physical ingredients are cosmic rays \citep[e.g.][]{Pfrommer:17} and thermal conduction \citep[e.g.][]{Dubois:16, Kannan:17}.
Modelling these processes in a realistic setting would require the inclusion of a cold and dusty phase in radiation-hydrodynamic simulations and should be revisited in the future.

Additionally, we had to take the crude approximation that all gas has a high dust opacity regardless of its temperature.
In reality, we expect dust grains in radiation pressure-driven outflows to be destroyed through thermal sputtering in shock-heated gas.
Photoevaporation of dust should also be important, though at much smaller spatial scales than considered in this study since the equilibrium temperature, given by
\begin{equation}
\begin{split}
T_{\rm r} \, = \, & \left( \frac{L}{4 \pi \sigma_{\rm SB} R^2} \right)^{1/4} \\ \, \approx \, & 35 \left( \frac{L}{10^{46} \, \mathrm{erg \, s^{-1}}} \right)^{1/4} \left( \frac{R}{1 \, \mathrm{kpc}} \right)^{-1/2} \, \rm K ,
\end{split}
\end{equation}
where $\sigma_{\rm SB}$ is the Stefan-Boltzmann constant, is much lower than the typical dust sublimation temperature.

Dust destruction in outflowing gas should decrease the ability of radiation pressure to continue to accelerate it.
However, the possibility that dust survives in unresolved, though resilient, clumps of cold gas potentially entrained in the outflow \citep{McCourt:15, McCourt:16} cannot yet be ruled out particularly if the outflowing gas cools rapidly.
Our simulations therefore implicitly assume that surviving cold clumps of dusty gas pervade the outflow and continue to feel the radiation force.
Also in the context of the analytic models reviewed in Section~\ref{sec_analyticalsolutions}, assumptions such as a high dust-to-gas ratio throughout the galactic halo are undoubtedly optimistic and play in favour of radiation pressure as an AGN feedback model. 
We have attempted to relax this assumption in this paper, but nevertheless stress that more realistic simulations including a cold gas phase and dust physics would be required for more definite statements.

Outflows in our simulations are remarkably spherical and do not appear to develop an instability even in the presence of radiative cooling.
A serious concern is that more realistic clumpy gas distributions could result in IR radiation leakage \citep[e.g.][]{Krumholz:12, Rosdahl:15a, Bieri:17} and, therefore, a decrease in the coupling efficiency. 
While the fragmentation of gas into a small number of clumps certainly leads to IR radiation losses, we here point out that the extent to which the efficiency of IR trapping drops in inhomogeneous media is very sensitive to the details of the initial density distribution. In particular, in media in which the typical distance between cold clouds is smaller than the photon mean free path, a scenario possibly in line with the `fog' picture recently proposed by \citet{McCourt:16}, IR trapping could be efficient \citep[see][for an analogous argument, though for Lyman-$\alpha$ radiation]{Gronke:16}. Interestingly, the simulations performed by \citet{McCourt:16} also suggest that highly fragmented media may accelerate more efficiently. Perhaps the action of radiation pressure on such media is the key to both accelerating the gas rapidly and ensuring high IR trapping efficiencies.

The ability to trap IR radiation in inhomogeneous media depends also on the adopted radiative transfer method, with more accurate (less diffusive) strategies using the `variable Eddington tensor' (VET) method \citep[e.g.][]{Davis:14, Zhang:16} or Monte Carlo methods \citep[e.g.][]{Tsang:15} typically resulting in higher IR trapping efficiencies than both M1 (used here) or `flux limited diffusion' methods.

There are various ways in which additional effects could further reduce the ability of radiation pressure to regulate star formation and black hole accretion.
The setup investigated in this paper consists of a gaseous halo in hydrostatic equilibrium.
Real massive galaxies at high redshift are, however, likely to be fed by rapid streams of cold gas. 
Thus, in addition to overcoming the gravitational potential, radiation pressure would have to be strong enough to revert the inflows \citep{Costa:14}.

\section{Conclusions}
\label{sec_conclusions}

By recasting the equation of motion of radiatively-driven dusty shells in isothermal and NFW haloes into dimensionless form, we have confirmed that there is a critical AGN luminosity above which a galactic outflow can be launched. 
Multi-scattered IR radiation pressure can significantly accelerate mass-loaded outflows to speeds well above $1000 \, \rm km \, s^{-1}$ but only at small $< 1 \, \rm kpc$ scales where gas is optically thick in the IR. 

If the AGN lifetime is taken into account, however, we estimate that radiation pressure-driven shells stall at a scale of about $10\% \-- 20 \%$ of the transparency radius $R_{\rm \tau}$ in isothermal haloes. For halo parameters specific to massive dark matter haloes, i.e. $\sigma \, = \, (250 \-- 300) \, \rm km \, s^{-1}$, the stalling radius is $< 10 \, \rm kpc$.
Ejecting a majority of baryons from NFW haloes is highly unlikely, though strictly not impossible, and would require a combination of long AGN lifetimes, high AGN luminosities and high baryonic concentrations.
The additional force that is required to revert cosmological inflows in a more realistic setting is likely to reduce the efficiency of radiation pressure-driving even further, however, so that radiation pressure on dust is likely to play a role only in the innermost regions of galactic haloes.

While unlikely to prevent catastrophic cooling in massive dark matter haloes, outflows driven by radiation pressure on dust might be strong enough to quench star formation and black hole accretion in massive compact galaxies at high redshift.
The simultaneous suppression of star formation and black hole accretion could in principle result in a $M_{\rm BH} \propto \sigma^4$ scaling close to that which is observed, though the inclusion of IR multi-scattering complicates the scaling.
If radiatively-driven outflows do indeed stall within their parent halo, it is possible that current observations of ubiquitous cold gas in the circumgalactic medium of quasar host haloes, in fact, probe the former cold reservoir of a central compact and gas-rich galaxy dispersed across its halo due to AGN radiation pressure.

We have also carefully compared the predictions of these widely used analytic models to those of numerical simulations performed with the radiation-hydrodynamic code {\sc Ramses-RT}.
Our strategy was to adopt the same assumptions as made in the analytic models, i.e. we followed the hydrodynamic response of the gas component in static isolated dark matter haloes to both single- and multi-scattering AGN radiation pressure.
We performed a large number of simulations in which an AGN was placed at the centre of a gaseous NFW halo, emitting at a constant luminosity.
The solutions to the analytic models were generally found to be well reproduced in our idealised simulations, confirming that the code {\sc Ramses-RT} is able to tackle the problem of IR radiation pressure on dust accurately.

Specifically, we confirm that $\dot{P} \approx \tau_{\rm IR} L/c$ as long as the IR optical depth is mild ($\tau_{\rm IR} \lesssim 10$).
Highly IR optically thick ($\tau_{\rm IR} \gtrsim 50$) shells, on the other hand, propagate at a lower speed than predicted in the analytic models. In particular, the momentum flux $\dot{P} \ll \tau_{\rm IR} L/c$ despite the absence of low density channels that can lead to radiation leakage. We also find that the discrepancy between analytic and numeric solutions is exacerbated by increasing the AGN luminosity. 
The cause of this disagreement is the fundamental limitation for momentum transfer between trapped IR radiation and optically thick gas that occurs when the shell outflow time is comparable or shorter than the radiation diffusion time.
In cases in which the diffusion times are too long, the trapped radiation cannot keep up with the outflowing shell and the feedback coupling efficiency is rapidly suppressed.
We show that when the analytic model provides a poor description of the evolution of the simulated shells due to a high $\tau_{\rm IR}$, a simple modification in the analytic model can recover the numerical solutions much more accurately. We suggest that the term governing the momentum thrust $\tau_{\rm IR} L/c$ should be modified to $\tau_{\rm IR} L/c (1 - e^{-t_{\rm flow} / t_{\rm trap}})$ so that the diffusion timescale and rapid suppression in coupling efficiency when this is similar to the flow time are taken into account.

Despite its limitations, we find that the upper limit to efficient momentum transfer through AGN IR radiation pressure in our simulations is high ($\gtrsim 50 L/c)$. Encouragingly, the coupling efficiency below such extreme optical depths can be sufficiently high to (1) explain existing observations that indicate high momentum boost factors ($\approx 20$) and (2) quench star formation and black hole accretion in massive galaxies. The greatest uncertainty in the feasibility of this mechanism remains the demand for optically thick gas with a high covering fraction around the AGN.
A crucial test to this model will be to apply it to fully radiation-hydrodynamic cosmological simulations.

\section{Acknowledgements}

We thank the anonymous referee for a detailed report.
TC gratefully acknowledges Wasilij Barsukow, Taysun Kimm, Andrew King, Joop Schaye and Benny Tsang for enlightening discussions. We thank Romain Teyssier for developing the code {\sc Ramses} and for making it publicly available. TC is supported by a NOVA Fellowship. JR was funded by the European Research Council under the European Union's Seventh Framework Programme (FP7/2007-2013) / ERC Grant agreement
278594-GasAroundGalaxies. The authors further acknowledge support by the ERC ADVANCED Grant 320596 `The Emergence of Structure during the epoch of Reionization' (PI: M. G. Haehnelt), the ERC Starting Grant 638707 `Black holes and their host galaxies: coevolution across cosmic time' (PI: D. Sijacki) and by the ORAGE project from the Agence Nationale de la Recherche under grant ANR-14-CE33-0016-03. The simulations were performed on the following: the Darwin Supercomputer of the University of Cambridge High Performance Computing Service (http://www.hpc.cam.ac.uk/), provided by Dell Inc. using Strategic Research Infrastructure Funding from the Higher Education Funding Council for England and funding from the Science and Technology Facilities Council, the COSMA Data Centric system based at Durham University, operated by the Institute for Computational Cosmology on behalf of the STFC DiRAC HPC Facility (www.dirac.ac.uk). This equipment was funded by a BIS National E-infrastructure capital grant ST/K00042X/1, STFC capital grant ST/K00087X/1, DiRAC Operations grant ST/K003267/1 and Durham University. DiRAC is part of the National E-Infrastructure. Various simulations were performed on the Dutch national e-infrastructure with the support of SURF Cooperative. This work was sponsored by NWO Exacte Wetenschappen (Physical Sciences) for the use of supercomputer facilities, with financial support from the Nederlandse Organisatie voor Wetenschappelijk Onderzoek (NWO).   

\bibliographystyle{mn2e} 
\bibliography{references}

\begin{thebibliography}{149}
\expandafter\ifx\csname natexlab\endcsname\relax\def\natexlab#1{#1}\fi

\bibitem[{{Aalto} {et~al}\mbox{.}(2012){Aalto}, {Muller}, {Sakamoto},
  {Gallagher}, {Mart{\'{\i}}n}, \& {Costagliola}}]{Aalto:12}
{Aalto} S., {Muller} S., {Sakamoto} K., {Gallagher} J.~S., {Mart{\'{\i}}n} S.,
  {Costagliola} F., 2012, \aap, 546, A68

\bibitem[{{Agertz} {et~al}\mbox{.}(2013){Agertz}, {Kravtsov}, {Leitner}, \&
  {Gnedin}}]{Agertz:13}
{Agertz} O., {Kravtsov} A.~V., {Leitner} S.~N., {Gnedin} N.~Y., 2013, \apj,
  770, 25

\bibitem[{{Angl{\'e}s-Alc{\'a}zar}
  {et~al}\mbox{.}(2017){Angl{\'e}s-Alc{\'a}zar}, {Dav{\'e}},
  {Faucher-Gigu{\`e}re}, {{\"O}zel}, \& {Hopkins}}]{Angles-Alcazar:17}
{Angl{\'e}s-Alc{\'a}zar} D., {Dav{\'e}} R., {Faucher-Gigu{\`e}re} C.-A.,
  {{\"O}zel} F., {Hopkins} P.~F., 2017, \mnras, 464, 2840

\bibitem[{{Aumer} {et~al}\mbox{.}(2013){Aumer}, {White}, {Naab}, \&
  {Scannapieco}}]{Aumer:13}
{Aumer} M., {White} S.~D.~M., {Naab} T., {Scannapieco} C., 2013, \mnras, 434,
  3142

\bibitem[{{Banerji} {et~al}\mbox{.}(2017){Banerji}, {Carilli}, {Jones}, {Wagg},
  {McMahon}, {Hewett}, {Alaghband-Zadeh}, \& {Feruglio}}]{Banerji:17}
{Banerji} M., {Carilli} C.~L., {Jones} G., {Wagg} J., {McMahon} R.~G., {Hewett}
  P.~C., {Alaghband-Zadeh} S., {Feruglio} C., 2017, \mnras, 465, 4390

\bibitem[{{Barai} {et~al}\mbox{.}(2016){Barai}, {Murante}, {Borgani},
  {Gaspari}, {Granato}, {Monaco}, \& {Ragone-Figueroa}}]{Barai:16}
{Barai} P., {Murante} G., {Borgani} S., {Gaspari} M., {Granato} G.~L., {Monaco}
  P., {Ragone-Figueroa} C., 2016, \mnras, 461, 1548

\bibitem[{{Barro} {et~al}\mbox{.}(2016){Barro}, {Kriek},
  {P{\'e}rez-Gonz{\'a}lez}, {Trump}, {Koo}, {Faber}, {Dekel}, {Primack}, {Guo},
  {Kocevski}, {Mu{\~n}oz-Mateos}, {Rujopakarn}, \& {Seth}}]{Barro:16}
{Barro} G. {et~al.}, 2016, \apjl, 827, L32

\bibitem[{{Bieri} {et~al}\mbox{.}(2017){Bieri}, {Dubois}, {Rosdahl}, {Wagner},
  {Silk}, \& {Mamon}}]{Bieri:17}
{Bieri} R., {Dubois} Y., {Rosdahl} J., {Wagner} A., {Silk} J., {Mamon} G.~A.,
  2017, \mnras, 464, 1854

\bibitem[{{Bischetti} {et~al}\mbox{.}(2017){Bischetti}, {Piconcelli}, {Vietri},
  {Bongiorno}, {Fiore}, {Sani}, {Marconi}, {Duras}, {Zappacosta}, {Brusa},
  {Comastri}, {Cresci}, {Feruglio}, {Giallongo}, {La Franca}, {Mainieri},
  {Mannucci}, {Martocchia}, {Ricci}, {Schneider}, {Testa}, \&
  {Vignali}}]{Bischetti:17}
{Bischetti} M. {et~al.}, 2017, \aap, 598, A122

\bibitem[{{Booth} \& {Schaye}(2009)}]{Booth:09}
{Booth} C.~M., {Schaye} J., 2009, \mnras, 398, 53

\bibitem[{{Bower} {et~al}\mbox{.}(2006){Bower}, {Benson}, {Malbon}, {Helly},
  {Frenk}, {Baugh}, {Cole}, \& {Lacey}}]{Bower:06}
{Bower} R.~G., {Benson} A.~J., {Malbon} R., {Helly} J.~C., {Frenk} C.~S.,
  {Baugh} C.~M., {Cole} S., {Lacey} C.~G., 2006, \mnras, 370, 645

\bibitem[{{Cano-D{\'{\i}}az} {et~al}\mbox{.}(2012){Cano-D{\'{\i}}az},
  {Maiolino}, {Marconi}, {Netzer}, {Shemmer}, \& {Cresci}}]{Cano-Diaz:12}
{Cano-D{\'{\i}}az} M., {Maiolino} R., {Marconi} A., {Netzer} H., {Shemmer} O.,
  {Cresci} G., 2012, \aap, 537, L8

\bibitem[{{Cappi} {et~al}\mbox{.}(2009){Cappi}, {Tombesi}, {Bianchi}, {Dadina},
  {Giustini}, {Malaguti}, {Maraschi}, {Palumbo}, {Petrucci}, {Ponti},
  {Vignali}, \& {Yaqoob}}]{Cappi:09}
{Cappi} M. {et~al.}, 2009, \aap, 504, 401

\bibitem[{{Carniani} {et~al}\mbox{.}(2015){Carniani}, {Marconi}, {Maiolino},
  {Balmaverde}, {Brusa}, {Cano-D{\'{\i}}az}, {Cicone}, {Comastri}, {Cresci},
  {Fiore}, {Feruglio}, {La Franca}, {Mainieri}, {Mannucci}, {Nagao}, {Netzer},
  {Piconcelli}, {Risaliti}, {Schneider}, \& {Shemmer}}]{Carniani:15}
{Carniani} S. {et~al.}, 2015, \aap, 580, A102

\bibitem[{{Carniani} {et~al}\mbox{.}(2016){Carniani}, {Marconi}, {Maiolino},
  {Balmaverde}, {Brusa}, {Cano-D{\'{\i}}az}, {Cicone}, {Comastri}, {Cresci},
  {Fiore}, {Feruglio}, {La Franca}, {Mainieri}, {Mannucci}, {Nagao}, {Netzer},
  {Piconcelli}, {Risaliti}, {Schneider}, \& {Shemmer}}]{Carniani:16}
{Carniani} S. {et~al.}, 2016, \aap, 591, A28

\bibitem[{{Ceverino} {et~al}\mbox{.}(2014){Ceverino}, {Klypin}, {Klimek},
  {Trujillo-Gomez}, {Churchill}, {Primack}, \& {Dekel}}]{Ceverino:14}
{Ceverino} D., {Klypin} A., {Klimek} E.~S., {Trujillo-Gomez} S., {Churchill}
  C.~W., {Primack} J., {Dekel} A., 2014, \mnras, 442, 1545

\bibitem[{{Chattopadhyay} {et~al}\mbox{.}(2012){Chattopadhyay}, {Sharma},
  {Nath}, \& {Ryu}}]{Chattopadhyay:12}
{Chattopadhyay} I., {Sharma} M., {Nath} B.~B., {Ryu} D., 2012, \mnras, 423,
  2153

\bibitem[{{Choi} {et~al}\mbox{.}(2012){Choi}, {Ostriker}, {Naab}, \&
  {Johansson}}]{Choi:12}
{Choi} E., {Ostriker} J.~P., {Naab} T., {Johansson} P.~H., 2012, \apj, 754, 125

\bibitem[{{Churazov} {et~al}\mbox{.}(2005){Churazov}, {Sazonov}, {Sunyaev},
  {Forman}, {Jones}, \& {B{\"o}hringer}}]{Churazov:05}
{Churazov} E., {Sazonov} S., {Sunyaev} R., {Forman} W., {Jones} C.,
  {B{\"o}hringer} H., 2005, \mnras, 363, L91

\bibitem[{{Cicone} {et~al}\mbox{.}(2015){Cicone}, {Maiolino}, {Gallerani},
  {Neri}, {Ferrara}, {Sturm}, {Fiore}, {Piconcelli}, \& {Feruglio}}]{Cicone:15}
{Cicone} C. {et~al.}, 2015, \aap, 574, A14

\bibitem[{{Cicone} {et~al}\mbox{.}(2014){Cicone}, {Maiolino}, {Sturm},
  {Graci{\'a}-Carpio}, {Feruglio}, {Neri}, {Aalto}, {Davies}, {Fiore},
  {Fischer}, {Garc{\'{\i}}a-Burillo}, {Gonz{\'a}lez-Alfonso},
  {Hailey-Dunsheath}, {Piconcelli}, \& {Veilleux}}]{Cicone:14}
{Cicone} C. {et~al.}, 2014, \aap, 562, A21

\bibitem[{{Ciotti} \& {Ostriker}(2001)}]{Ciotti:01}
{Ciotti} L., {Ostriker} J.~P., 2001, \apj, 551, 131

\bibitem[{{Costa} {et~al}\mbox{.}(2017){Costa}, {Rosdahl}, {Sijacki}, \&
  {Haehnelt}}]{Costa:17}
{Costa} T., {Rosdahl} J., {Sijacki} D., {Haehnelt} M.~G., 2017, \mnras \,
  (submitted), arXiv:1709.08638

\bibitem[{{Costa} {et~al}\mbox{.}(2014{\natexlab{a}}){Costa}, {Sijacki}, \&
  {Haehnelt}}]{Costa:14}
{Costa} T., {Sijacki} D., {Haehnelt} M.~G., 2014{\natexlab{a}}, \mnras, 444,
  2355

\bibitem[{{Costa} {et~al}\mbox{.}(2015){Costa}, {Sijacki}, \&
  {Haehnelt}}]{Costa:15}
{Costa} T., {Sijacki} D., {Haehnelt} M.~G., 2015, \mnras, 448, L30

\bibitem[{{Costa} {et~al}\mbox{.}(2014{\natexlab{b}}){Costa}, {Sijacki},
  {Trenti}, \& {Haehnelt}}]{Costa:14a}
{Costa} T., {Sijacki} D., {Trenti} M., {Haehnelt} M.~G., 2014{\natexlab{b}},
  \mnras, 439, 2146

\bibitem[{{Cresci} {et~al}\mbox{.}(2015){Cresci}, {Marconi}, {Zibetti},
  {Risaliti}, {Carniani}, {Mannucci}, {Gallazzi}, {Maiolino}, {Balmaverde},
  {Brusa}, {Capetti}, {Cicone}, {Feruglio}, {Bland-Hawthorn}, {Nagao}, {Oliva},
  {Salvato}, {Sani}, {Tozzi}, {Urrutia}, \& {Venturi}}]{Cresci:15}
{Cresci} G. {et~al.}, 2015, \aap, 582, A63

\bibitem[{{Croton} {et~al}\mbox{.}(2006){Croton}, {Springel}, {White}, {De
  Lucia}, {Frenk}, {Gao}, {Jenkins}, {Kauffmann}, {Navarro}, \&
  {Yoshida}}]{Croton:06}
{Croton} D.~J. {et~al.}, 2006, \mnras, 365, 11

\bibitem[{{Davis} {et~al}\mbox{.}(2014){Davis}, {Jiang}, {Stone}, \&
  {Murray}}]{Davis:14}
{Davis} S.~W., {Jiang} Y.-F., {Stone} J.~M., {Murray} N., 2014, \apj, 796, 107

\bibitem[{{Debuhr} {et~al}\mbox{.}(2011){Debuhr}, {Quataert}, \&
  {Ma}}]{Debuhr:11}
{Debuhr} J., {Quataert} E., {Ma} C.-P., 2011, \mnras, 412, 1341

\bibitem[{{Debuhr} {et~al}\mbox{.}(2012){Debuhr}, {Quataert}, \&
  {Ma}}]{Debuhr:12}
{Debuhr} J., {Quataert} E., {Ma} C.-P., 2012, \mnras, 420, 2221

\bibitem[{{Detmers} {et~al}\mbox{.}(2011){Detmers}, {Kaastra}, {Steenbrugge},
  {Ebrero}, {Kriss}, {Arav}, {Behar}, {Costantini}, {Branduardi-Raymont},
  {Mehdipour}, {Bianchi}, {Cappi}, {Petrucci}, {Ponti}, {Pinto}, {Ratti}, \&
  {Holczer}}]{Detmers:11}
{Detmers} R.~G. {et~al.}, 2011, \aap, 534, A38

\bibitem[{{Di Matteo} {et~al}\mbox{.}(2008){Di Matteo}, {Colberg}, {Springel},
  {Hernquist}, \& {Sijacki}}]{DiMatteo:08}
{Di Matteo} T., {Colberg} J., {Springel} V., {Hernquist} L., {Sijacki} D.,
  2008, \apj, 676, 33

\bibitem[{{Di Matteo} {et~al}\mbox{.}(2005){Di Matteo}, {Springel}, \&
  {Hernquist}}]{DiMatteo:05}
{Di Matteo} T., {Springel} V., {Hernquist} L., 2005, \nat, 433, 604

\bibitem[{{Draine}(1995)}]{Draine:95}
{Draine} B.~T., 1995, \apss, 233, 111

\bibitem[{{Dubois} \& {Commer{\c c}on}(2016)}]{Dubois:16}
{Dubois} Y., {Commer{\c c}on} B., 2016, \aap, 585, A138

\bibitem[{{Dubois} {et~al}\mbox{.}(2012{\natexlab{a}}){Dubois}, {Devriendt},
  {Slyz}, \& {Teyssier}}]{Dubois:12}
{Dubois} Y., {Devriendt} J., {Slyz} A., {Teyssier} R., 2012{\natexlab{a}},
  \mnras, 420, 2662

\bibitem[{{Dubois} {et~al}\mbox{.}(2011){Dubois}, {Devriendt}, {Teyssier}, \&
  {Slyz}}]{Dubois:11}
{Dubois} Y., {Devriendt} J., {Teyssier} R., {Slyz} A., 2011, \mnras, 417, 1853

\bibitem[{{Dubois} {et~al}\mbox{.}(2012{\natexlab{b}}){Dubois}, {Pichon},
  {Haehnelt}, {Kimm}, {Slyz}, {Devriendt}, \& {Pogosyan}}]{Dubois:12b}
{Dubois} Y., {Pichon} C., {Haehnelt} M., {Kimm} T., {Slyz} A., {Devriendt} J.,
  {Pogosyan} D., 2012{\natexlab{b}}, \mnras, 423, 3616

\bibitem[{{Duffy} {et~al}\mbox{.}(2010){Duffy}, {Schaye}, {Kay}, {Dalla
  Vecchia}, {Battye}, \& {Booth}}]{Duffy:10}
{Duffy} A.~R., {Schaye} J., {Kay} S.~T., {Dalla Vecchia} C., {Battye} R.~A.,
  {Booth} C.~M., 2010, \mnras, 405, 2161

\bibitem[{{Fabian}(1999)}]{Fabian:99}
{Fabian} A.~C., 1999, \mnras, 308, L39

\bibitem[{{Fabian}(2012)}]{Fabian:12}
{Fabian} A.~C., 2012, \araa, 50, 455

\bibitem[{{Fabian} \& {Iwasawa}(1999)}]{Fabian:99b}
{Fabian} A.~C., {Iwasawa} K., 1999, \mnras, 303, L34

\bibitem[{{Fabian} {et~al}\mbox{.}(2008){Fabian}, {Vasudevan}, \&
  {Gandhi}}]{Fabian:08}
{Fabian} A.~C., {Vasudevan} R.~V., {Gandhi} P., 2008, \mnras, 385, L43

\bibitem[{{Faucher-Gigu{\`e}re} \& {Quataert}(2012)}]{Faucher-Giguere:12}
{Faucher-Gigu{\`e}re} C.-A., {Quataert} E., 2012, \mnras, 425, 605

\bibitem[{{Ferland} {et~al}\mbox{.}(1998){Ferland}, {Korista}, {Verner},
  {Ferguson}, {Kingdon}, \& {Verner}}]{Ferland:98}
{Ferland} G.~J., {Korista} K.~T., {Verner} D.~A., {Ferguson} J.~W., {Kingdon}
  J.~B., {Verner} E.~M., 1998, \pasp, 110, 761

\bibitem[{{Fiore} {et~al}\mbox{.}(2017){Fiore}, {Feruglio}, {Shankar},
  {Bischetti}, {Bongiorno}, {Brusa}, {Carniani}, {Cicone}, {Duras}, {Lamastra},
  {Mainieri}, {Marconi}, {Menci}, {Maiolino}, {Piconcelli}, {Vietri}, \&
  {Zappacosta}}]{Fiore:17}
{Fiore} F. {et~al.}, 2017, \aap, 601, A143

\bibitem[{{Gan} {et~al}\mbox{.}(2014){Gan}, {Yuan}, {Ostriker}, {Ciotti}, \&
  {Novak}}]{Gan:14}
{Gan} Z., {Yuan} F., {Ostriker} J.~P., {Ciotti} L., {Novak} G.~S., 2014, \apj,
  789, 150

\bibitem[{{Gaspari} {et~al}\mbox{.}(2011){Gaspari}, {Melioli}, {Brighenti}, \&
  {D'Ercole}}]{Gaspari:11}
{Gaspari} M., {Melioli} C., {Brighenti} F., {D'Ercole} A., 2011, \mnras, 411,
  349

\bibitem[{{Gayley} {et~al}\mbox{.}(1995){Gayley}, {Owocki}, \&
  {Cranmer}}]{Gayley:95}
{Gayley} K.~G., {Owocki} S.~P., {Cranmer} S.~R., 1995, \apj, 442, 296

\bibitem[{{Genzel} {et~al}\mbox{.}(2014){Genzel}, {F{\"o}rster Schreiber},
  {Rosario}, {Lang}, {Lutz}, {Wisnioski}, {Wuyts}, {Wuyts}, {Bandara},
  {Bender}, {Berta}, {Kurk}, {Mendel}, {Tacconi}, {Wilman}, {Beifiori},
  {Brammer}, {Burkert}, {Buschkamp}, {Chan}, {Carollo}, {Davies}, {Eisenhauer},
  {Fabricius}, {Fossati}, {Kriek}, {Kulkarni}, {Lilly}, {Mancini}, {Momcheva},
  {Naab}, {Nelson}, {Renzini}, {Saglia}, {Sharples}, {Sternberg}, {Tacchella},
  \& {van Dokkum}}]{Genzel:14}
{Genzel} R. {et~al.}, 2014, \apj, 796, 7

\bibitem[{{Glazebrook} {et~al}\mbox{.}(2017){Glazebrook}, {Schreiber},
  {Labb{\'e}}, {Nanayakkara}, {Kacprzak}, {Oesch}, {Papovich}, {Spitler},
  {Straatman}, {Tran}, \& {Yuan}}]{Glazebrook:17}
{Glazebrook} K. {et~al.}, 2017, \nat, 544, 71

\bibitem[{{Gnedin} \& {Abel}(2001)}]{Gnedin:01}
{Gnedin} N.~Y., {Abel} T., 2001, \na, 6, 437

\bibitem[{{Gofford} {et~al}\mbox{.}(2011){Gofford}, {Reeves}, {Turner},
  {Tombesi}, {Braito}, {Porquet}, {Miller}, {Kraemer}, \&
  {Fukazawa}}]{Gofford:11}
{Gofford} J. {et~al.}, 2011, \mnras, 414, 3307

\bibitem[{{Gronke} {et~al}\mbox{.}(2016){Gronke}, {Dijkstra}, {McCourt}, \&
  {Oh}}]{Gronke:16}
{Gronke} M., {Dijkstra} M., {McCourt} M., {Oh} S.~P., 2016, \apjl, 833, L26

\bibitem[{{G{\"u}ltekin} {et~al}\mbox{.}(2009){G{\"u}ltekin}, {Richstone},
  {Gebhardt}, {Lauer}, {Tremaine}, {Aller}, {Bender}, {Dressler}, {Faber},
  {Filippenko}, {Green}, {Ho}, {Kormendy}, {Magorrian}, {Pinkney}, \&
  {Siopis}}]{Gultekin:09}
{G{\"u}ltekin} K. {et~al.}, 2009, \apj, 698, 198

\bibitem[{{Haehnelt} {et~al}\mbox{.}(1998){Haehnelt}, {Natarajan}, \&
  {Rees}}]{Haehnelt:98}
{Haehnelt} M.~G., {Natarajan} P., {Rees} M.~J., 1998, \mnras, 300, 817

\bibitem[{{Hambrick} {et~al}\mbox{.}(2011){Hambrick}, {Ostriker}, {Naab}, \&
  {Johansson}}]{Hambrick:11}
{Hambrick} D.~C., {Ostriker} J.~P., {Naab} T., {Johansson} P.~H., 2011, \apj,
  738, 16

\bibitem[{{Hennawi} {et~al}\mbox{.}(2006){Hennawi}, {Prochaska}, {Burles},
  {Strauss}, {Richards}, {Schlegel}, {Fan}, {Schneider}, {Zakamska}, {Oguri},
  {Gunn}, {Lupton}, \& {Brinkmann}}]{Hennawi:06}
{Hennawi} J.~F. {et~al.}, 2006, \apj, 651, 61

\bibitem[{{Henriques} {et~al}\mbox{.}(2015){Henriques}, {White}, {Thomas},
  {Angulo}, {Guo}, {Lemson}, {Springel}, \& {Overzier}}]{Henriques:15}
{Henriques} B.~M.~B., {White} S.~D.~M., {Thomas} P.~A., {Angulo} R., {Guo} Q.,
  {Lemson} G., {Springel} V., {Overzier} R., 2015, \mnras, 451, 2663

\bibitem[{{Hopkins} {et~al}\mbox{.}(2011){Hopkins}, {Quataert}, \&
  {Murray}}]{Hopkins:11}
{Hopkins} P.~F., {Quataert} E., {Murray} N., 2011, \mnras, 417, 950

\bibitem[{{Hopkins} {et~al}\mbox{.}(2012{\natexlab{a}}){Hopkins}, {Quataert},
  \& {Murray}}]{Hopkins:12}
{Hopkins} P.~F., {Quataert} E., {Murray} N., 2012{\natexlab{a}}, \mnras, 421,
  3522

\bibitem[{{Hopkins} {et~al}\mbox{.}(2012{\natexlab{b}}){Hopkins}, {Quataert},
  \& {Murray}}]{Hopkins:12a}
{Hopkins} P.~F., {Quataert} E., {Murray} N., 2012{\natexlab{b}}, \mnras, 421,
  3488

\bibitem[{{Hopkins} {et~al}\mbox{.}(2016){Hopkins}, {Torrey},
  {Faucher-Gigu{\`e}re}, {Quataert}, \& {Murray}}]{Hopkins:16}
{Hopkins} P.~F., {Torrey} P., {Faucher-Gigu{\`e}re} C.-A., {Quataert} E.,
  {Murray} N., 2016, \mnras, 458, 816

\bibitem[{{Husemann} {et~al}\mbox{.}(2016){Husemann}, {Scharw{\"a}chter},
  {Bennert}, {Mainieri}, {Woo}, \& {Kakkad}}]{Husemann:16}
{Husemann} B., {Scharw{\"a}chter} J., {Bennert} V.~N., {Mainieri} V., {Woo}
  J.-H., {Kakkad} D., 2016, \aap, 594, A44

\bibitem[{{Ishibashi} \& {Fabian}(2015)}]{Ishibashi:15}
{Ishibashi} W., {Fabian} A.~C., 2015, \mnras, 451, 4612

\bibitem[{{Ishibashi} \& {Fabian}(2016{\natexlab{a}})}]{Ishibashi:16}
{Ishibashi} W., {Fabian} A.~C., 2016{\natexlab{a}}, \mnras, 463, 1291

\bibitem[{{Ishibashi} \& {Fabian}(2016{\natexlab{b}})}]{Ishibashi:16a}
{Ishibashi} W., {Fabian} A.~C., 2016{\natexlab{b}}, \mnras, 457, 2864

\bibitem[{{Ishiki} \& {Okamoto}(2017)}]{Ishiki:16}
{Ishiki} S., {Okamoto} T., 2017, \mnras, 466, L123

\bibitem[{{Kaastra} {et~al}\mbox{.}(2000){Kaastra}, {Mewe}, {Liedahl},
  {Komossa}, \& {Brinkman}}]{Kaastra:00}
{Kaastra} J.~S., {Mewe} R., {Liedahl} D.~A., {Komossa} S., {Brinkman} A.~C.,
  2000, \aap, 354, L83

\bibitem[{{Kannan} {et~al}\mbox{.}(2017){Kannan}, {Vogelsberger}, {Pfrommer},
  {Weinberger}, {Springel}, {Hernquist}, {Puchwein}, \& {Pakmor}}]{Kannan:17}
{Kannan} R., {Vogelsberger} M., {Pfrommer} C., {Weinberger} R., {Springel} V.,
  {Hernquist} L., {Puchwein} E., {Pakmor} R., 2017, \apjl, 837, L18

\bibitem[{{Kauffmann} \& {Haehnelt}(2000)}]{Kauffmann:00}
{Kauffmann} G., {Haehnelt} M., 2000, \mnras, 311, 576

\bibitem[{{Khandai} {et~al}\mbox{.}(2015){Khandai}, {Di Matteo}, {Croft},
  {Wilkins}, {Feng}, {Tucker}, {DeGraf}, \& {Liu}}]{Khandai:15}
{Khandai} N., {Di Matteo} T., {Croft} R., {Wilkins} S., {Feng} Y., {Tucker} E.,
  {DeGraf} C., {Liu} M.-S., 2015, \mnras, 450, 1349

\bibitem[{{Kim} {et~al}\mbox{.}(2011){Kim}, {Wise}, {Alvarez}, \&
  {Abel}}]{Kim:11}
{Kim} J.-h., {Wise} J.~H., {Alvarez} M.~A., {Abel} T., 2011, \apj, 738, 54

\bibitem[{{King}(2003)}]{King:03}
{King} A., 2003, \apjl, 596, L27

\bibitem[{{King}(2005)}]{King:05}
{King} A., 2005, \apjl, 635, L121

\bibitem[{{King} \& {Pounds}(2014)}]{King:14}
{King} A.~R., {Pounds} K.~A., 2014, \mnras, 437, L81

\bibitem[{{Kormendy} \& {Ho}(2013)}]{Kormendy:13}
{Kormendy} J., {Ho} L.~C., 2013, \araa, 51, 511

\bibitem[{{Krumholz} \& {Matzner}(2009)}]{Krumholz:09}
{Krumholz} M.~R., {Matzner} C.~D., 2009, \apj, 703, 1352

\bibitem[{{Krumholz} \& {Thompson}(2012)}]{Krumholz:12}
{Krumholz} M.~R., {Thompson} T.~A., 2012, \apj, 760, 155

\bibitem[{{Krumholz} \& {Thompson}(2013)}]{Krumholz:13}
{Krumholz} M.~R., {Thompson} T.~A., 2013, \mnras, 434, 2329

\bibitem[{{Larkin} \& {McLaughlin}(2016)}]{Larkin:16}
{Larkin} A.~C., {McLaughlin} D.~E., 2016, \mnras, 462, 1864

\bibitem[{{Li} \& {Bryan}(2014)}]{Li:14}
{Li} Y., {Bryan} G.~L., 2014, \apj, 789, 54

\bibitem[{{Liu} {et~al}\mbox{.}(2013){Liu}, {Zakamska}, {Greene}, {Nesvadba},
  \& {Liu}}]{Liu:13}
{Liu} G., {Zakamska} N.~L., {Greene} J.~E., {Nesvadba} N.~P.~H., {Liu} X.,
  2013, \mnras, 430, 2327

\bibitem[{{Martizzi} {et~al}\mbox{.}(2012){Martizzi}, {Teyssier}, {Moore}, \&
  {Wentz}}]{Martizzi:12}
{Martizzi} D., {Teyssier} R., {Moore} B., {Wentz} T., 2012, \mnras, 422, 3081

\bibitem[{{McCarthy} {et~al}\mbox{.}(2010){McCarthy}, {Schaye}, {Ponman},
  {Bower}, {Booth}, {Dalla Vecchia}, {Crain}, {Springel}, {Theuns}, \&
  {Wiersma}}]{McCarthy:10}
{McCarthy} I.~G. {et~al.}, 2010, \mnras, 406, 822

\bibitem[{{McConnell} \& {Ma}(2013)}]{McConnell:13}
{McConnell} N.~J., {Ma} C.-P., 2013, \apj, 764, 184

\bibitem[{{McCourt} {et~al}\mbox{.}(2016){McCourt}, {Oh}, {O'Leary}, \&
  {Madigan}}]{McCourt:16}
{McCourt} M., {Oh} S.~P., {O'Leary} R.~M., {Madigan} A.-M., 2016, \mnras
  \,(submitted) \, arXiv:1610.01164

\bibitem[{{McCourt} {et~al}\mbox{.}(2015){McCourt}, {O'Leary}, {Madigan}, \&
  {Quataert}}]{McCourt:15}
{McCourt} M., {O'Leary} R.~M., {Madigan} A.-M., {Quataert} E., 2015, \mnras,
  449, 2

\bibitem[{{McKernan} {et~al}\mbox{.}(2007){McKernan}, {Yaqoob}, \&
  {Reynolds}}]{McKernan:07}
{McKernan} B., {Yaqoob} T., {Reynolds} C.~S., 2007, \mnras, 379, 1359

\bibitem[{{McQuillin} \& {McLaughlin}(2012)}]{McQuillin:12}
{McQuillin} R.~C., {McLaughlin} D.~E., 2012, \mnras, 423, 2162

\bibitem[{{Morganti} {et~al}\mbox{.}(2015){Morganti}, {Oosterloo}, {Oonk},
  {Frieswijk}, \& {Tadhunter}}]{Morganti:15}
{Morganti} R., {Oosterloo} T., {Oonk} J.~B.~R., {Frieswijk} W., {Tadhunter} C.,
  2015, \aap, 580, A1

\bibitem[{{Murray} {et~al}\mbox{.}(2005){Murray}, {Quataert}, \&
  {Thompson}}]{Murray:05}
{Murray} N., {Quataert} E., {Thompson} T.~A., 2005, \apj, 618, 569

\bibitem[{{Namekata} \& {Umemura}(2016)}]{Namekata:16}
{Namekata} D., {Umemura} M., 2016, \mnras, 460, 980

\bibitem[{{Navarro} {et~al}\mbox{.}(1997){Navarro}, {Frenk}, \&
  {White}}]{Navarro:97}
{Navarro} J.~F., {Frenk} C.~S., {White} S.~D.~M., 1997, \apj, 490, 493

\bibitem[{{Novak} {et~al}\mbox{.}(2012){Novak}, {Ostriker}, \&
  {Ciotti}}]{Novak:12}
{Novak} G.~S., {Ostriker} J.~P., {Ciotti} L., 2012, \mnras, 427, 2734

\bibitem[{{Owocki} {et~al}\mbox{.}(2004){Owocki}, {Gayley}, \&
  {Shaviv}}]{Owocki:04}
{Owocki} S.~P., {Gayley} K.~G., {Shaviv} N.~J., 2004, \apj, 616, 525

\bibitem[{{Pfrommer} {et~al}\mbox{.}(2017){Pfrommer}, {Pakmor}, {Schaal},
  {Simpson}, \& {Springel}}]{Pfrommer:17}
{Pfrommer} C., {Pakmor} R., {Schaal} K., {Simpson} C.~M., {Springel} V., 2017,
  \mnras, 465, 4500

\bibitem[{{Pounds} {et~al}\mbox{.}(2003){Pounds}, {Reeves}, {King}, {Page},
  {O'Brien}, \& {Turner}}]{Pounds:03}
{Pounds} K.~A., {Reeves} J.~N., {King} A.~R., {Page} K.~L., {O'Brien} P.~T.,
  {Turner} M.~J.~L., 2003, \mnras, 345, 705

\bibitem[{{Prochaska} {et~al}\mbox{.}(2014){Prochaska}, {Lau}, \&
  {Hennawi}}]{Prochaska:14}
{Prochaska} J.~X., {Lau} M.~W., {Hennawi} J.~F., 2014, \apj, 796, 140

\bibitem[{{Proga} \& {Kallman}(2004)}]{Proga:04}
{Proga} D., {Kallman} T.~R., 2004, \apj, 616, 688

\bibitem[{{Reeves} {et~al}\mbox{.}(2009){Reeves}, {O'Brien}, {Braito}, {Behar},
  {Miller}, {Turner}, {Fabian}, {Kaspi}, {Mushotzky}, \& {Ward}}]{Reeves:09}
{Reeves} J.~N. {et~al.}, 2009, \apj, 701, 493

\bibitem[{{Reeves} {et~al}\mbox{.}(2013){Reeves}, {Porquet}, {Braito},
  {Gofford}, {Nardini}, {Turner}, {Crenshaw}, \& {Kraemer}}]{Reeves:13}
{Reeves} J.~N., {Porquet} D., {Braito} V., {Gofford} J., {Nardini} E., {Turner}
  T.~J., {Crenshaw} D.~M., {Kraemer} S.~B., 2013, \apj, 776, 99

\bibitem[{{Rosdahl} {et~al}\mbox{.}(2013){Rosdahl}, {Blaizot}, {Aubert},
  {Stranex}, \& {Teyssier}}]{Rosdahl:13}
{Rosdahl} J., {Blaizot} J., {Aubert} D., {Stranex} T., {Teyssier} R., 2013,
  \mnras, 436, 2188

\bibitem[{{Rosdahl} {et~al}\mbox{.}(2015){Rosdahl}, {Schaye}, {Teyssier}, \&
  {Agertz}}]{Rosdahl:15b}
{Rosdahl} J., {Schaye} J., {Teyssier} R., {Agertz} O., 2015, \mnras, 451, 34

\bibitem[{{Rosdahl} \& {Teyssier}(2015)}]{Rosdahl:15a}
{Rosdahl} J., {Teyssier} R., 2015, \mnras, 449, 4380

\bibitem[{{Rosen} \& {Bregman}(1995)}]{Rosen:95}
{Rosen} A., {Bregman} J.~N., 1995, \apj, 440, 634

\bibitem[{{Roth} {et~al}\mbox{.}(2012){Roth}, {Kasen}, {Hopkins}, \&
  {Quataert}}]{Roth:12}
{Roth} N., {Kasen} D., {Hopkins} P.~F., {Quataert} E., 2012, \apj, 759, 36

\bibitem[{{Rupke} \& {Veilleux}(2013)}]{Rupke:13}
{Rupke} D.~S.~N., {Veilleux} S., 2013, \apj, 768, 75

\bibitem[{{Sazonov} {et~al}\mbox{.}(2005){Sazonov}, {Ostriker}, {Ciotti}, \&
  {Sunyaev}}]{Sazonov:05}
{Sazonov} S.~Y., {Ostriker} J.~P., {Ciotti} L., {Sunyaev} R.~A., 2005, \mnras,
  358, 168

\bibitem[{{Scannapieco} \& {Oh}(2004)}]{Scannapieco:04}
{Scannapieco} E., {Oh} S.~P., 2004, \apj, 608, 62

\bibitem[{{Schawinski} {et~al}\mbox{.}(2015){Schawinski}, {Koss}, {Berney}, \&
  {Sartori}}]{Schawinski:15}
{Schawinski} K., {Koss} M., {Berney} S., {Sartori} L.~F., 2015, \mnras, 451,
  2517

\bibitem[{{Schaye} {et~al}\mbox{.}(2015){Schaye}, {Crain}, {Bower}, {Furlong},
  {Schaller}, {Theuns}, {Dalla Vecchia}, {Frenk}, {McCarthy}, {Helly},
  {Jenkins}, {Rosas-Guevara}, {White}, {Baes}, {Booth}, {Camps}, {Navarro},
  {Qu}, {Rahmati}, {Sawala}, {Thomas}, \& {Trayford}}]{Schaye:15}
{Schaye} J. {et~al.}, 2015, \mnras, 446, 521

\bibitem[{{Scoville} {et~al}\mbox{.}(2017){Scoville}, {Murchikova}, {Walter},
  {Vlahakis}, {Koda}, {Vanden Bout}, {Barnes}, {Hernquist}, {Sheth}, {Yun},
  {Sanders}, {Armus}, {Cox}, {Thompson}, {Robertson}, {Zschaechner}, {Tacconi},
  {Torrey}, {Hayward}, {Genzel}, {Hopkins}, {van der Werf}, \&
  {Decarli}}]{Scoville:17}
{Scoville} N. {et~al.}, 2017, \apj, 836, 66

\bibitem[{{Shankar} {et~al}\mbox{.}(2016){Shankar}, {Bernardi}, {Sheth},
  {Ferrarese}, {Graham}, {Savorgnan}, {Allevato}, {Marconi}, {L{\"a}sker}, \&
  {Lapi}}]{Shankar:16}
{Shankar} F. {et~al.}, 2016, \mnras, 460, 3119

\bibitem[{{Sijacki} \& {Springel}(2006)}]{Sijacki:06}
{Sijacki} D., {Springel} V., 2006, \mnras, 366, 397

\bibitem[{{Sijacki} {et~al}\mbox{.}(2007){Sijacki}, {Springel}, {Di Matteo}, \&
  {Hernquist}}]{Sijacki:07}
{Sijacki} D., {Springel} V., {Di Matteo} T., {Hernquist} L., 2007, \mnras, 380,
  877

\bibitem[{{Sijacki} {et~al}\mbox{.}(2015){Sijacki}, {Vogelsberger}, {Genel},
  {Springel}, {Torrey}, {Snyder}, {Nelson}, \& {Hernquist}}]{Sijacki:15}
{Sijacki} D., {Vogelsberger} M., {Genel} S., {Springel} V., {Torrey} P.,
  {Snyder} G.~F., {Nelson} D., {Hernquist} L., 2015, \mnras, 452, 575

\bibitem[{{Silk} \& {Nusser}(2010)}]{Silk:10}
{Silk} J., {Nusser} A., 2010, \apj, 725, 556

\bibitem[{{Silk} \& {Rees}(1998)}]{Silk:98}
{Silk} J., {Rees} M.~J., 1998, \aap, 331, L1

\bibitem[{{Springel} {et~al}\mbox{.}(2005b){Springel}, {Di Matteo}, \&
  {Hernquist}}]{Springel:05c}
{Springel} V., {Di Matteo} T., {Hernquist} L., 2005b, \mnras, 361, 776

\bibitem[{{Straatman} {et~al}\mbox{.}(2014){Straatman}, {Labb{\'e}}, {Spitler},
  {Allen}, {Altieri}, {Brammer}, {Dickinson}, {van Dokkum}, {Inami},
  {Glazebrook}, {Kacprzak}, {Kawinwanichakij}, {Kelson}, {McCarthy},
  {Mehrtens}, {Monson}, {Murphy}, {Papovich}, {Persson}, {Quadri}, {Rees},
  {Tomczak}, {Tran}, \& {Tilvi}}]{Straatman:14}
{Straatman} C.~M.~S. {et~al.}, 2014, \apjl, 783, L14

\bibitem[{{Sturm} {et~al}\mbox{.}(2011){Sturm}, {Gonz{\'a}lez-Alfonso},
  {Veilleux}, {Fischer}, {Graci{\'a}-Carpio}, {Hailey-Dunsheath}, {Contursi},
  {Poglitsch}, {Sternberg}, {Davies}, {Genzel}, {Lutz}, {Tacconi}, {Verma},
  {Maiolino}, \& {de Jong}}]{Sturm:11}
{Sturm} E. {et~al.}, 2011, \apjl, 733, L16

\bibitem[{{Szomoru} {et~al}\mbox{.}(2012){Szomoru}, {Franx}, \& {van
  Dokkum}}]{Szomoru:12}
{Szomoru} D., {Franx} M., {van Dokkum} P.~G., 2012, \apj, 749, 121

\bibitem[{{Tadhunter} {et~al}\mbox{.}(2014){Tadhunter}, {Morganti}, {Rose},
  {Oonk}, \& {Oosterloo}}]{Tadhunter:14}
{Tadhunter} C., {Morganti} R., {Rose} M., {Oonk} J.~B.~R., {Oosterloo} T.,
  2014, \nat, 511, 440

\bibitem[{{Teyssier}(2002)}]{Teyssier:02}
{Teyssier} R., 2002, \aap, 385, 337

\bibitem[{{Teyssier} {et~al}\mbox{.}(2011){Teyssier}, {Moore}, {Martizzi},
  {Dubois}, \& {Mayer}}]{Teyssier:11}
{Teyssier} R., {Moore} B., {Martizzi} D., {Dubois} Y., {Mayer} L., 2011,
  \mnras, 414, 195

\bibitem[{{Thompson} {et~al}\mbox{.}(2015){Thompson}, {Fabian}, {Quataert}, \&
  {Murray}}]{Thompson:15}
{Thompson} T.~A., {Fabian} A.~C., {Quataert} E., {Murray} N., 2015, \mnras,
  449, 147

\bibitem[{{Toft} {et~al}\mbox{.}(2007){Toft}, {van Dokkum}, {Franx}, {Labbe},
  {F{\"o}rster Schreiber}, {Wuyts}, {Webb}, {Rudnick}, {Zirm}, {Kriek}, {van
  der Werf}, {Blakeslee}, {Illingworth}, {Rix}, {Papovich}, \&
  {Moorwood}}]{Toft:07}
{Toft} S. {et~al.}, 2007, \apj, 671, 285

\bibitem[{{Tombesi} {et~al}\mbox{.}(2012){Tombesi}, {Cappi}, {Reeves}, \&
  {Braito}}]{Tombesi:12}
{Tombesi} F., {Cappi} M., {Reeves} J.~N., {Braito} V., 2012, \mnras, 422, L1

\bibitem[{{Tombesi} {et~al}\mbox{.}(2015){Tombesi}, {Mel{\'e}ndez}, {Veilleux},
  {Reeves}, {Gonz{\'a}lez-Alfonso}, \& {Reynolds}}]{Tombesi:15}
{Tombesi} F., {Mel{\'e}ndez} M., {Veilleux} S., {Reeves} J.~N.,
  {Gonz{\'a}lez-Alfonso} E., {Reynolds} C.~S., 2015, \nat, 519, 436

\bibitem[{{Tremaine} {et~al}\mbox{.}(2002){Tremaine}, {Gebhardt}, {Bender},
  {Bower}, {Dressler}, {Faber}, {Filippenko}, {Green}, {Grillmair}, {Ho},
  {Kormendy}, {Lauer}, {Magorrian}, {Pinkney}, \& {Richstone}}]{Tremaine:02}
{Tremaine} S. {et~al.}, 2002, \apj, 574, 740

\bibitem[{{Treu}(2010)}]{Treu:10}
{Treu} T., 2010, \araa, 48, 87

\bibitem[{{Tsang} \& {Milosavljevi{\'c}}(2015)}]{Tsang:15}
{Tsang} B.~T.-H., {Milosavljevi{\'c}} M., 2015, \mnras, 453, 1108

\bibitem[{{van den Bosch}(2016)}]{vandenBosch:16}
{van den Bosch} R.~C.~E., 2016, \apj, 831, 134

\bibitem[{{van Dokkum} {et~al}\mbox{.}(2014){van Dokkum}, {Bezanson}, {van der
  Wel}, {Nelson}, {Momcheva}, {Skelton}, {Whitaker}, {Brammer}, {Conroy},
  {F{\"o}rster Schreiber}, {Fumagalli}, {Kriek}, {Labb{\'e}}, {Leja},
  {Marchesini}, {Muzzin}, {Oesch}, \& {Wuyts}}]{vanDokkum:14}
{van Dokkum} P.~G. {et~al.}, 2014, \apj, 791, 45

\bibitem[{{Vernaleo} \& {Reynolds}(2006)}]{Vernaleo:06}
{Vernaleo} J.~C., {Reynolds} C.~S., 2006, \apj, 645, 83

\bibitem[{{Vogelsberger} {et~al}\mbox{.}(2013){Vogelsberger}, {Genel},
  {Sijacki}, {Torrey}, {Springel}, \& {Hernquist}}]{Vogelsberger:13}
{Vogelsberger} M., {Genel} S., {Sijacki} D., {Torrey} P., {Springel} V.,
  {Hernquist} L., 2013, \mnras, 436, 3031

\bibitem[{{Volonteri} {et~al}\mbox{.}(2016){Volonteri}, {Dubois}, {Pichon}, \&
  {Devriendt}}]{Volonteri:16}
{Volonteri} M., {Dubois} Y., {Pichon} C., {Devriendt} J., 2016, \mnras, 460,
  2979

\bibitem[{{Wagner} {et~al}\mbox{.}(2013){Wagner}, {Umemura}, \&
  {Bicknell}}]{Wagner:13}
{Wagner} A.~Y., {Umemura} M., {Bicknell} G.~V., 2013, \apjl, 763, L18

\bibitem[{{Weinberger} {et~al}\mbox{.}(2017){Weinberger}, {Springel},
  {Hernquist}, {Pillepich}, {Marinacci}, {Pakmor}, {Nelson}, {Genel},
  {Vogelsberger}, {Naiman}, \& {Torrey}}]{Weinberger:17}
{Weinberger} R. {et~al.}, 2017, \mnras, 465, 3291

\bibitem[{{Williams} {et~al}\mbox{.}(2017){Williams}, {Maiolino}, {Krongold},
  {Carniani}, {Cresci}, {Mannucci}, \& {Marconi}}]{WIlliams:17}
{Williams} R.~J., {Maiolino} R., {Krongold} Y., {Carniani} S., {Cresci} G.,
  {Mannucci} F., {Marconi} A., 2017, \mnras, 467, 3399

\bibitem[{{Wilson} {et~al}\mbox{.}(2014){Wilson}, {Rangwala}, {Glenn},
  {Maloney}, {Spinoglio}, \& {Pereira-Santaella}}]{Wilson:14}
{Wilson} C.~D., {Rangwala} N., {Glenn} J., {Maloney} P.~R., {Spinoglio} L.,
  {Pereira-Santaella} M., 2014, \apjl, 789, L36

\bibitem[{{Wyithe} \& {Loeb}(2003)}]{Wyithe:03}
{Wyithe} J.~S.~B., {Loeb} A., 2003, \apj, 595, 614

\bibitem[{{Xie} {et~al}\mbox{.}(2017){Xie}, {Yuan}, \& {Ho}}]{Xie:17}
{Xie} F.-G., {Yuan} F., {Ho} L.~C., 2017, \apj, 844, 42

\bibitem[{{Zakamska} {et~al}\mbox{.}(2016){Zakamska}, {Hamann}, {P{\^a}ris},
  {Brandt}, {Greene}, {Strauss}, {Villforth}, {Wylezalek}, {Alexandroff}, \&
  {Ross}}]{Zakamska:16}
{Zakamska} N.~L. {et~al.}, 2016, \mnras, 459, 3144

\bibitem[{{Zhang} \& {Davis}(2017)}]{Zhang:16}
{Zhang} D., {Davis} S.~W., 2017, \apj, 839, 54

\bibitem[{{Zubovas} \& {King}(2012)}]{Zubovas:12}
{Zubovas} K., {King} A., 2012, \apjl, 745, L34

\bibitem[{{Zubovas} \& {King}(2014)}]{Zubovas:14}
{Zubovas} K., {King} A.~R., 2014, \mnras, 439, 400

\end{thebibliography}

\begin{figure*}
\centering
\includegraphics[scale = 0.40]{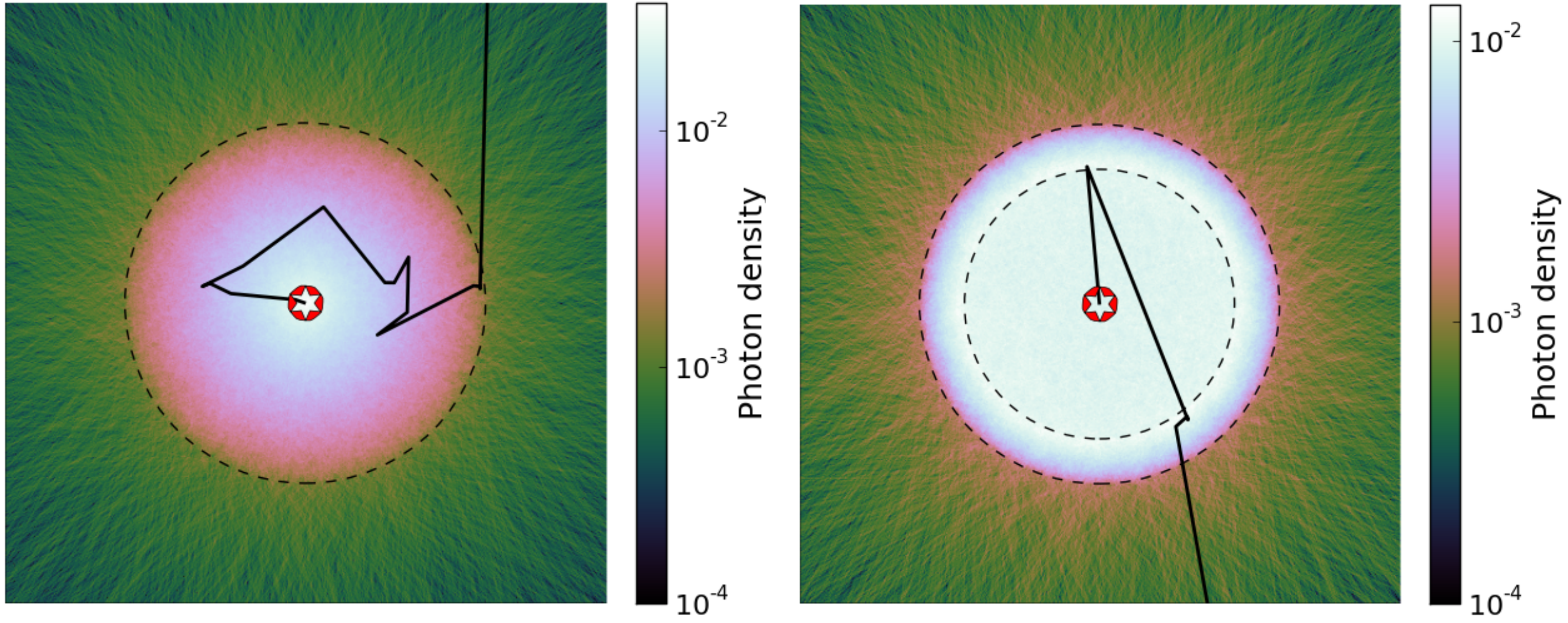}
\caption{Monte Carlo description of 2D transport of IR photons through a gas sphere (left-hand panel) and shell (right-hand panel). Dashed circles indicate the outer boundary of the sphere (left-hand panel) and the inner and outer boundaries of the shell (right-hand panel). The IR optical depth is $\tau_{\rm IR} \,=\, 5$ in both cases. Solid black lines indicate single Monte Carlo realisations and illustrate possible photon trajectories. In the case of a spherical configuration, IR photons undergo a random walk until they escape.  In the case of a shell, the photons stream freely within the optically thin cavity bounded by the inner surface. When inside the shell, photons travel through diffusion. In colour, we show the probability for a given cell to be visited by a photon, obtained by stacking 30,000 possible photon paths.}
\label{fig_diffusion_schematic}
\end{figure*}

\begin{figure}
\centering
\includegraphics[scale = 0.45]{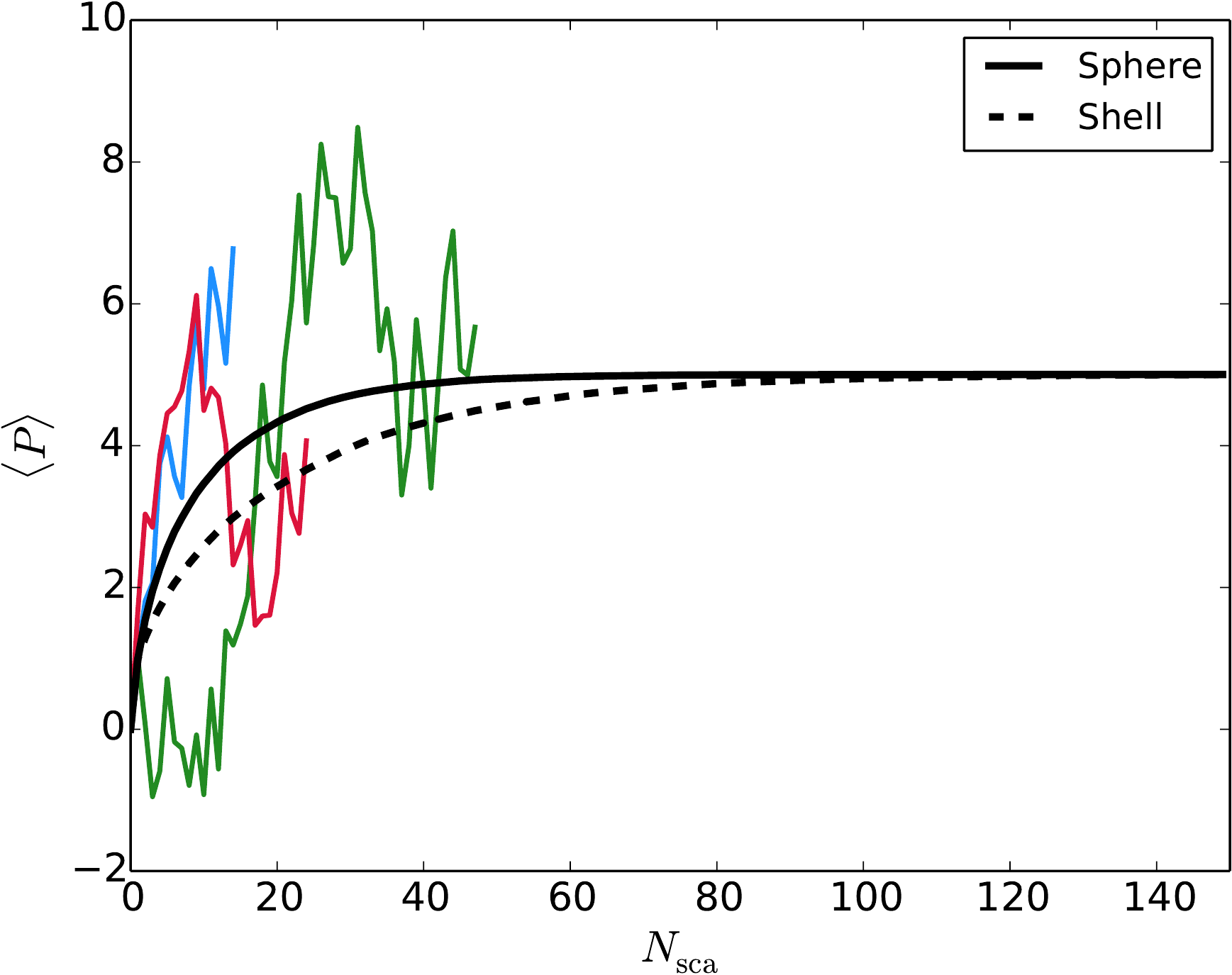}
\caption{Cumulative momentum imparted after a number of scatterings $N_{\rm sca}$ for a sphere (solid black line) and shell (dashed black line) with $\tau_{\rm IR} \, =\, 5$ averaged over 30,000 Monte Carlo realisations. The momentum per photon converges to $\tau_{\rm IR}$, as expected. Each photon is assumed to have momentum of $p_{\rm ph} \, = \, 1$. The coloured lines illustrate net cumulative momenta imparted by three individual photons in the case of a shell geometry. Lines are truncated at the time at which photons escape the sphere.}
\label{fig_diffusion_momentum}
\end{figure}

\appendix
\section{IR radiation pressure in Monte Carlo simulations}
\label{sec_appendix1}

In this Appendix, we review the transport of IR photons through optically thick gas and build an intuition of how this process leads to a net outward force.
Individual IR photons must scatter multiple times, undergoing a random walk, before they can escape the optically thick region. 
The left-hand panel of Fig.~\ref{fig_diffusion_schematic} provides a schematic illustration of the paths traversed by IR photons as they diffuse through a 2D optically thick ($\tau_{\rm IR} \,=\, 5$) sphere (left-hand side) and shell (right-hand side). Both sphere and shell are assumed to have an outer radius of $30 \%$ of the box width with the shell having an inner radius equal to $0.75$ times the outer radius, i.e. $22. 5\%$ the box size.
The problem is, otherwise, scale-free.

In order to compute the paths of the IR photons, we develop a simple Monte Carlo algorithm; the density field is discretised on a $2048 \times 2048$ Cartesian grid and, at a given step, (1) the optical depth $\tau_{\rm IR}$ travelled by the photon is drawn from an exponential distribution function $e^{-\tau_{\rm IR}}$, (2) the photon is emitted into a random direction and advanced by one grid cell. If the optical depth traversed by the photon exceeds $\tau_{\rm IR}$, the photon is absorbed. Otherwise, the photon is displaced to another cell along its original direction. The process is repeated until the photon escapes the volume. 

The black lines on panels in Fig.~\ref{fig_diffusion_schematic} show one possible photon path realisation in both sphere and shell configurations. 
In the case of the sphere, the photon undergoes the classical random-walk; it scatters multiple times until it reaches its outskirts and finally escapes.
In the case of the shell, IR photons may also free-stream within the optically thin cavity.
We then use a stack of 30,000 such IR photon trajectories and compute the number of times each given cell is visited by a photon. The coloured field shown in both panels in Fig.~\ref{fig_diffusion_schematic} shows the resulting density field.
The photon density is highest at the centre of the sphere and close to the inner boundary of the shell, where radiation is most efficiently trapped.
Note that in the latter, the photon density field is homogeneous within the central hollow space, which contains IR photons that were absorbed inside the shell but scattered in an inward (random) direction. A similar result, though using simulations performed with the moment method, is shown in Fig.~\ref{fig_shellsolutions}.

In the case of the sphere, where IR photon transport is a pure random walk up until the photon reaches the boundary, we expect that the mean displacement is $\langle d \rangle \propto \sqrt{N_{\rm sca}} l_{\rm mfp}$, where $N_{\rm sca}$ is the number of scattering events and $l_{\rm mfp}$ is the mean free path of the IR photon. 
Clearly a number of scatterings $\langle N_{\rm sca} \rangle \sim (R / l_{\rm mfp})^2 \sim \tau_{\rm IR}^2$ are required before the photons escape the sphere.
As the trapped IR radiation diffuses through the optically thick medium, it transfers momentum to the gas.
In the case of a spherical configuration, the cumulative momentum $P$ imparted by a single photon, with momentum $p_{\rm ph}$ also follows a random walk.
In particular we can write
\begin{equation}
P \, \approx \, \sqrt{N_{\rm sca}} p_{\rm ph}  \, = \, \tau_{\rm IR} p_{\rm ph} \, ,
\end{equation}
since $\sim \tau_{\rm IR}^2$ scatters are, on average, required for escape. 
Since it propagates in a random direction, part of the imparted momentum ultimately cancels out and the net momentum transferred is certainly less than the number of scatters times $p_{\rm ph}$.
This is a common point of confusion since it is often \emph{incorrectly} stated that IR photons scatter $\tau_{\rm IR}$ times, as opposed to $\tau_{\rm IR}^2$. 

We verify these basic expectations using our Monte Carlo code.
In Fig.~\ref{fig_diffusion_momentum}, we show the average cumulative momentum imparted to the optically thick sphere (solid line) and shell (dashed line) after a number of scatterings $N_{\rm sca}$.
The average was taken over 30,000 Monte Carlo realisations.
The coloured lines show examples of the momentum imparted by individual photons in the case of a shell geometry.
Despite strong fluctuations in the cumulative momentum imparted, the net momentum transferred by a large number of photons increases steadily with time \citep[cf.][]{Gayley:95}, as expected.
In both shell and sphere cases, the average momentum converges to $\tau_{\rm IR} \, = \, 5$.
The main impact of geometry appears to be the time-to-convergence, which is somewhat slower in the case of a shell configuration, likely due to the fact that the process of diffusion has to be started from scratch every time a photon leaks into the inner optically thin cavity.

\end{document}